\documentclass[11pt]{article}
\usepackage{amsfonts,amssymb,amsmath}
\usepackage{slashed}
\usepackage{color}
\usepackage{pdfsync}
\usepackage{accents}
\usepackage{yfonts}   
\usepackage{cancel}
\usepackage{soul}
\usepackage{verbatim}
\newcommand{\C}{\mathbb C}

\renewcommand{\H}{\mathbb H}
\newcommand{\N}{\mathbb N}
\newcommand{\M}{\mathbb M}

\newcommand{\R}{\mathbb R}
\newcommand{\Z}{\mathbb Z}
\newcommand{\T}{\mathbb T}
\newcommand{\e}{\boldsymbol e}
\newcommand{\Q}{\boldsymbol Q}
\newcommand{\eps}{\varepsilon}
\newcommand{\HHH}{\mathcal H}
\newcommand{\SSS}{\mathcal S}

\newcommand\mb[1]{\mbox{\rm#1}}
\newcommand{\tr}{\mbox{tr}}
\newcommand{\ds}{\displaystyle}
\newcommand{\qu}{\overline}
\newcommand\qqu[1]{\accentset{\rule{.4em}{.8pt}}{#1}}
\newcommand{\nop[1]}{\mathop{:\!#1\!\!:}}
\newcommand{\be}{\begin{equation}}
\newcommand{\ee}{\end{equation}}
\definecolor{purple}{rgb}{0.6,0,0.6}
\definecolor{bl}{rgb}{0.2,0.1,0.9}
\definecolor{tur}{rgb}{0.1,0.9,0.9}
\definecolor{gr}{rgb}{0.2,0.7,0.2}
\definecolor{pink}{rgb}{1,0,0.8}

\newcounter{save}
\newcounter{rmk}

\numberwithin{equation}{section}
\newcommand\Dtorus{\operatorname{D_4-torus}}

\title{\mbox{}\hfill {\small DCPT-16/45}\\[30pt]
The Conway Moonshine Module is a Reflected K3 Theory}
\vspace{ -1cm}
\author{\\ \Large{Anne Taormina\footnote{anne.taormina@durham.ac.uk}\;\; and Katrin Wendland\footnote{katrin.wendland@math.uni-freiburg.de}}\\ [5pt]
 \normalsize{$^*$Centre for Particle Theory, Department of Mathematical Sciences} \\ \normalsize{University of Durham, Durham, DH1 3LE, U.K. }\\[5pt]
 \normalsize{$^{\dagger}$Mathematics Institute, University of Freiburg}\\
 \normalsize{D-79104 Freiburg, Germany.  }}
 \date{\vspace{-5ex}}
 
\begin{document}
 
\maketitle
\begin{center}
{\small {\sc{subject class}: [2010]{Primary 81T40, Secondary  11F22, 14J28, 17B69, 20C35}}}
\end{center}
\setlength{\parindent}{20pt}
\parskip 0.10cm
\begin{abstract}
\noindent Recently, Duncan and Mack-Crane established an isomorphism, 
as Virasoro modules at central charges $c=12$, between the space 
of states of the Conway Moonshine Module and the space of states 
 of  a special 
K3 theory that was extensively studied 
some time ago by Gaberdiel, Volpato and 
the two authors. 
In the present work, we lift this result to the level of modules of 
the extensions of  these Virasoro algebras
to $N=4$ super Virasoro algebras. Moreover,
we relate the super vertex operator algebra and module structure of the
Conway Moonshine Module to the operator product expansion
of this special K3 theory by a procedure  we call {\em reflection}. 
This procedure can be applied to certain superconformal 
field theories, transforming all fields to holomorphic
ones.  It also allows to describe  certain
superconformal field theories within the language
of super vertex operator algebras.
We discuss  reflection  and its limitations in 
general, and we argue that through reflection, the Conway Moonshine
Module inherits from the K3 theory a richer structure  
than anticipated so far. 
The comparison between the 
Conway Moonshine Module and the K3 theory is considerably 
facilitated by exploiting the free fermion description as well
as the lattice vertex operator algebra description of both theories.
We include an explicit construction of cocycles for the relevant
charge lattices, which are half integral. 
The transition from 
the K3 theory to the 
Conway Moonshine Module via reflection promotes the latter 
to the role of a medium that  collects 
the symmetries of  K3 
theories  from distinct points of the 
moduli space, thus uncovering a version of symmetry surfing in this context.
\end{abstract}
\clearpage

\tableofcontents
\section*{Introduction}
Very few would have anticipated in 1978 that the innocuous decomposition 
$196884=196883+1$ would spark so much interest in the mathematics 
community, culminating with the award of a Fields medal to Borcherds twenty 
years later. The defining moment was John McKay's observation that 
$196883$ is the dimension of the smallest non trivial irreducible 
representation of the Monster group $\mathbb{M}$, and that $196884$ 
is the coefficient of the linear term 
in the Fourier expansion of the 
$j$-function, a Hauptmodul for ${\rm SL}(2,\Z)$. What had appeared to be a mere coincidence 
at first turned out to be part of an intriguing pattern. Not only did all the Fourier 
coefficients of the $j$-function, bar the constant term, coincide with 
dimensions of representations of $\mathbb{M}$, exhibiting $j$ as a graded
dimension of some $\M$-module $V^\natural$, but 
{\em all} the so-called McKay-Thompson 
series \cite{th79b}, which are graded characters for  arbitrary 
elements  of 
$\mathbb{M}$, are themselves Hauptmoduln of 
genus zero subgroups $\Gamma \subset {\rm SL}(2,\R)$. This  
phenomenon was
coined {\em Monstrous Moonshine} by Conway and Norton. 
The connections between the largest sporadic group 
and modular functions have deep roots and manifest themselves 
in string theory for example. To this day though, and despite the 
beautiful construction of the Monster Module by Frenkel, 
Lepowsky and Meurman \cite{flm84} using conformal field theory 
techniques, as well as its realization in terms of superstrings
\cite{dgh88a}, there has been little use of these connections in string theory.

A game changing event took place in 2010, when Eguchi, Ooguri and 
Tachikawa \cite{eot10} noticed an intriguing connection between 
the elliptic 
genus $\mathcal{E}_{\rm K3}$  of  K3, and the 
sporadic group $M_{24}$. The arena is that of closed superstrings 
propagating on K3 surfaces, where the existence of a 
worldsheet $N=(4,4)$ superconformal symmetry is well established 
and intimately related to the hyperk\"ahler structure of K3. One of us, in collaboration with
Eguchi, Ooguri and Yang, used 
techniques pioneered by Witten to 
calculate the elliptic genus of  K3 exploiting the structure of string 
partition functions based on orbifolds and Gepner models in \cite{eoty89}. Such 
partition functions are expressed in terms of massless and massive, 
unitary irreducible characters of the $N=4$ superconformal algebra at 
central charge $c=6$ \cite{egta87,egta88a,egta88b}, and it follows 
that  $\mathcal{E}_{\rm K3}$ may be decomposed into an infinite sum of such 
characters. The coefficients of all massive $N=4$ characters appearing 
in that sum  were conjectured in \cite{eot10} to be the dimensions 
of representations of $M_{24}$. 
In fact, an imprint of this relation had been anticipated previously
by Govindarajan and Krishna \cite{gokr09,go11} in their studies of Borcherds-Kac-Moody Lie
superalgebras obtained from dyon spectra in $\Z_N$-CHL orbifolds.
The conjecture became more specific after the work of 
Cheng \cite{ch10}, Gaberdiel, Hohenegger and Volpato \cite{ghv10a,ghv10b},
and Eguchi and Hikami \cite{eghi11}, who determined the
expected twining genera.
This {\em Mathieu Moonshine} was 
mathematically proven by Gannon \cite{ga12}, but his approach 
does not shed much light on the
role of $M_{24}$ in string theory. This remains an open question, 
which in our eyes, is well worth studying. Indeed, investigations so far suggest that a full understanding of an 
$M_{24}$ action in this context requires new conceptual thinking.

There have been several lines of attack to probe Mathieu Moonshine. 
As mentioned above, shortly after the observation by Eguchi, Ooguri and Tachikawa,
and building on Thompson's idea of twist \cite{th79b}, several 
groups constructed twining elliptic genera that were proven 
to be graded characters of an infinite-dimensional $M_{24}$-module \cite{ga12}. 
In another development, a new family of moonshines was 
discovered, of which Mathieu Moonshine is a member. Dubbed 
{\em Umbral Moonshine}, this family of connections between certain mock 
modular forms and automorphism groups of Niemeier lattices 
has opened the door to some 
fruitful collaborations that bridge mathematics and theoretical 
physics \cite{cdh12,cdh14,dgo15}. 
In the meantime,  we  investigated
the geometric symmetries of strings propagating on K3 surfaces of 
Kummer type. Using  lattice techniques to introduce the concept of 
symmetry surfing the moduli space of  K3 theories, we showed
how the overarching  group $\Z_2^4 \colon A_8$ emerges 
from {\em symmetry surfing} \cite{tawe11,tawe13}. 
The basic idea, at this level, is an application of Kondo's beautiful 
strategy of proof \cite{ko98} of Mukai's seminal
classification result \cite{mu88}
for symplectic automorphisms of K3 surfaces:
the lattice of integral cohomology of K3, which enters crucially in 
the construction of the moduli space of K3 theories \cite{asmo94,nawe00},
is replaced by an even, self-dual lattice of the same rank, which thereby
serves as a medium to collect symmetries from distinct points of the moduli
space. To move beyond  generating
the relevant groups, in order  to construct
the expected representations, one needs to leave the comfort zone of
lattice techniques. 
In the 
context of $\Z_2$-orbifolds of toroidal conformal field theories, this has
been achieved in \cite{tawe12,gakepa16}, providing further evidence in
favour of the idea of symmetry surfing. 

Our present investigation picks up more recent
efforts to establish connections between 
certain conformal field theories and certain
super vertex operator algebras. Indeed, 
we aim
at clarifying the relationship between
the K3 theory studied in \cite{gtvw14} and
the Conway Moonshine Module of Duncan and 
Mack-Crane \cite{du07,dmc15}. As Virasoro modules at central
charge $c=12$, Duncan and Mack-Crane showed that the spaces
of states of the two models agree. However, while the former is
an $N=(4,4)$ superconformal field theory at 
central charges $c=\qu c=6$, the latter is a super
vertex operator algebra at central charge $c=12$, together
with its (unique, up to isomorphism, irreducible) canonically 
twisted supermodule.
We show that the two are related\footnote{up to
exchanging the roles of bosons and fermions in the Ramond 
sector of the Conway Moonshine Module, while
 accounting for an extra
factor of $(-1)$  introduced by hand in the Ramond sector partition
functions between
\cite[(8.7)]{dmc15} and \cite[(9.10), (9.14)]{dmc15}} by a procedure which we
call {\em reflection}. Only very 
special superconformal field theories
allow reflection, which transforms all fields to holomorphic ones.
The operator product expansion (OPE) of the K3 theory thereby induces
the super vertex operator algebra and admissible module structure
on the Conway Moonshine Module, and more, since in the K3 theory,
an OPE  is defined between {\em any} pair of fields.
The reflection procedure provides a bridge between conformal field
theory and vertex operator algebra techniques. This bridge may be
used in both directions, hopefully allowing some of the experts in 
vertex operator algebras to enter the world of K3 theories, or 
even more general superconformal field theories.

By our interpretation of the Conway Moonshine Module as
the image of a K3 theory under reflection, 
the modular properties of
the partition function and its  building
blocks receive a 
natural explanation from superstring theory. It would be interesting 
to know whether all the genus zero properties of Conway Moonshine
can be traced back to K3. On the level of lattices, reflection
is an implementation of the techniques mentioned above, 
where the K3 lattice with signature $(4,20)$ is replaced by an 
even, self-dual positive definite lattice of the same rank. This allows
us to reveal  the proposal of \cite{dmc15}, that is, the realisation of all symmetries
of K3 theories as automorphisms of the Conway Moonshine
Module, as an incarnation of symmetry surfing by means of 
lattice techniques \cite{tawe11,tawe13}.  It also means that 
we do not expect a 
construction of an $M_{24}$ vertex operator algebra 
that explains Mathieu Moonshine
from the
Conway Moonshine Module, since the latter
disregards the twist of \cite{tawe12,gakepa16}.

The work \cite{cdr17} by
Creutzig, Duncan and Riedler  complements ours, with some overlap
with our results. To clarify the relation between the Conway
Moonshine Module and the K3 theory studied in \cite{gtvw14}, they
introduce the notion of a {\em potential bulk SCFT}, which in our
language
amounts to the image of a SCFT under reflection, viewed
as a module of its chiral-antichiral algebra. They find sufficient conditions
for such a potential bulk SCFT to agree with (adequately) {\em nice}
super vertex operator algebras, and they provide examples
where these conditions hold. 
Given that the potential bulk SCFT obtained
from the above-mentioned K3 theory is among these examples, they have
in particular, independently from us, extended the identification with the 
Conway Moonshine Module to the level of modules of a supersymmetric
extension of the previously studied two copies of the Virasoro algebra. 
\vskip 0.5cm

The present work is organised as follows. In Section 1, we start by revisiting the K3
theory based on the $\Z_2$-orbifold of the $D_4$-torus 
theory, whose symmetry group $\Z_2^8\colon \M_{20}$ is  
one of the largest  symmetry groups of K3 theories preserving 
$N=(4,4)$ supersymmetry \cite{gtvw14}. As a K3 sigma model, this
theory is built on the tetrahedral Kummer surface studied in 
depth in \cite{tawe11,tawe12,tawe13}.
With a view to compare 
 this theory to the 
Conway Moonshine Module later on, we pay particular attention 
to its description in terms of twenty-four free Majorana fermions, twelve  
left- and twelve right-movers, and highlight its underlying affine current 
algebra\footnote{We use the 
notation $\widehat{\mathfrak{g}}_1^n=\widehat{\mathfrak{g}}_1^{\oplus n}$throughout.} 
$\left(\widehat{\mathfrak{so}}(4)_{1,L} \oplus \widehat{\mathfrak{so}}(4)_{1,R}\right)^3
\subset
\widehat{\mathfrak{so}}(8)_{1}^3$. One of the three summands 
$\widehat{\mathfrak{so}}(4)_{1,L} \oplus \widehat{\mathfrak{so}}(4)_{1,R}$ is 
the affine algebra arising from the  fermionic superpartners  of the four left 
and four right-moving $U(1)$ currents of the 
bosonic $D_4$-torus theory. We make a 
choice of a left- (resp. right-)moving $U(1)$ current whose zero mode 
generates the Cartan subalgebra of an affine 
$\widehat{\mathfrak{su}}(2)_{1,L} \subset  \widehat{\mathfrak{so}}(4)_{1,L}$ 
(resp. 
$\widehat{\mathfrak{su}}(2)_{1,R} \subset  \widehat{\mathfrak{so}}(4)_{1,R}$), which is determined by
the left- (resp. right)-moving 
$N=4$ superconformal algebra\footnote{In fact, this is the {\em small} $N=4$
superconformal algebra of \cite{aetal76}. We simply call it {\em the
$N=4$ superconformal algebra} to untangle the terminology.} 
at central charge $c=6$ (resp. $\qu c=6$). 
We use standard conformal field theory techniques to identify the spectrum 
of our special K3 theory in terms of the vacuum, vector and two spinor 
representations of $\widehat{\mathfrak{so}}(8)_1$ and to give an 
elegant description in terms of lattice vertex operator algebras. 
In addition, we
 provide  the full partition function, including its  explicit dependence 
on the charges measured by the zero modes of the $U(1)$ currents 
described above.
The underlying charge lattice $\Gamma$
of our K3 theory is half integral. Our presentation includes
the   classification of all equivalence classes of 
cocycles for a certain type of half integral  lattices, in particular  $\Gamma$.
Moreover, in each such equivalence class of cocycles we 
explicitly construct 
a representative which obeys all required compatibility conditions with 
the real structure on our space of states.

Section \ref{supernatural} is devoted to a recapitulation of  
the Conway Moonshine Module 
 presented in \cite{du07,dmc14, dmc15}.
 We also offer a description in terms of lattice vertex operator algebras.
In addition, we explain  how  to obtain $U(1)$ currents 
that allow an interpretation
as images of the $U(1)$ currents in the K3 theory, under reflection, and we 
 present the 
partition function for the Conway Moonshine Module with its dependence 
on the corresponding charges.

In Section  \ref{flip}, we determine some
necessary and sufficient conditions for non-holomorphic 
superconformal field theories 
to allow {\em reflection}, that is, a mathematically consistent transformation 
of  all  fields  to  holomorphic  ones.
We show that reflection amounts to a complex conjugation for the anti-holomorphic
parameters of the OPE 
when restricted to pairs
of fields that create {\em real} states.\footnote{This procedure
is reminiscent of that used by Harvey and Moore in their 
definition of BPS algebras \cite{hamo96}. However, we differ
in the following crucial point: throughout our work,
we parametrize fields $\phi(z,\qu z)$
in the complex plane rather than 
on the cylinder. Therefore, on a formal level
complex conjugation of the right-moving degrees of freedom 
amounts to restricting 
to $z=\qu z$. On the other hand, 
the prescription given in \cite[\S9]{hamo96}, which also 
changes the right-moving  fields to left-moving  fields, 
amounts to enforcing $z=\qu z^{-1}$, i.e. using a complex
conjugation on the cylinder. Although the latter may be
a natural choice, 
our construction, 
in the context of lattice vertex operator algebras,
entails the change in signature 
of the charge lattice 
required
to make contact with the super vertex operator algebras.}
The real structure on the space of 
states of the original superconformal field theory 
is thus found to play a crucial
role\footnote{As is customary, we denote the real structure
on $\C$ by $z\mapsto\qu z$ for $z\in\C$, and we write
$i=\exp\left({i\pi\over2}\right)$ for our choice of $\sqrt{-1}$
throughout.}. We show that our reflection procedure,
if applicable,
yields the structure of a  super
vertex operator algebra on the Neveu-Schwarz sector $\H^{\rm NS}$ 
and that of an admissible $\H^{\rm NS}$-module on the Ramond sector of the theory. 
We furthermore discuss  how reflection induces an
additional structure on the resulting super vertex operator algebra and 
admissible module.
 
In Section \ref{map}, we show  that
the Conway Moonshine Module emerges 
via reflection of the  K3 
theory with $\Z_2^8\colon \M_{20}$ symmetry. 
In this case, we show that reflection amounts to replacing the lattice  of
signature $(6,6)$, which governs the lattice vertex operator algebra 
description of the K3 theory, by a lattice of signature $(12,0)$.
Thereby, we interpret the ideas of \cite{dmc15}, namely to realize  
all symmetries
of K3 theories as automorphisms of the Conway Moonshine Module,
in terms of symmetry surfing. Although this yields a natural action of
$M_{24}$ on
the Conway Moonshine Module that extends the action of the geometric symmetry group 
of the K3 theory, this cannot explain Mathieu Moonshine.
Indeed, the twist that had already been observed in \cite{tawe12} is not 
implemented in the Conway Moonshine Module.
Three appendices summarise, respectively, our approach to 
superconformal field theory, technical background concerning 
cocycles on half integral lattices,
and some useful identities for Jacobi theta
functions.


\section{A K3 theory with $\Z_2^8\colon\M_{20}$ symmetry}\label{gtvw}
In this section, we present the basic ingredients of the K3 theory which is
central to this work\footnote{For a summary of relevant notions from
(super-)conformal field theory, see Appendix \ref{cft}.}, namely
the {\em K3 theory based on the $\Z_2$-orbifold of the $D_4$-torus theory}.
This model accounts 
for one of the largest possible discrete symmetry groups of K3 theories preserving $N=(4,4)$ 
supersymmetry, namely $\Z_2^8\colon\M_{20}$. That some K3 theory would possess 
this symmetry was predicted in the very interesting classification paper \cite{ghv12}. 
As a consequence, it was a sound enterprise to determine and study
such a model  further, constructing its symmetries explicitly  in order to
gain further insight in relation to the $M_{24}$ Moonshine phenomenon. The model was first investigated in \cite{nawe00}, where
it was denoted $(\widetilde{2})^4$ in reference
to the fact that it can be constructed as a Gepner type model. It was 
studied extensively in \cite{we01,gtvw14} from several perspectives, one of which  involves 
a description\footnote{The authors of \cite{dmc15} seem unaware of the fact that the free fermion 
description of the K3 theory predates their own account.}
in terms of $12$ left-moving Majorana fermions $\psi_j(z)$ and $12$ 
right-moving Majorana fermions $\qu{\psi}_j(\qu{z}),\,j\in\{1,\dots, 12\}$.

The description of this model 
in terms of free fermions, as given in  \cite{gtvw14}, is summarized below since
 it is central in making contact with the recent works of Duncan and Mack-Crane 
\cite{dmc14,dmc15}. 
 For later convenience, we include a detailed discussion of the 
 OPEs in this model, fixing  in particular  the delicate choices of
phase factors.
%
\subsection{Bosonic $D_4$-torus model}\label{bosonicD4torus}
The {\em bosonic} $D_4$-torus model, which we consider 
as a starting point to the construction of our K3 theory,
is a toroidal theory  based 
on the $4$-dimensional torus $\T=\R^4/L$, where $L$ is the $D_4$-lattice 
$L_{D_4} \subset \R^4$ and where the $B$-field and the metric are chosen 
in such a way that the generic left-moving affine $\widehat{\mathfrak u}(1)^4$ algebra 
is enhanced to the affine algebra $\widehat{\mathfrak{so}}(8)_1$,
and analogously for the right-movers
(see \cite[\S2]{gtvw14} for details). 
The lattice of $\widehat{\mathfrak u}(1)^4_L\oplus \widehat{\mathfrak u}(1)^4_R$ 
charges, which completely determines the $D_4$-torus model, equals
\be\label{Ddcharge}
\begin{array}{rcl}
\ds\Gamma_{d,d}
&:=& 
\left\{ 
({\Q};\qu{\Q}) \in \left(\Z^d\oplus\Z^d \right) \cup 
\left(\,\left({\textstyle{1\over2}}+\Z\right)^d\ \times \left({\textstyle{1\over2}}+\Z\right)^d\right)
\right.\\
&&\ds
\hphantom{({\Q};\qu{\Q}) \in \left(\Z^d\oplus\Z^d \right) \cup \left({\textstyle{1\over2}}+\Z\right)^d }
\left| \textstyle\sum\limits_{k=1}^d \left(Q_k+\qu Q_k\right)\equiv 0 \mod 2 \right\}
\end{array}
\ee
with $d=4$, c.f.\ \cite[(2.11)]{gtvw14}. 
{The lattice $\Gamma_{d,d}\subset\R^{d,d}=\R^d\oplus\R^d$ is 
equipped with the symmetric bilinear form
\be\label{signatureform}
\forall ({\Q};\qu{\Q}),\, ({\Q}^\prime;\qu{\Q}^\prime)\in\Gamma_{d,d}\colon\qquad
({\Q};\qu{\Q}) \bullet ({\Q}^\prime;\qu{\Q}^\prime)
:= {\Q}\cdot {\Q}^\prime - \qu{\Q}\cdot \qu{\Q}^\prime,
\ee
where here and in the following, ${\Q}\cdot{\Q}^\prime\in\R$ denotes the standard scalar product of
${\Q},\,{\Q}^\prime\in\R^d$.}
This yields {an even, self-dual lattice $\Gamma_{d,d}$ in general, and}
the space of states
of the bosonic $D_4$-torus model as 
\be\label{Fockrep}
\HHH_{\Dtorus} = \bigoplus_{\gamma\in\Gamma_{4,4}} \HHH_\gamma
\ee
with $\HHH_\gamma,\, \gamma=({\Q};\qu{\Q})\in\Gamma_{4,4}$,
the Fock space representation of $4$ left-moving and $4$ right-moving free bosons,
built on a ground state $\upsilon_\gamma$ of 
{$\widehat{\mathfrak u}(1)^4_L\oplus \widehat{\mathfrak u}(1)^4_R$  charge $\gamma$ and}
conformal dimensions $(h;\qu h) ={1\over2} ({\Q}\cdot{\Q};\qu{\Q}\cdot\qu{\Q})$.
We generally choose such $\upsilon_\gamma\in\HHH_\gamma$ with 
$\upsilon_\gamma^\ast=\upsilon_{-\gamma}$  as unit vectors with
respect to the scalar
product $\langle\cdot,\cdot\rangle$ on $\HHH_{\Dtorus}$. Moreover,
we assume that $\upsilon_0=\Omega$ is the vacuum.

By construction, the space $\HHH_{\Dtorus}$
decomposes into four sectors, since
\be\label{deco}
\begin{array}{rcl}
\ds
\makebox[0pt]{$\Gamma_{4,4}$} 
&=\!& \ds\bigcup_{a=0}^3 \left( \gamma^{( a)}_{D_4} + \Gamma^{(0)}_{D_4} \right),
\quad \mb{ with }\\
&&  \Gamma^{(0)}_{D_4} :=
\left\{ ({\Q};\qu{\Q}) \in \Z^4\oplus\Z^4 \left| 
\sum\limits_{k=1}^4 Q_k \equiv  \sum\limits_{k=1}^4\qu Q_k \equiv 0 \mod 2 \right. \right\},\\[8pt]
&&
\textstyle
\gamma^{(0)}_{D_4} :=0,\,
\gamma^{(1)}_{D_4} := ( \e_4; {\e}_4) , \,
\gamma^{(2)}_{D_4} := {1\over2} \sum\limits_{k=1}^4 \left( {\e}_k; {\e}_k\right),\,
\gamma^{(3)}_{D_4} := \gamma^{(1)}_{D_4}+ \gamma^{(2)}_{D_4},
\end{array}
\ee
{where here and in the following, $({\e}_k)_{k\in\{1,\ldots,d\}}$ denotes the
standard basis of $\R^d$.}
\medskip\par

\subsubsection{Holomorphic and anti-holomorphic currents and fermionization}\label{Walgebras}
The four holomorphic currents of the model
in the Cartan subalgebra of $\widehat{\mathfrak{so}}(8)_1$, 
$j_k(z)$ with $k\in\{1,\ldots,4\}$, obey the OPEs
\be\label{OPEj}
j_k(z)j_\ell(w)\sim \frac{\delta_{k\ell}}{(z-w)^2}
\quad \forall k,\ell\in\{1,\ldots,4\}.
\ee
Analogously, in the right-moving sector, one has four anti-holomorphic 
$U(1)$ currents $\qu\jmath_k(\qu{z}),\, k\in\{1,\ldots,4\}$.
\smallskip\par

One may fermionize the theory  by the
{\em Frenkel-Kac-Segal} construction
\cite{frka80,se81,gool84}. To do so, one introduces
eight free 
left-moving Majorana fermions $\psi_i(z),\, i\in\{5,\ldots 12\}$ 
and eight free right-moving Majorana fermions 
$\qu \psi_i(\qu z),\, i\in\{5,\ldots 12\}$, with OPEs
$$
\psi_i(z)\psi_j(w)\sim \frac{\delta_{ij}}{z-w}, \quad
\qu\psi_i(\qu z)\qu\psi_j(\qu w)\sim \frac{\delta_{ij}}{\qu z-\qu w},
\qquad \forall i,j\in\{5,\ldots,12\},
$$
all with coupled spin structures.
In terms of the free holomorphic Dirac fermions 
\be\label{Dirac}
\textstyle
\quad
x_k := \frac{1}{\sqrt{2}} ( \psi_{k+4} + i \psi_{k+8} )\ , \quad 
x_k^\ast := {1\over\sqrt2} ( \psi_{{k+4}} - i \psi_{{k+8}} ),\qquad k\in\{1,\ldots, 4\},
\ee
which satisfy the OPEs
\be \label{OPEx} 
x_k(z) \, x_k^\ast(w)\sim {1\over z-w}\sim x_k^\ast(z) \, x_k(w),\qquad k\in\{1,\ldots, 4\},
\ee
the four  left-moving $U(1)$ currents are given by
\be\label{u1currents}
j_k(z) = \nop[x_k(z) x_k^\ast(z)] = - i\nop[\psi_{{k+4}}(z)\psi_{{k+8}}(z)]\, ,
\ee
as can be checked by calculating their OPEs \eqref{OPEj} with the help of 
\eqref{OPEx}. Introducing
$j_k(z)=i\partial \phi_k(z)$, one may identify 
\be \label{cocycles}
\quad
x_k(z)=\nop[\exp (i\phi_k(z))]c_k, \quad
x^*_k(z)=\nop[\exp (-i\phi_k(z))] c_{-k},
\qquad k\in\{1,\ldots ,4\},
\ee
where $c_k$ and $c_{-k}$ are cocycle factors that ensure that the fermions of 
different species anticommute, as we shall discuss in 
greater detail below.

The analogous construction holds for the right-moving sector, 
through the introduction of four right-moving Dirac fermions
\be \label{antiholcocycles}
\textstyle
\qu x_k := {1\over\sqrt2} ( \qu { \psi}_{{k+4}} + i \qu{ \psi}_{{k+8}} ) , \quad 
\qu x_k^\ast := {1\over\sqrt2} ( \qu\psi_{{k+4}} - i \qu\psi_{{k+8}} ) , \quad 
k\in\{1,\ldots ,4\},
\ee
as well as $\qu\jmath_k(\qu z) := \nop[\qu x_k(\qu z)\qu x_k^\ast(\qu z)] = i\qu\partial\qu\phi_k(\qu z)$
with
$$
\quad
i\qu x_k(\qu z)=\nop[\exp (i\qu\phi_k(\qu z))]\qu c_k, \quad
i\qu x^*_k(\qu z)=\nop[\exp (-i\qu\phi_k(\qu z))] \qu c_{-k},
\qquad k\in\{1,\ldots ,4\}.
$$
It is then straightforward to express the $24$ currents of 
$\widehat{\mathfrak{so}}(8)_1$ associated with the roots of 
$D_4$, namely the vectors $\pm {\e}_j \pm {\e}_k,\, 1 \le j < k \le 4$, in terms 
of the Dirac fermions  \eqref{cocycles}: possibly up to cocycle factors,
these $24$ conformal weight 
$(h;\qu h)=(1;0)$-fields may  be realized as
\begin{equation}\label{currentident}
\begin{array}{rclrcr}
V_{\boldsymbol{e}_j+\boldsymbol{e}_{k}}(z) &\!\!\!\!=\!\!\!\!& \nop[x_j(z) x_k(z)] ,\qquad&
V_{\boldsymbol{-e}_{j}-\boldsymbol{e}_{k}}(z) &\!\!\!\!=\!\!\!\!& \nop[x_j^\ast(z) x_k^\ast(z)] ,\\[5pt]
V_{\boldsymbol{e}_{j}-\boldsymbol{e}_{k}}(z)&\!\!\!\!=\!\!\!\!&  \nop[x_j(z) x_k^\ast(z)] ,\qquad&
V_{\boldsymbol{-e}_{j}+\boldsymbol{e}_{k}}(z)&\!\!\!\!=\!\!\!\!& \nop[x_j^\ast(z)x_k(z) ] .
\end{array}
\end{equation}
In other words, $\Gamma^{(0)}_{D_4}$ contains the charge vectors 
$(\pm {\e}_j\pm {\e}_k;0)$ that
are responsible for the extended $\widehat{\mathfrak{so}}(8)_1$ symmetry of the model.
\medskip\par

\subsubsection{General momentum-winding fields}\label{bosonicfieldcontent}
The currents \eqref{currentident}
are  special momentum-winding fields 
with left and right $\widehat{\mathfrak u}(1)^4$ charges 
$(\boldsymbol{Q};\qu{\boldsymbol{Q}})$. The latter, a priori, 
are vectors of the charge 
lattice $\Gamma_{4,4} \subset \R^{4,4}$ of the $D_4$-torus model
as in \eqref{Ddcharge}.  The momentum-winding {field
for any $({\Q};\qu{\Q})\in\Gamma_{4,4}$}
 may be written as  
\be\label{vdef}
\textstyle
V_{({\Q};\qu{\Q})}(z,{\qu z})
:=\nop[{\exp\left[ i\sum\limits_{k=1}^4{Q}_k\phi_k(z) + i\sum\limits_{k=1}^4{ \qu Q}_k \qu\phi_k(\qu z)\right]}] 
\; c_{({\Q};\qu{\Q})},
\ee
with $c_{({\Q};\qu{\Q})}$ denoting appropriate {\em cocycle factors}
\cite{frka80,se81,gool84}. 
This means that for every $\gamma\in\Gamma_{4,4}$, we have a linear
operator $c_\gamma$ on $\H$, where
$(c_\gamma)_{\mid\HHH_{\gamma^\prime}}=
\eps(\gamma,\gamma^\prime)\cdot\mb{id}_{\HHH_{\gamma^\prime}}$
for all charge vectors $\gamma,\,\gamma^\prime\in\Gamma_{4,4}$,
with {\em cocycles} 
$$
\eps\colon \Gamma_{4,4}\times\Gamma_{4,4} \longrightarrow \{ \pm1\}.
$$
Here, as is common in the physics literature,
the term {\em cocycles} more precisely refers to 
$2$-cocycles on $\Gamma_{4,4}$ with values in 
$\{\pm1\}$ that obey the additional symmetry requirement
\eqref{cocycleswap} with respect to the bilinear form
\eqref{signatureform} on $\Gamma_{4,4}$.
In  Appendix \ref{halfcoc} we review the definition of such
cocycles. Since $\Gamma_{4,4}$ is an integral lattice,
their explicit construction, also given in Appendix \ref{halfcoc},
is well-known.

In the notations of \eqref{vdef}, 
the bosonic $(h;\qu h)=(1;0)$-fields \eqref{currentident} 
have $\boldsymbol{Q}=\pm \boldsymbol{e}_j \pm \boldsymbol{e}_k$ and 
$\boldsymbol{\qu Q}=0$, and we write
$$
V_{\pm\boldsymbol{e}_j\pm\boldsymbol{e}_{k}}(z) := V_{(\pm\boldsymbol{e}_j\pm\boldsymbol{e}_{k};{\bf 0})}(z,\qu z).
$$

According to \eqref{deco}, the four cosets in $\Gamma_{4,4}/\Gamma^{(0)}_{D_4}$,
namely $\gamma^{(a)}_{D_4}+\Gamma^{(0)}_{D_4}$ 
for $a\in\{0,\ldots,3\}$, induce the decomposition of the space of states of the bosonic $D_4$-torus 
model  into representations of the left and 
right-moving $\widehat{\mathfrak s \mathfrak o}(8)_1$-algebras as
$$
 {\mathcal H}_{\Dtorus } = ({\mathcal H}_{L,0} \otimes {\mathcal H}_{R,0}) \oplus 
({\mathcal H}_{L,v} \otimes {\mathcal H}_{R,v}) \oplus ({\mathcal H}_{L,s} \otimes {\mathcal H}_{R,s}) \oplus ({\mathcal H}_{L,c} \otimes {\mathcal H}_{R,c}),
$$
where ${\mathcal H}_{L,0}$ is the left-moving $\widehat{\mathfrak s \mathfrak o}(8)_1$ 
vacuum representation, 
while ${\mathcal H}_{L,v}, {\mathcal H}_{L,s}$ and ${\mathcal H}_{L,c}$  
are the vector and the two spinor representations, 
respectively. The   ${\mathcal H}_{R,\bullet}$ denote the 
corresponding right-moving representations. 
Hence 
$$
\HHH_{L,0}\otimes \HHH_{R,0} 
= \bigoplus_{\gamma\in\Gamma^{(0)}_{D_4}} \HHH_\gamma
$$
with the notations of \eqref{deco} {and \eqref{Fockrep}}.
The vector representation 
$$
{\mathcal H}_{L,v} \otimes {\mathcal H}_{R,v}
= \bigoplus_{\gamma\in\gamma^{(1)}_{D_4}+\Gamma^{(0)}_{D_4}} \HHH_\gamma
$$ 
is generated by  OPEs  of the 
left and right-moving
currents with the {\em vector} 
${(h;\qu h):=}(\frac{1}{2}; \frac{1}{2})$ winding-momentum fields 
$V_{{(}\boldsymbol{Q};\qu{\Q}{)}}(z,\qu z)$, where
$$
{\bf Q}
=\pm {\e}_i\ ,\quad \qu{\Q}=\pm {\e}_j. 
$$
The spinor representations  $\HHH_{L,s}\otimes \HHH_{R,s},\; 
\HHH_{L,c}\otimes \HHH_{R,c}$
are analogously 
generated by   the OPEs  of the 
left and right-moving
currents with the {\em spin} $(h;\qu h):=(\frac{1}{2}; \frac{1}{2})$ winding-momentum fields 
$V_{\boldsymbol{Q};\boldsymbol{\qu Q}}(z,\qu z)$ for
\begin{align}
\textstyle
{\Q} = {1\over2} \sum\limits_{j=1}^4 \eps_j {\e}_j\ , \quad 
\qu{\Q} = {1\over2} \sum\limits_{k=1}^4 \delta_k {\e}_k\ , 
   \label{bosgen} \nonumber \\ 
\textstyle
\mbox{where } \eps_j,\, \delta_k\in\{\pm1\} \mbox{ and } 
\sum\limits_{k=1}^4 (Q_k+\qu Q_k)\equiv 0\mod 2\ .
\nonumber \end{align}
In fact,  
$$
\HHH_{L,s}\otimes \HHH_{R,s} 
= \bigoplus_{\gamma\in\gamma^{(2)}_{D_4}+\Gamma^{(0)}_{D_4}} \HHH_\gamma \, ,
\qquad
\HHH_{L,c}\otimes \HHH_{R,c} 
= \bigoplus_{\gamma\in\gamma^{(3)}_{D_4}+\Gamma^{(0)}_{D_4}} \HHH_\gamma\, ,
$$
i.e.\ 
$\sum\limits_{k=1}^4 Q_k$ and $\sum\limits_{k=1}^4 \qu Q_k$ are both even for 
${\mathcal H}_{L,s} \otimes {\mathcal H}_{R,s}$ 
and both odd for ${\mathcal H}_{L,c} \otimes {\mathcal H}_{R,c}$.\\[10pt]

Fermionizing  the bosonic $\Dtorus$ theory 
 as mentioned  earlier allows us 
 to extend the definition of the momentum-winding fields
in \eqref{vdef} to include {\em fermionic fields}
$V_{\gamma}(z,\qu z)$ with
$\gamma\in(\Z^4\oplus\Z^4)
\setminus \Gamma_{4,4}\,\subset
 (\Gamma_0^{D_4})^\ast$,
 where 
$$
\Gamma_0^{D_4}:= (\Z^4\oplus\Z^4)\;\cap\;\Gamma_{4,4}.
$$
Indeed\footnote{For later convenience, we use slightly
different normalizations than the ones
given in \cite{gtvw14}.}, 
in \eqref{cocycles} we have already presented special cases of such fermionic
fields, namely
\be\label{bosonization}
\begin{array}{rclrcl}
x_k(z)&=&V_{\e_k}(z),&  x_k^\ast (z)&=&V_{-\e_k}(z);\\[5pt]
{i\qu x_k(\qu z)}&=&{V_{(0;\e_k)}(z,\qu z),}& {
 i\qu x_k^\ast (\qu z)}&=&V_{(0;-\e_k)}(z,\qu z),\quad
 k\in\{1,\ldots,4\},
 \end{array}
\ee
where $c_{(\pm \e_k;0)}:=c_{\pm k}$ and $c_{(0;\pm \e_k)}:=\qu c_{\pm k}$.
Following \cite{gool86,gnos86,gnors87}, we may actually extend further to
$\HHH_{\gamma^\prime}$ with 
$\gamma^\prime\in\Gamma_{D_4}$,
$$
\Gamma_{D_4}:=
\Gamma_{4,4}\cup\left((0;\e_4)+\Gamma_{4,4} \right)
= (\Gamma_0^{D_4})^\ast,
$$
a half integral lattice. 
The lattice $\Gamma_0^{D_4}=\left(\Gamma_{D_4}\right)^\ast$ 
is  a sublattice of  $\Gamma_{D_4}$ of index $4$, 
and as such, $(\Gamma_0^{D_4},\Gamma_{D_4})$ form a
$\Z_2$ lattice pair $(\Gamma_0,\Gamma)$ of the type used in 
Appendix \ref{halfcoc}. In particular, Appendix \ref{halfcoc} includes
an explicit construction of cocycles $\eps$ for this half integral lattice. 
These cocycles are bimultiplicative in the sense of \eqref{epsbimulti},
and they are in the special gauge \eqref{specialgauge}. They 
take values in the group
of eighth roots of unity in $\C^\ast$.
To clear notations, we write the charge lattice,
extended to include fermions, as
\be
\label{deco2}
\begin{array}{rcl}
\ds
\makebox[0pt]{$\Gamma_{D_4}$} 
&=\!& \ds\bigcup_{a=0}^3 \left( \widetilde{\gamma}^{( a)}_{D_4} + \Gamma_0^{D_4} \right),
\quad \mb{ with } \qquad
\Delta:=\left\{ \widetilde{\gamma}^{(1)}_{D_4}, \widetilde{\gamma}^{(2)}_{D_4},\widetilde{\gamma}^{(3)}_{D_4}  \right\},\\
&&
\textstyle
\widetilde{\gamma}^{(0)}_{D_4} :=0,\,
\widetilde{\gamma}^{(1)}_{D_4} := (0; {\e}_4) , \,
\widetilde{\gamma}^{(2)}_{D_4}:=\gamma^{(2)}_{D_4} = {1\over2} \sum\limits_{k=1}^4 \left( {\e}_k; {\e}_k\right),\,
\widetilde{\gamma}^{(3)}_{D_4} := \widetilde{\gamma}^{(1)}_{D_4}+ \widetilde{\gamma}^{(2)}_{D_4}.
\end{array}
\ee
For any $\gamma,\,\gamma^\prime\in\Gamma_{D_4}$,
the OPEs between 
momentum-winding fields
$V_\gamma,\, V_{\gamma^\prime}$
are given by
\be\label{leftrightmovingOPE}
\begin{array}{rcl}
V_\gamma(z,\qu z) V_{\gamma^\prime}(w,\qu w)
&\!\!\!\sim\!\!\!& 
(z-w)^{\boldsymbol{Q}\cdot \boldsymbol{Q}^\prime}\,
(\qu z-\qu w)^{\boldsymbol{\qu Q}\cdot \boldsymbol{\qu Q^\prime}}\,
\varepsilon\left( \gamma,\gamma^\prime \right)
V_{\gamma+\gamma^\prime}(w,\qu w)\\[5pt]
&&\;
\times \left\{ \vphantom{1\over2} 1+(z-w) \sum\limits_{k=1}^4{Q_k}j_k (w)
+(\qu z-\qu w) \sum\limits_{k=1}^4{\qu Q_k}\qu\jmath_k(\qu w)  + \cdots\right\}.\\[5pt]
\end{array}
\ee
Note that in these OPEs, 
apart from integral powers of
$(z-w),\, (\qu z-\qu w)$ and $|\qu z-\qu w|$, odd integral
powers of $(z-w)^{\pm{1\over2}}$ and $(\qu z-\qu w)^{\pm{1\over2}}$
occur iff
$$
\gamma\in\delta+\Gamma_0^{D_4}, \quad 
\gamma^\prime\in\delta^\prime+\Gamma_0^{D_4}
\quad \mbox{ and }\quad 
\delta,\delta^\prime\in  \Delta, \quad
\delta\neq\delta^\prime, 
$$
or equivalently, $\gamma\bullet\gamma^\prime\in{1\over2}+\Z$.
Then, implementation of \eqref{leftrightmovingOPE} in an 
$n$-point function affords the restriction of the domain of definition
to some contractible open $U\subset\C^n\setminus\cup_{i\neq j}\{z\in\C^n|z_i=z_j\}$.
An unambiguous formulation of such an extension of 
\eqref{leftrightmovingOPE} states for all $\gamma,\,\gamma^\prime\in \Gamma_{D_4}$:
\be\label{moreleftrightmovingOPE}
\begin{array}{ll}
\quad\\
\;\;V_\gamma(z,\qu z) \upsilon_{\gamma^\prime} 
\sim
z^{\boldsymbol{Q}\cdot \boldsymbol{Q}^\prime}\,
\qu z^{\boldsymbol{\qu Q}\cdot \boldsymbol{\qu Q^\prime}}
\varepsilon\left( \gamma,\gamma^\prime \right) \left\{ 1+z \smash{\sum\limits_{k=1}^4Q_k a^{(k)}_{-1}}
+\qu z \smash{\sum\limits_{k=1}^4} \qu Q_k \qu a^{(k)}_{-1} + \cdots\right\}
\upsilon_{\gamma+\gamma^\prime},
\end{array}
\ee
where $z\in U$ with $U\subset\C^\ast$ a contractible open subset,
and where $a_n^{(k)},\, \qu a_n^{(k)}$ with $k\in\{1,\ldots,4\}$,
$n\in\Z$, denote the modes of $j_k(z),\,\qu\jmath_k(\qu z)$.

Using \eqref{moreleftrightmovingOPE}, 
one checks that the coboundary condition
\eqref{cocycledef} ensures
{\em associativity} of the OPE,
$$
\forall \alpha,\,\beta,\,\gamma\in\Gamma_{D_4}\colon\qquad
V_\alpha(z,\qu z) \left( V_\beta(w,\qu w)\upsilon_\gamma \right)
\sim \left(V_\alpha(z,\qu z) V_\beta(w,\qu w)\right) \upsilon_\gamma.
$$
The additional symmetry condition \eqref{cocycleswap} ensures {\em semilocality}
\begin{align*}
\forall a\in\{1,\,2,\,3\}, \;
\forall \alpha,\,\beta\in\Gamma_0^{D_4}\cup 
\left(\widetilde{\gamma}^{(a)}_{D_4}+\Gamma_0^{D_4}\right)\colon
\\
V_\alpha(z,\qu z)  V_\beta(w,\qu w) 
&\sim (-1)^{(\alpha\bullet\alpha)\cdot (\beta\bullet\beta)}  V_\beta(w,\qu w)V_\alpha(z,\qu z).
\end{align*}
In other words, semilocality is only required to hold between
$V_\alpha(z,\qu z)$  and $V_\beta(w,\qu w)$ if $\alpha\bullet\beta\in\Z$.
Indeed, this condition
cannot be imposed on all pairs
$\alpha,\,\beta\in\Gamma_{D_4}$,
since the above-mentioned square root cuts obstruct
semilocality.  

The choice of special gauge \eqref{specialgauge} for the cocycles
ensures that the OPE \eqref{leftrightmovingOPE}
is compatible with the real structure on the space of states 
according to \eqref{realcomp}. Indeed, for 
$\alpha=({\boldsymbol Q};\qu{\boldsymbol Q})\in\Gamma_{D_4}$
we see 
from \eqref{moreleftrightmovingOPE} that
the condition $\eps(\alpha,0)=1$ ensures that the
field $V_\alpha(z,\qu z)$ creates the state
$\upsilon_\alpha$ from the vacuum. Furthermore,
$\eps(-\alpha,\alpha)=1$ amounts to the hermiticity 
condition that ensures that
$(V_\alpha(z,\qu z))^\dagger 
= \qu z^{-{\boldsymbol Q}\cdot{\boldsymbol Q}} z^{-\qu{\boldsymbol Q}\cdot\qu{\boldsymbol Q}}
V_{-\alpha}(\qu z^{-1},z^{-1})$ is compatible with our requirements
 $\upsilon_\alpha^\ast=\upsilon_{-\alpha}$ and $\upsilon_0=\Omega$.
 More generally, by \eqref{daggercondition} we have
$$
\forall \alpha,\,\beta \in\Gamma_{D_4}\colon\;\;
\qu{<V_{\alpha+\beta}(w,\qu w)V_{-\alpha}(x, \qu x)V_{-\beta}(z,\qu z)> }
= < \left(V_{-\beta}(z,\qu z)\right)^\dagger  \left(V_{-\alpha}(x, \qu x)\right)^\dagger 
\left(V_{\alpha+\beta}(w,\qu w)\right)^\dagger >,
$$
such that the above requirement for the Hermitian conjugate fields 
together with \eqref{leftrightmovingOPE} yield the last equation in
\eqref{specialgauge}.
\medskip\par

\subsection{Supersymmetric $D_4$-torus model}\label{SUSYD4torus}
The {\em supersymmetric} $D_4$-torus mo\-del is obtained by adjoining
$d=4$ free Majorana fermions $(\psi_k(z),\qu\psi_k(\qu z)), \,k\in\{1,\ldots,\,4\}$, related to  
the ${\rm U}(1)$ currents $j_k(z)$ and their right-moving counterparts 
by 
world-sheet supersymmetry. 
Similarly to \eqref{Dirac}, it is more convenient
to work with the Dirac fermions
\be\label{Diracnew}
\textstyle
\chi_j := \frac{1}{\sqrt{2}} ( \psi_{2j-1} + i \psi_{2j} )\ , \quad 
\chi_j^\ast := {1\over\sqrt2} ( \psi_{{2j-1}} - i \psi_{{2j}} ),\qquad j\in\{1,2\},
\ee
and their right-moving counterparts, all of
which have coupled spin structures. Hence these Dirac fermions give rise to 
the affine symmetry
\be\label{affer}
\widehat{\mathfrak{so}}(8)_1 \supset
\widehat{\mathfrak{so}}(4)_{1,L}\oplus \widehat{\mathfrak{so}}(4)_{1,R} 
\cong \widehat{\mathfrak s\mathfrak u}(2)_1^4 \ .
\ee
Details of the construction of the corresponding currents in terms of 
the four Majorana fermions may be found in \cite[\S2]{gtvw14}. 
This model enjoys 
extended left- and right-moving worldsheet 
supersymmetry. We choose a particular left- (resp. right-) moving 
$N=4$ superconformal algebra at central charge $c=6$ 
(resp. $\qu c=6$),  which 
comes with an affine $\widehat{\mathfrak{su}}(2)_{1,L} \subset \widehat{\mathfrak{so}}(4)_{1,L}$ 
(resp. $\widehat{\mathfrak{su}}(2)_{1,R} \subset \widehat{\mathfrak{so}}(4)_{1,R}$) 
for $\widehat{\mathfrak{so}}(4)_{1,L}$  and  $\widehat{\mathfrak{so}}(4)_{1,R}$
 in \eqref{affer}. Our choice of $U(1)$
currents
\be\label{gtvwu1choice}
\textstyle
J :=  \nop[\chi_1\chi_1^\ast]+\nop[\chi_2\chi_2^\ast],
\qquad
\qu J :=  \nop[\qu\chi_1^\ast\qu\chi_1]+\nop[\qu\chi_2^\ast\qu\chi_2],
\ee
whose zero modes generate the Cartan subalgebras of 
the above-mentioned $\widehat{\mathfrak{su}}(2)_{1,L}$ and 
$\widehat{\mathfrak{su}}(2)_{1,R}$, is of particular importance in what follows.

Altogether, the total affine symmetry of the supersymmetric $D_4$-torus model is 
$$
\widehat{\mathfrak{so}}(8)_{1} \oplus \widehat{\mathfrak{so}}(16)_{1}    
\supset
\left(\widehat{\mathfrak{so}}(4)_{1,L} \oplus \widehat{\mathfrak{so}}(8)_{1,L}    \right)
\oplus
\left( \widehat{\mathfrak{so}}(4)_{1,R} \oplus \widehat{\mathfrak{so}}(8)_{1,R}   \right)
. 
$$
\medskip\par

The pair $\chi_k,\, \chi_k^\ast,\, \qu\chi_k,\, \qu\chi^\ast_k$, 
$k\in\{1,2\}$,
of two left- and two right-moving
Dirac fermions, all with coupled spin structures, 
gives rise to a {\em fermionic CFT} at central charges 
$c=2,\, \qu c=2$, three copies of which
suffice to give a complete
description of the supersymmetric $D_4$-torus model, as was done in
\cite[\S 3]{gtvw14}  and shall be recalled shortly. 
As a preparation, we first give a description of this 
fermionic CFT by means of toroidal
 momentum-winding fields
as in Section 
\ref{bosonicD4torus}, along the lines of
\cite[Appendix D]{gtvw14},
including the fermionic
contributions. Though the fermionic
CFT at  central charges $c=2,\, \qu c=2$ possesses
neither worldsheet nor spacetime
supersymmetry, Neveu-Schwarz and Ramond sectors 
are well-defined by means of the fermion boundary
conditions.
By the above,
the supersymmetric $D_4$-torus mo\-del is the 
tensor product of the bosonic $D_4$-torus mo\-del
of Section \ref{bosonicD4torus} and this fermionic CFT.
\smallskip\par

For each $U(1)$ current $\mathfrak{j}$ in the $\widehat{\mathfrak{so}}(8)_1$
current algebra of \eqref{affer}, similarly to 
\eqref{u1currents}, \eqref{antiholcocycles}, we may introduce
$\mathfrak{j}=i\partial\varphi$. Thus we {\em bosonize} by writing
\be\label{newu1currents}
\mathfrak{j}_k:=
-i\nop[\psi_{2k-1}\psi_{2k}] = i\partial\varphi_{k}, \quad 
\qqu{\smash{\mathfrak{j}}\vphantom{\jmath}}_k:=-i\nop[\qu\psi_{2k-1}\qu\psi_{2k}] = i\qu\partial\qu\varphi_{k}
\quad \mb{ for } \quad k\in\{1,2\},
\ee and we 
recover
$$
\begin{array}{rclrcl}
\chi_k(z) &=&  \nop[\exp (i\varphi_k(z))]c_k, \quad
&\chi^*_k(z)&=&\nop[\exp (-i\varphi_k(z))] c_{-k},\\[5pt]
{i}\qu\chi_k(\qu z) &=&  \nop[\exp (i\qu\varphi_k(\qu z))] c_{k+2}, \quad
&{i}\qu\chi^*_k(\qu z)&=&\nop[\exp (-i\qu\varphi_k(\qu z))] c_{-(k+2)}.\\
  \end{array}
 $$
Analogously to \eqref{bosonization}, all contributions from the free fermions
$\chi_k,\, \chi_k^\ast,\, \qu\chi_k,\, \qu\chi^\ast_k$, 
$k\in\{1,2\}$, are now generated by fields $V_{(\Q;\qu\Q)}(z;\qu z)$ as in 
\eqref{vdef}, where
$(\Q;\qu\Q)\in\widetilde\Gamma_{2,2}\subset \R^{2,2}$,
a lattice equipped with the symmetric bilinear form $\bullet$ 
that was introduced in \eqref{signatureform}. This half integral lattice, which extends the charge lattice 
$\Gamma_{2,2}$ given by \eqref{Ddcharge} for $d=2$, is needed to accommodate 
fermionic fields in the same way as was presented in Subsection \ref{bosonicfieldcontent}. One has
\be  \label{gammat22}
\begin{array}{rcl}
\ds
\makebox[0pt]{$\widetilde\Gamma_{2,2}$} 
&:=\!& \ds\Gamma_{2,2}\cup \left(\,(0;{\e}_2)+\Gamma_{2,2}\,\right)
=\left(\Z^2\oplus\Z^2\right)\cup({1\over2}+\Z)^2\times ({1\over2}+\Z)^2
\;=\; \ds\bigcup_{i=0}^3 \left( \widetilde\gamma^{(i)} + \widetilde\Gamma_0\right)
\quad \mb{ with }\\
&&  \widetilde\Gamma_0  :=
\left\{ ({\Q};\qu{\Q}) \in \Z^2\oplus\Z^2 \left| 
\sum\limits_{k=1}^2 (Q_k+\qu Q_k)\equiv 0 \mod 2 \right. \right\},\\[15pt]
&&
\textstyle
\widetilde\gamma^{(0)} :=0,\,
\widetilde\gamma^{(1)} := (0; \e_2), \,
\widetilde\gamma^{(2)} := {1\over2}  \sum\limits_{k=1}^2 (\e_k; \e_k),\,
\widetilde\gamma^{(3)} 
:=\widetilde\gamma^{(1)}+\widetilde\gamma^{(2)}.
\end{array}
\ee
The above charge lattice $\widetilde\Gamma_{2,2}$ with
its sublattice $\widetilde\Gamma_0$ yields another example of a
$\Z_2$ lattice pair $(\widetilde\Gamma_0,\widetilde\Gamma_{2,2})$
of the type used in Appendix \ref{halfcoc}. 
Hence analogously to the discussion in Section \ref{bosonicfieldcontent},
a general winding-momentum field creating a ground state 
$\upsilon_\gamma \in\HHH_\gamma$,
$\gamma=(\Q;\qu\Q)\in\widetilde\Gamma_{2,2}$, has the form \eqref{vdef}, 
$$
\textstyle
V_{\gamma}(z,{\qu z})
:=\nop[{\exp\left[ i\sum\limits_{k=1}^2{Q}_k\varphi_k(z) + i\sum\limits_{k=1}^2{ \qu Q}_k \qu\varphi_k(\qu z)\right]}] 
\; c_{\gamma}.
$$
Consistent cocycles governing the cocycle factors $c_\gamma$, 
$\gamma\in\widetilde\Gamma_{2,2}$,  with  the additional
symmetry and gauge requirements \eqref{cocycleswap}, \eqref{epsbimulti}, \eqref{specialgauge},
are constructed in 
our Appendix \ref{halfcoc}.
We refer to the end of Section \ref{bosonicfieldcontent} for the justification
of these requirements.

Since all spin structures in our 
pair $\chi_k,\, \chi_k^\ast,\, \qu\chi_k,\, \qu\chi^\ast_k$, 
$k\in\{1,2\}$,
of two left- and two right-moving
Dirac fermions are coupled, the space of states arising from the standard
Fock space representations of these fermions 
decomposes into the contributions from the
vacuum, vector, spinor and antispinor representations
of $\widehat{\mathfrak{so}}(8)_1\supset 
\widehat{\mathfrak{so}}(4)_{1,L} \oplus \widehat{\mathfrak{so}}(4)_{1,R}$
labelled $0,\, v,\, s,\, c$, above.  Similarly to the discussion in 
Section \ref{bosonicD4torus} and with notations as in \eqref{Fockrep}, 
we collect these contributions in sectors $\HHH_{\SSS}$ with $\SSS\in\{0,\, v,\, s,\, c\}$ 
and find
\be\label{so8deco}
\begin{array}{rclrcl}
\HHH_0 &=& \bigoplus\limits_{\mbox{$\scriptstyle\gamma\in\widetilde\Gamma_0$}} \HHH_\gamma, \qquad
&\HHH_v &=& \bigoplus\limits_{\gamma\in\widetilde\gamma^{(1)}
+\widetilde\Gamma_0} \HHH_\gamma, \\[15pt]
\HHH_s &=& \bigoplus\limits_{\gamma\in\widetilde\gamma^{(2)}
+\widetilde\Gamma_0} \HHH_\gamma, \qquad
&\HHH_c &=& \bigoplus\limits_{\gamma\in\widetilde\gamma^{(3)}
+\widetilde\Gamma_0} \HHH_\gamma.
\end{array}
\ee
The bosonic sector of this model is $\HHH_0\oplus\HHH_s$,
while $\HHH_v\oplus\HHH_c$ yields the fermions.
The Neveu-Schwarz sector is $\HHH_0\oplus\HHH_v$, while
the Ramond sector is $\HHH_s\oplus\HHH_c$.
\medskip\par

Generalizing the definition \eqref{Diracnew} of the Dirac fermions $\chi_\ell,
\qu\chi_\ell$, to include $\ell\in\{3,\ldots,6\}$, the sectors of the
bosonic $D_4$-torus theory of Section \ref{bosonicD4torus} arise
as 
\be\label{sectorsagree}
\HHH_{L,\SSS}\otimes \HHH_{R, \SSS}\cong 
\HHH_\SSS\otimes\HHH_\SSS,
\qquad \SSS\in\{ 0,\,v,\,s,\,c\},
\ee
where $\otimes$ denotes a fermionic tensor product, 
whenever needed.
 The sector 
$\HHH_{L,\SSS} \otimes \HHH_{R, \SSS}$ is governed by the
lattice $\Gamma_{D_4}$, which yields the charge lattice with
respect to the zero modes of 
$(j_1,\,\ldots,\,j_{4};$ $\qu\jmath_1,\,\ldots,\,\qu\jmath_{4})$
where as in \eqref{u1currents},
$j_k= -i\nop[\psi_{k+4}\psi_{k+8}]$,  $k\in\{1,\,\ldots,\,4\}$, 
and similarly for $\qu\jmath_{k}$. 
Our choice of $U(1)$ currents on 
$\HHH_\SSS\otimes\HHH_\SSS$ is 
$$
(\mathfrak{j}_3,\ldots,\mathfrak{j}_6; \qqu{\smash{\mathfrak{j}}\vphantom{\jmath}}_3,\ldots, 
\qqu{\smash{\mathfrak{j}}\vphantom{\jmath}}_6)=
( -i\nop[\psi_{5}\psi_{6}],\ldots,\,-i\nop[\psi_{11}\psi_{12}]\; ; \;
-i\nop[\qu\psi_{5}\qu\psi_{6}],\ldots,\,-i\nop[\qu\psi_{11}\qu\psi_{12}] ),
$$
as in \eqref{newu1currents}.
\subsection{$\Z_2$-orbifold of the supersymmetric $D_4$-torus model}\label{D4orbifold}
In order to obtain a K3 theory, we now consider a 
{\em $\mathbb{Z}_2$-orbifold} of the supersymmetric $D_4$-torus model. 
The group $\Z_2$ acts in the 
usual manner on the fields
{of the bosonic $D_4$-model}, i.e.\ it maps 
$j_k(z)\mapsto -j_k(z)$, $\qu\jmath_k(\qu z)\mapsto -\qu\jmath_k(\qu z)$, 
$k\in\{1,\ldots,4\}$, and 
$V_{\gamma} \mapsto 
V_{-\gamma}$ 
for all $\gamma\in\Gamma_{4,4}$. This action is induced by the 
transformation that  leaves $\psi_{{5}}(z),\ldots, \psi_{{8}}(z)$ invariant, 
while mapping $\psi_i(z)\mapsto -\psi_i(z)$ where $i\in\{{9,\ldots,12}\}$, as 
can be checked by inspection of \eqref{u1currents}.
In other words, we have $x_k(z)\leftrightarrow x_k^\ast(z)$, and analogously 
for the right-moving fermions. 
Note that the $\Z_2$-orbifold action on the eight Majorana fermions
$\psi_i(z),\, i\in\{5,\ldots 12\}$, and their anti-holomorphic counterparts,
which before orbifolding had coupled 
spin structures as demanded by  the 
$\widehat{\mathfrak{so}}(16)_{1}$  symmetry, decouples the boundary conditions 
of the first four Majorana spinors from the last four. Therefore,
the $\Z_2$-orbifold action breaks the $\widehat{\mathfrak{so}}(16)_{1}$ 
symmetry of the supersymmetric $D_4$-torus model to 
$\widehat{\mathfrak{so}}(8)_{1}\oplus\widehat{\mathfrak{so}}(8)_{1}$.
\smallskip\par

On the fermions $\psi_k(z)$ and $ \qu \psi_k(\qu z)$ for 
$k\in\{1,\ldots,4\}$, which are the supersymmetric partners of the 
$U(1)$ currents $j_k(z)$ and $\qu\jmath_k(\qu z)$, the group $\Z_2$ acts as $\psi_k \mapsto - \psi_k$ and
likewise for the right-movers, $\overline\psi_k \mapsto -\overline\psi_k$. 
In particular, the orbifold  leaves the 
$\widehat{\mathfrak{so}}(8)_{1}$
algebra in \eqref{affer} invariant, since it is generated by all bilinear fermion combinations,
whose $\widehat{\mathfrak{so}}(4)_{1,L}$
currents are given in  \cite[(2.13) - (2.16)]{gtvw14}. 

Altogether, the orbifold thus has an affine current algebra of type
$$
\left(\widehat{\mathfrak{so}}(4)_{1,L} \oplus \widehat{\mathfrak{so}}(4)_{1,R}\right)^3
\subset
\widehat{\mathfrak{so}}(8)_{1}^3 . 
$$
The untwisted sector of the $\Z_2$-orbifold is  
generated by the $\Z_2$-invariant $(h;\qu h)=(1;0)$-fields with $\C$-basis 
$$
\mbox{for } j<k,\qquad
V_{({\e}_j+{\e}_k;0)}(z) + V_{(-{\e}_j-{\e}_k;0)}(z),\qquad 
V_{({\e}_j-{\e}_k;0)}(z)+V_{(-{\e}_j+{\e}_k;0)}(z) ,
$$
along with the $\Z_2$-invariant $(h;\qu h)=({1\over2};{1\over2})$-fields
which are of the form $V_{\gamma}(z,\qu z)+ 
V_{-\gamma}(z,\qu z)$.
\medskip\par

In the twisted sector, the twisted ground states of our $\Z_2$-orbifold amount to the
Ramond ground states for pairs of free Dirac fermions
$\chi_k,\, \chi_k^\ast,\, \qu\chi_k,\, \qu\chi^\ast_k$ with {$k\in\{1,2\}$,} $k\in\{3,4\}$ and
$k\in\{5,6\}$, respectively. Hence our K3  theory allows an elegant free fermion
description with respect to the 
$\left(\widehat{\mathfrak{so}}(4)_{1,L} \oplus \widehat{\mathfrak{so}}(4)_{1,R}\right)^3$
current algebra (c.f.\ \cite[\S3.2 and Appendix D]{gtvw14})
introduced above:
the spin structures of left- and right-movers within each of the three 
summands $\widehat{\mathfrak{so}}(4)_{1,L} \oplus \widehat{\mathfrak{so}}(4)_{1,R}$
are coupled; the contributions of each of these summands  to the 
Neveu-Schwarz sector, according to \cite[(C.3), (C.4)]{gtvw14}, are
\be\label{C3}
({\rm NS}, {\rm NS}, {\rm NS})\quad({\rm NS}, {\rm R}, {\rm R})\quad
({\rm R}, {\rm NS}, {\rm R})\quad ({\rm R}, {\rm R}, {\rm NS}) \ ,
\ee
and those  to the Ramond-sector come from
\be\label{C4}
({\rm R}, {\rm NS}, {\rm NS})\quad({\rm R}, {\rm R}, {\rm R})\quad
({\rm NS}, {\rm NS}, {\rm R})\quad ({\rm NS}, {\rm R}, {\rm NS}) \ .
\ee
In terms of the vacuum, vector, spinor and antispinor representations
of $\widehat{\mathfrak{so}}(8)_1\supset 
\widehat{\mathfrak{so}}(4)_{1,L} \oplus \widehat{\mathfrak{so}}(4)_{1,R}$
{let us denote by
$\HHH_{\SSS_1\SSS_2\SSS_3}$ {with $\SSS_k\in\{0,\, v,\, s,\, c\}$}
the threefold (fermionic) tensor product 
{of the respective $\HHH_{S_k}$ of \eqref{so8deco},}
according to the
three entries in each triplet of \eqref{C3}, \eqref{C4}. 
Then}
\eqref{C3} means that the Neveu-Schwarz 
sector of the theory has the following bosonic and fermionic spaces of states:
\be\label{NSspaceofstates}
\begin{array}{rcl}
\HHH_{\rm bos}^{\rm NS}
&=& \HHH_{000} \oplus \HHH_{0ss}\oplus \HHH_{s0s} \oplus \HHH_{ss0},\\[5pt]
\HHH_{\rm ferm}^{\rm NS}
&=& \HHH_{vvv} \oplus \HHH_{vcc}\oplus \HHH_{cvc} \oplus \HHH_{ccv}.
\end{array}
\ee
{Analogously}, by \eqref{C4},  the Ramond sector
of the theory has the following bosonic and fermionic spaces of states:
\be\label{Rspaceofstates}
\begin{array}{rcl}
\HHH_{\rm bos}^{\rm R}
&=& \HHH_{s00} \oplus \HHH_{sss}\oplus \HHH_{00s} \oplus \HHH_{0s0},\\[5pt]
\HHH_{\rm ferm}^{\rm R}
&=& \HHH_{cvv} \oplus \HHH_{ccc}\oplus \HHH_{vvc} \oplus \HHH_{vcv}.
\end{array}
\ee
This is in accord with \cite[(11.15), (11.16)]{dmc15}.

The explicit form \eqref{leftrightmovingOPE} of the OPE 
now
confirms that OPEs are well-defined without
square root cuts between any two fields corresponding to
states in $\HHH_{\rm bos}^{\rm NS}\oplus \HHH_{\rm bos}^{\rm R}$,
and also between 
any two fields corresponding to
states in $\HHH_{\rm bos}^{\rm NS}\oplus \HHH_{\rm ferm}^{\rm NS}$,
as they should.
\smallskip\par

\noindent
The charge lattice $\Gamma$ governing this theory is
most conveniently described as a sublattice
\be\label{tripletbeforereflection}
\textstyle
\Gamma
\subset \widetilde\Gamma_{2,2}\oplus\widetilde\Gamma_{2,2}\oplus\widetilde\Gamma_{2,2}
\quad\subset\quad\R^{6,6},
\ee
with $\widetilde\Gamma_{2,2}$ as in \eqref{gammat22},
equipped with the symmetric bilinear form $\bullet$ 
that was introduced in \eqref{signatureform}.
Here, each  
of the three
identical summands $\widetilde\Gamma_{2,2}$
in the overlattice governs the charges of one of the sectors 
$\SSS_1,\,\SSS_2,\,\SSS_3$ in $\HHH_{\SSS_1\SSS_2\SSS_3}$. 

Now from \eqref{NSspaceofstates}, \eqref{Rspaceofstates} and 
recalling  \eqref{so8deco}, one reads
that the bosonic sector $\HHH^{\rm NS}_{\rm bos}\oplus \HHH^{\rm R}_{\rm bos}$
is governed by the even self-dual lattice
$$
\begin{array}{rcl}
\Gamma_{\rm bos} &:=&
\left(\widetilde\Gamma_0\cup(\widetilde\gamma^{(2)}+\widetilde\Gamma_0)\right)
\oplus \left(\widetilde\Gamma_0\cup(\widetilde\gamma^{(2)}+\widetilde\Gamma_0)\right)
\oplus \left(\widetilde\Gamma_0\cup(\widetilde\gamma^{(2)}+\widetilde\Gamma_0)\right)\\[5pt]
&\cong&\Gamma_{2,2}\oplus\Gamma_{4,4}=\Gamma_{6,6}
\end{array}
$$
and thus agrees, as a bosonic conformal field theory, with 
 the bosonic sector of the 
toroidal superconformal field theory on the standard torus $\R^4/\Z^4$ with 
vanishing B-field. This was in fact already shown in \cite[Rem.~3.8]{nawe00}.
Using the notations of \eqref{gammat22} and in keeping with the decomposition 
\eqref{tripletbeforereflection} into contributions from the three summands 
$\widetilde\Gamma_{2,2}$, we set
$$
\gamma^{(0)}:=0, \quad
\gamma^{(1)}:= \left(\widetilde\gamma^{(1)},\widetilde\gamma^{(1)},\widetilde\gamma^{(1)}  \right), \quad
\gamma^{(2)}:=\left(\widetilde\gamma^{(2)},\widetilde\gamma^{(2)},\widetilde\gamma^{(2)}  \right), \quad
\gamma^{(3)}:=\gamma^{(1)}+\gamma^{(2)},
$$
and find that the charge lattice of our K3 theory is half-integral,
$$
\Gamma=\Gamma_{\rm{bos}}\cup\left( \gamma^{(1)} + \Gamma_{\rm bos} \right).
$$
The lattice $\Gamma$
meets all the assumptions on the lattice $\Gamma$
 of Appendix \ref{halfcoc}, with
 $ \Gamma_0 =\Gamma^\ast$. By our construction 
 in Appendix \ref{halfcoc}, we thus obtain well-defined 
cocycles obeying  the additional symmetry and 
gauge requirements \eqref{cocycleswap}, \eqref{epsbimulti}, 
\eqref{specialgauge}.
\subsection{Partition function}\label{partifunc}
The free fermion description given above is  convenient in order to determine the 
partition function of the theory and -- by means of the elliptic genus -- to confirm
that it  is a K3  theory. In fact, by the results of 
\cite{eoty89}, the usual $\Z_2$-orbifold of every supersymmetric $(d=4)$-dimensional
torus model has the elliptic genus of K3 and thus
is indeed a {\em K3 theory} by definition, see \cite{we14}.

In the following, we calculate the various contributions
to the partition function that can be read from \eqref{NSspaceofstates},
\eqref{Rspaceofstates}. We use the standard notations for Jacobi 
theta functions, which we also summarize in Appendix \ref{theta} for the
reader's convenience.

By \eqref{NSspaceofstates}, the contributions to the partition function
$$
Z_{\widetilde{\rm NS}}(\tau,z)
= \tr_{\HHH_{\rm bos}^{\rm NS}} \left( y^{J_0} \qu y^{\qu J_0} q^{L_0-1/4} \qu q^{\qu L_0-1/4} \right)
- \tr_{\HHH_{\rm ferm}^{\rm NS}} \left( y^{J_0} \qu y^{\qu J_0} q^{L_0-1/4} \qu q^{\qu L_0-1/4} \right)
$$
from the Neveu-Schwarz sector as defined in \eqref{partifuncdef}, in terms of the different ingredients
to \eqref{C3}, are given by
\begin{eqnarray}
 \qquad\qquad
 ({\rm NS}, {\rm NS}, {\rm NS}):
 && \textstyle
 {1\over4}
 \left(
   \left| {\vartheta_3(\tau)\over\eta(\tau)}\right|^{8}
   + \left| {\vartheta_4(\tau)\over\eta(\tau)}\right|^{8} \right)
   \cdot \left| {\vartheta_4(\tau,z)\over\eta(\tau)}\right|^4\label{sectornsnsns}\\
   &&\qquad\qquad\qquad\qquad\textstyle
+ {1\over2} \left| {\vartheta_3(\tau)\vartheta_4(\tau)\over\eta^2(\tau)}\right|^4
\cdot \left| {\vartheta_3(\tau,z)\over\eta(\tau)}\right|^4\, \nonumber\\
 ({\rm NS}, {\rm R}, {\rm R}):
 &&\textstyle
  {1\over4} \left| {\vartheta_2(\tau)\over\eta(\tau)}\right|^{8} 
    \cdot \left| {\vartheta_4(\tau,z)\over\eta(\tau)}\right|^4,
 \label{sectornsrr}\\
({\rm R}, {\rm NS}, {\rm R})
&+& ({\rm R}, {\rm R}, {\rm NS}):\label{sectorrnsr}\\
&&\textstyle
{1\over2} \left| {\vartheta_2(\tau)\vartheta_3(\tau)\over\eta^2(\tau)}\right|^4
\cdot \left| {\vartheta_1(\tau,z)\over\eta(\tau)}\right|^4 
+ 
{1\over2} \left| {\vartheta_2(\tau)\vartheta_4(\tau)\over\eta^2(\tau)}\right|^4
\cdot \left| {\vartheta_2(\tau,z)\over\eta(\tau)}\right|^4 \nonumber
\ .
\end{eqnarray}
Analogously, by \eqref{Rspaceofstates}, \eqref{C4}, the 
contributions to the partition function
$$
Z_{\widetilde R}(\tau,z)
= \tr_{\HHH_{\rm bos}^{\rm R}} \left( y^{J_0} \qu y^{\qu J_0} q^{L_0-1/4} \qu q^{\qu L_0-1/4} \right)
- \tr_{\HHH_{\rm ferm}^{\rm R}} \left( y^{J_0} \qu y^{\qu J_0} q^{L_0-1/4} \qu q^{\qu L_0-1/4} \right)
$$
from the Ramond sector are
\begin{eqnarray}
 \qquad({\rm R}, {\rm NS}, {\rm NS}):
 && \textstyle
 {1\over4}
 \left(
   \left| {\vartheta_3(\tau)\over\eta(\tau)}\right|^{8}
   + \left| {\vartheta_4(\tau)\over\eta(\tau)}\right|^{8} \right)
   \cdot \left| {\vartheta_1(\tau,z)\over\eta(\tau)}\right|^4\label{sectorrnsns}\\
   &&\qquad\qquad\qquad\qquad\textstyle
+ {1\over2} \left| {\vartheta_3(\tau)\vartheta_4(\tau)\over\eta^2(\tau)}\right|^4
\cdot \left| {\vartheta_2(\tau,z)\over\eta(\tau)}\right|^4, \nonumber\\
 ({\rm R}, {\rm R}, {\rm R}):
 &&\textstyle
  {1\over4} \left| {\vartheta_2(\tau)\over\eta(\tau)}\right|^{8} 
    \cdot \left| {\vartheta_1(\tau,z)\over\eta(\tau)}\right|^4,
 \label{sectorrrr}\\
({\rm NS}, {\rm NS}, {\rm R})
&+& ({\rm NS}, {\rm R}, {\rm NS}):\label{sectornsnsr}\\
&&\textstyle
{1\over2} \left| {\vartheta_2(\tau)\vartheta_3(\tau)\over\eta^2(\tau)}\right|^4
\cdot \left| {\vartheta_4(\tau,z)\over\eta(\tau)}\right|^4 
+ 
{1\over2} \left| {\vartheta_2(\tau)\vartheta_4(\tau)\over\eta^2(\tau)}\right|^4
\cdot \left| {\vartheta_3(\tau,z)\over\eta(\tau)}\right|^4 \nonumber
\ .
\end{eqnarray}
Altogether, the four parts of the partition function of \eqref{partifuncdef} are given by
\begin{eqnarray*}
\textstyle
Z_{{\rm NS}}(\tau, z) 
&=& \textstyle{1\over2}\left(
  {1\over2}\sum\limits_{k=2}^4 \left| {\vartheta_k(\tau)\over\eta(\tau)}\right|^{8}
\cdot \left| {\vartheta_3(\tau,z)\over\eta(\tau)}\right|^4
+ \left| {\vartheta_3(\tau)\vartheta_4(\tau)\over\eta^2(\tau)}\right|^4
\cdot \left| {\vartheta_4(\tau,z)\over\eta(\tau)}\right|^4 \right.\nonumber\\
&&\textstyle
\hfill\left.\;\; +   
\left| {\vartheta_2(\tau)\vartheta_3(\tau)\over\eta^2(\tau)}\right|^4
\cdot \left| {\vartheta_2(\tau,z)\over\eta(\tau)}\right|^4 
+ 
\left| {\vartheta_2(\tau)\vartheta_4(\tau)\over\eta^2(\tau)}\right|^4
\cdot \left| {\vartheta_1(\tau,z)\over\eta(\tau)}\right|^4 
\right),\\
Z_{\widetilde{\rm NS}}(\tau, z) 
&=& \textstyle{1\over2}\left(
  {1\over2}\sum\limits_{k=2}^4 \left| {\vartheta_k(\tau)\over\eta(\tau)}\right|^{8}
\cdot \left| {\vartheta_4(\tau,z)\over\eta(\tau)}\right|^4
+ \left| {\vartheta_3(\tau)\vartheta_4(\tau)\over\eta^2(\tau)}\right|^4
\cdot \left| {\vartheta_3(\tau,z)\over\eta(\tau)}\right|^4 \right.\nonumber\\
&&\textstyle
\hfill\left.\;\; +   
\left| {\vartheta_2(\tau)\vartheta_3(\tau)\over\eta^2(\tau)}\right|^4
\cdot \left| {\vartheta_1(\tau,z)\over\eta(\tau)}\right|^4 
+ 
\left| {\vartheta_2(\tau)\vartheta_4(\tau)\over\eta^2(\tau)}\right|^4
\cdot \left| {\vartheta_2(\tau,z)\over\eta(\tau)}\right|^4 
\right),
\\
Z_{{\rm R}}(\tau, z) 
&=&\textstyle {1\over2}\left(
  {1\over2}\sum\limits_{k=2}^4 \left| {\vartheta_k(\tau)\over\eta(\tau)}\right|^{8}
\cdot \left| {\vartheta_2(\tau,z)\over\eta(\tau)}\right|^4
+ \left| {\vartheta_3(\tau)\vartheta_4(\tau)\over\eta^2(\tau)}\right|^4
\cdot \left| {\vartheta_1(\tau,z)\over\eta(\tau)}\right|^4 \right.\nonumber\\
&&\textstyle
\hfill\left.\;\; +   
\left| {\vartheta_2(\tau)\vartheta_3(\tau)\over\eta^2(\tau)}\right|^4
\cdot \left| {\vartheta_3(\tau,z)\over\eta(\tau)}\right|^4 
+ 
\left| {\vartheta_2(\tau)\vartheta_4(\tau)\over\eta^2(\tau)}\right|^4
\cdot \left| {\vartheta_4(\tau,z)\over\eta(\tau)}\right|^4 
\right),
\\
Z_{\widetilde{\rm R}}(\tau, z) 
&=& \textstyle{1\over2}\left(
  {1\over2}\sum\limits_{k=2}^4 \left| {\vartheta_k(\tau)\over\eta(\tau)}\right|^{8}
\cdot \left| {\vartheta_1(\tau,z)\over\eta(\tau)}\right|^4
+ \left| {\vartheta_3(\tau)\vartheta_4(\tau)\over\eta^2(\tau)}\right|^4
\cdot \left| {\vartheta_2(\tau,z)\over\eta(\tau)}\right|^4 \right.\nonumber\\
&&\textstyle
\hfill\left.\;\; +   
\left| {\vartheta_2(\tau)\vartheta_3(\tau)\over\eta^2(\tau)}\right|^4
\cdot \left| {\vartheta_4(\tau,z)\over\eta(\tau)}\right|^4 
+ 
\left| {\vartheta_2(\tau)\vartheta_4(\tau)\over\eta^2(\tau)}\right|^4
\cdot \left| {\vartheta_3(\tau,z)\over\eta(\tau)}\right|^4 
\right).
\end{eqnarray*}
%
\section{The Conway Moonshine Module}\label{supernatural}
In this section, we summarize Duncan's construction of 
the {\em Conway Moonshine Module}\footnote{For the relevant definitions 
concerning super vertex operator algebras and their properties, we 
refer the reader to the literature, see e.g. \cite{ka96,frbe04,leli04}, as well as the very
accessible summary in \cite[\S5]{dmc15}.}
$V^{s\natural}\oplus V^{s\natural}_{\rm tw}$
\cite{du07,dmc14}. Section \ref{supernaturalconstruction}
closely follows the exposition in \cite[\S6]{dmc15} but accompanies it by
a description in terms of a lattice theory, while in Section \ref{u1choices},
we include some additional structure that we need for comparison to 
the K3 theory of Section \ref{gtvw}.

By \cite[Thm.~5.15]{du07},  
$V^{s\natural}$ 
is the unique self-dual, $C_2$-cofinite  super vertex operator algebra
of CFT type with central charge $c=12$, such that for the Virasoro zero mode
$L_{(0)}$ on $V^{s\natural}$, the kernel of $L_{(0)}-{1\over2}\mb{id}_{V^{s\natural}}$ is 
trivial.
Moreover, $V^{s\natural}_{\rm tw}$ is an irreducible canonically twisted 
$V^{s\natural}$-supermodule,
and as such, it is unique according to \cite[\S4]{dmc14}. 
\subsection{The construction of 
$V^{s\natural}\oplus V^{s\natural}_{\rm tw}$}\label{supernaturalconstruction}
Both $V^{s\natural}$ and $V^{s\natural}_{\rm tw}$ are obtained using
a standard construction  
\cite{ffr91,dmc14}
that attaches a super vertex operator algebra
$A(\mathfrak{a})$ and a canonically twisted module $A(\mathfrak{a})_{\rm tw}$
for it to any finite dimensional complex vector space $\mathfrak{a}$ equipped
with a non-degenerate symmetric bilinear form $(\cdot,\cdot)$. For later
convenience, we will always assume  that the dimension of $\mathfrak{a}$ is even.
Moreover, by a slight abuse of terminology, we will call a family
$(\upsilon_1,\ldots,\upsilon_k)$ of elements of $\mathfrak{a}$ 
{\em orthonormal}, iff for all $i,\,j\in\{1,\ldots,k\}$, we have
$(\upsilon_i,\upsilon_j)=\delta_{ij}$.

For every $n\in\Z$, one now introduces a copy 
$\mathfrak{a}_{(n+{1/2})}\cong\mathfrak{a}$ and sets
$$
\widehat{\mathfrak{a}}^- 
:= \bigoplus_{n<0} \mathfrak{a}_{(n+{1/2})},
\qquad
A(\mathfrak{a}):=\bigwedge (\widehat{\mathfrak{a}}^-)\Omega
\cong \bigwedge (\widehat{\mathfrak{a}}^-),
$$
where $\Omega$ denotes a 
choice of a {\em vacuum state}, such that in particular, 
for $w\in\bigwedge (\widehat{\mathfrak{a}}^-)$, 
$w(\Omega):=w\Omega$ yields the isomorphism 
$A(\mathfrak{a})\cong\bigwedge (\widehat{\mathfrak{a}}^-)$. 
The construction of the standard super vertex algebra
on the vector space $A(\mathfrak{a})$ involves a choice
of isomorphism $\mathfrak{a}\longrightarrow \mathfrak{a}_{(n+{1/2})}$
for every $n\in\Z$, denoted\footnote{As a warning to the bilingual
reader we remark that our $\upsilon_{(\nu)}$ are denoted $\upsilon{(\nu)}$ 
in the vertex algebra literature, while their $\upsilon_{(\nu)}$ relate
to our $\upsilon_{(\nu)}$ by a weight-dependent shift of $\nu$.}
$\upsilon\mapsto \upsilon_{(n+{1/2})}$.
The $\upsilon_{(n+{1/2})}$ are regarded as linear maps
on $A(\mathfrak{a})$, acting by left multiplication if $n<0$, and obeying 
$\upsilon_{(n+{1/2})}\Omega=0$ for all 
$\upsilon\in\mathfrak{a}$ and $n\in\N$ as well as
the Clifford algebra\footnote{Note that the normalization chosen by 
Duncan and Mack-Crane in  \cite{dmc14,dmc15} 
differs from ours and \cite[(2.41)]{ffr91} 
by a factor of  $-2$ on the right hand side of
\eqref{Clifford}.}
\be\label{Clifford}
\forall \upsilon,\varphi\in\mathfrak{a},\; \forall m,n\in\Z\colon
\qquad \upsilon_{(n+{1/2})} \varphi_{(m+{1/2})}
+ \varphi_{(m+{1/2})} \upsilon_{(n+{1/2})}
= \delta_{m+n+1,0}\cdot (\upsilon,\varphi) .
\ee
This uniquely
fixes the action of each $\upsilon_{(n+{1/2})}$
on $A(\mathfrak{a})$ and by \cite[Thm.~4.4.1]{frbe04} extends uniquely
to a super vertex algebra structure on $A(\mathfrak{a})$. 
In physics terminology,
the field associated to $\upsilon\in\mathfrak{a}$ is a free 
Majorana fermion. By
the standard Sugawara construction \cite{su68,so68}, 
applied to a maximal set of $U(1)$ currents with pairwise
trivial OPEs (c.f.\, \cite{frka80,se81,gool86}, for example), $A(\mathfrak{a})$ enjoys
the action of a Virasoro algebra at central charge $c={1\over2}\dim\mathfrak{a}$,
promoting $A(\mathfrak{a})$ to a super vertex {\em operator}  algebra.
The standard modes generating this Virasoro algebra are denoted
$L_{(n)},\, n\in\Z$, in the following.

The canonically twisted module of $A(\mathfrak{a})$ is similarly obtained
by introducing a copy 
$\mathfrak{a}_{(n)}\cong\mathfrak{a}$ 
and a choice of $\C$ vector space isomorphism
$\mathfrak{a}\longrightarrow \mathfrak{a}_{(n)}$,
$\upsilon\mapsto \upsilon_{(n)}$
for every $n\in\Z$. In addition, one chooses a polarization
$\mathfrak{a}=\mathfrak{a}^+\oplus\mathfrak{a}^-$ 
for $\mathfrak{a}$ with respect to $(\cdot, \cdot)$. Let 
$(\mathfrak{a}^-)_{(0)}$  denote the image of $\mathfrak{a}^-$ under
$\mathfrak{a}\longrightarrow \mathfrak{a}_{(0)}$,
$\upsilon\mapsto \upsilon_{(0)}$.
One then sets
$$
\widehat{\mathfrak{a}}^-_{\rm tw}
:= (\mathfrak{a}^-)_{(0)}\oplus\bigoplus_{n<0} \mathfrak{a}_{(n)},
\qquad
A(\mathfrak{a})_{\rm tw}:=\bigwedge (\widehat{\mathfrak{a}}^-_{\rm tw})\Omega_{\rm tw}
\cong \bigwedge (\widehat{\mathfrak{a}}^-_{\rm tw}),
$$
where $\Omega_{\rm tw}$ is a 
choice of a {\em twisted ground state}, and similarly to the above, 
for $w\in\bigwedge (\widehat{\mathfrak{a}}^-_{\rm tw})$, 
$w(\Omega_{\rm tw}):=w\Omega_{\rm tw}$. 
As above, the  $\upsilon_{(n)}$ are regarded as linear maps
on $A(\mathfrak{a})_{\rm tw}$, acting by left multiplication if $n<0$, and obeying 
$\upsilon_{(n)}\Omega_{\rm tw}=0$ 
if $n>0$, or $n=0$ and  $\upsilon\in\mathfrak{a}^+$,
as well as the
Clifford algebra \eqref{Clifford} in its incarnation
$$
\forall \upsilon,\varphi\in\mathfrak{a},\; \forall m,n\in\Z\colon
\qquad
\upsilon_{(n)}\varphi_{(m)}
+ \varphi_{(m)}\upsilon_{(n)}
= \delta_{m+n,0}\cdot (\upsilon,\varphi) .
$$
As for $A(\mathfrak{a})$, this uniquely 
fixes the action of each $\upsilon_{(n)}$
on $A(\mathfrak{a})_{\rm tw}$. 
According to \cite[\S2.2]{frsz04}, this
extends uniquely to a canonically twisted 
$A(\mathfrak{a})$-module structure on $A(\mathfrak{a})_{\rm tw}$.
\smallskip\par

The above construction ensures a natural action of the standard
Clifford algebra $\mb{Cliff}(\mathfrak{a})$ associated to
$\left( \mathfrak{a},(\cdot,\cdot) \right)$ on $A(\mathfrak{a})_{\rm tw}$,
such that an element represented within
the tensor algebra of $\mathfrak{a}$
by $\upsilon_1\otimes\cdots\otimes\upsilon_k$
with $\upsilon_1,\ldots,\upsilon_k\in\mathfrak{a}$, $k\in\N$,  acts by 
$(\upsilon_1)_{(0)}\circ\cdots\circ(\upsilon_k)_{(0)}$. The 
$\mb{Cliff}(\mathfrak{a})$-submodule $\boldsymbol{CM}$ of 
$A(\mathfrak{a})_{\rm tw}$ generated by $\Omega_{\rm tw}$
is the unique (up to isomorphism) non-trivial irreducible 
representation of $\mb{Cliff}(\mathfrak{a})$ 
\cite[(6.16)]{dmc15}. In addition, one chooses a 
{\em fermion number operator} $(-1)^F$ on $A(\mathfrak{a})$
and on $A(\mathfrak{a})_{\rm tw}$, where
$$
\textstyle
(-1)^F\upsilon_{(k)} +\upsilon_{(k)}(-1)^F =0 
\;\;\forall \upsilon\in\mathfrak{a},\; k\in{1\over2}\Z,
\quad
(-1)^F\Omega=\Omega,\; (-1)^F\Omega_{\rm tw}=\Omega_{\rm tw}.
$$
The algebra generated by the $\upsilon_{(0)}$ with $\upsilon\in\mathfrak{a}$
together with $(-1)^F$,  in the physics literature  is known as the
{\em fermionic zero mode algebra}. In \cite{dmc15},
$(-1)^F$ is obtained  by choosing a
lift of $-\mb{id}_{\mathfrak{a}}\in\mb{SO}(\mathfrak{a})$
to $\mb{Spin}(\mathfrak{a})$ which is compatible with the
polarization $\mathfrak{a}=\mathfrak{a}^+\oplus\mathfrak{a}^-$ of
$\mathfrak{a}$. The fermion number operator $(-1)^F$ induces a $\Z_2$-grading on 
$A(\mathfrak{a})$
and on $A(\mathfrak{a})_{\rm tw}$, such that
$$
A(\mathfrak{a})= A(\mathfrak{a})^0\oplus A(\mathfrak{a})^1,
\qquad
A(\mathfrak{a})_{\rm tw}=A(\mathfrak{a})_{\rm tw}^0\oplus A(\mathfrak{a})_{\rm tw}^1,
$$
where $A(\mathfrak{a})^j$ and $A(\mathfrak{a})_{\rm tw}^j$, 
with $j \in \{0,1\}$, are the $(-1)^j$ eigenspaces
of $(-1)^F$ on $A(\mathfrak{a})$, $A(\mathfrak{a})_{\rm tw}$. 
\smallskip\par

The super vertex operator algebra $V^{s\natural}$ and its canonically 
twisted module $V^{s\natural}_{\rm tw}$ are now obtained from 
$\mathfrak{a}\cong\C^{24}$ with the standard
bilinear form $(\cdot,\cdot)$ as
$$
V^{s\natural} := A(\mathfrak{a})^0\oplus A(\mathfrak{a})_{\rm tw}^1,
\qquad
V^{s\natural}_{\rm tw}:=A(\mathfrak{a})_{\rm tw}^0\oplus A(\mathfrak{a})^1,
$$
where according to \cite{dmc14} (c.f.\ \cite[Prop.~8.1]{dmc15}), the 
$A(\mathfrak{a})^0$-module structure of $V^{s\natural}$
extends uniquely to a super vertex operator algebra structure on 
$V^{s\natural}$, and the $A(\mathfrak{a})^0$-module structure of 
$V^{s\natural}_{\rm tw}$ extends uniquely to a canonically twisted
$V^{s\natural}$-module structure. 
As mentioned above, by 
\cite[Thm.~5.15]{du07}, $V^{s\natural}$ is the unique self-dual,
$C_2$-cofinite super vertex operator algebra with central
charge $c=12$ and trivial 
$\mb{ker}\left( L_{(0)} - {1\over2}\mb{id}_{V^{s\natural}}\right)$.
We call the subspace
$V^{s\natural}$, equipped with its structure as a 
super vertex operator algebra, the {\em Neveu-Schwarz sector}
of the Conway Moonshine Module,
and $V^{s\natural}_{\rm tw}$ its {\em Ramond sector}.

In physics terminology, every $\upsilon\in A(\mathfrak{a})^j$ or
$\upsilon\in A(\mathfrak{a})_{\rm tw}^j$, $j\in\{0,1\}$, is
a state in a free fermion theory (see, e.g., \cite{dowe08}
for a systematic description of free fermion theories
in the context of heterotic strings on Calabi-Yau three-folds),
obtained from $24$ free Majorana fermions with 
coupled spin structures. Analogously to the analysis 
of Section \ref{D4orbifold}, any
decomposition $\mathfrak{a}
=\mathfrak{b}_1\oplus\mathfrak{b}_2\oplus\mathfrak{b}_3$
with $\dim\mathfrak{b}_k=8$ for each $k\in\{1,2,3\}$ allows
a description of the contributions to $A(\mathfrak{a})^j$ and
$A(\mathfrak{a})_{\rm tw}^j$ in terms of threefold (fermionic)
tensor products $U_{\SSS_1\SSS_2\SSS_3}$, where 
$\SSS_k\in\{0,\,v,\,s,\,c\}$  for $k\in\{1,2,3\}$
labels the vacuum, vector,
spinor or antispinor representation of the 
 affine algebra $\widehat{\mathfrak{so}}(8)_1$ corresponding
to $\mathfrak{b}_k$:
\be\label{DMCsectordeco}
\begin{array}{rclcl}
V^{s\natural}_{\rm bos} &:=& A(\mathfrak{a})^0 \\[5pt]
&=& \left( A(\mathfrak{b}_1)\wedge A(\mathfrak{b}_2)\wedge A(\mathfrak{b}_3) \right)^0 
&=& U_{000}\oplus U_{0vv}\oplus U_{v0v}\oplus U_{vv0},\\[5pt]
V^{s\natural}_{\rm ferm} &:=& A(\mathfrak{a})_{\rm tw}^1 \\[5pt]
&=& \left( A(\mathfrak{b}_1)_{\rm tw}\wedge A(\mathfrak{b}_2)_{\rm tw}\wedge A(\mathfrak{b}_3)_{\rm tw} \right)^1
&=& U_{ccc}\oplus U_{css}\oplus U_{scs}\oplus U_{ssc},\\[5pt]
V^{s\natural}_{\rm tw, bos} &:=& A(\mathfrak{a})_{\rm tw}^0 \\[5pt]
&=& \left( A(\mathfrak{b}_1)_{\rm tw}\wedge A(\mathfrak{b}_2)_{\rm tw}\wedge A(\mathfrak{b}_3)_{\rm tw} \right)^0 
&=& U_{scc}\oplus U_{sss}\oplus U_{ccs}\oplus U_{csc},\\[5pt]
V^{s\natural}_{\rm tw, ferm} &:=& A(\mathfrak{a})^1\\[5pt]
&=& \left( A(\mathfrak{b}_1)\wedge A(\mathfrak{b}_2)\wedge A(\mathfrak{b}_3) \right)^1
&=& U_{v00}\oplus U_{vvv}\oplus U_{00v}\oplus U_{0v0},
\end{array}
\ee
c.f.\ \cite[(11.20),(11.21)]{dmc15}.
\smallskip\par
This also shows that the structure of 
$V^{s\natural}\oplus V^{s\natural}_{\rm tw}$ can be conveniently
encoded in terms of a lattice vertex operator algebra, by bosonization.
Indeed, analogously to the discussion in Section \ref{D4orbifold}, and
using the notations introduced there, we find
$$
V^{s\natural}\oplus V^{s\natural}_{\rm tw}
= \bigoplus_{\gamma\in\Gamma^{\rm refl}} \HHH_\gamma,
$$
where the relevant charge lattice is
$$
\textstyle
\Gamma^{\rm refl}
:=  \Z^{12}\cup ({1\over2}+\Z)^{12}\subset\R^{12},
$$
equipped with the Euclidean scalar product.
For later convenience we remark that 
$$
\Gamma^{\rm refl} \subset 
\widetilde\Gamma_{2,2}^{\rm refl}\oplus\widetilde\Gamma_{2,2}^{\rm refl}\oplus\widetilde\Gamma_{2,2}^{\rm refl}
\subset\R^{12},
$$
where 
$$
\textstyle
\widetilde\Gamma_{2,2}^{\rm refl}
:=  \Z^{4}\cup ({1\over2}+\Z)^{4}\subset\R^{4}.
$$
More precisely, with 
$\widetilde{\Gamma}^{{\rm refl}}_0
:=\{ {\bf Q} \in \mathbb{Z}^4 \vert \vphantom{\sum\limits^4}\smash{\sum\limits_{k=1}^4}
Q_k\equiv 0 \quad {\rm mod\,\, 2}\}$,
an index 4 sublattice of $\widetilde\Gamma_{2,2}^{\rm refl}\subset \mathbb{R}^4$, 
and  the 4-vectors
\be\label{ourreps}
\textstyle
\widetilde\gamma^{(0)} := 0,\quad
\widetilde\gamma^{(1)} :=  \e_4,\quad
\widetilde\gamma^{(2)} := {1\over2}\sum\limits_{k=1}^4(\e_k, \e_k),\quad
\widetilde\gamma^{(3)} := \widetilde\gamma^{(1)}+\widetilde\gamma^{(2)},
\ee
the cosets are 
$$
\forall a\in\{0,\ldots,3\}\colon \quad
\widetilde\Gamma_{a}^{\rm refl} := \widetilde\gamma^{(a)}+ \widetilde\Gamma_{0}^{\rm refl},
\qquad\mbox{ hence }\qquad \widetilde\Gamma_{2,2}^{\rm refl} 
\;=\;\bigcup\limits_{a=0}^3 \widetilde\Gamma_{a}^{\rm refl}.
$$
One has
$$
\Gamma^{\rm refl} \;=\; \Gamma_{\rm bos}^{\rm refl}\cup\Gamma_{\rm ferm}^{\rm refl}
$$
with
\begin{equation}\label{theirlattices}
\begin{array}{rcl}
\ds\Gamma_{\rm bos}^{\rm refl}
&:=&\ds \left( \widetilde\Gamma_{0}^{\rm refl}\cup \widetilde\Gamma_{1}^{\rm refl}\right)
\oplus  \left( \widetilde\Gamma_{0}^{\rm refl}\cup \widetilde\Gamma_{1}^{\rm refl}\right)
\oplus  \left( \widetilde\Gamma_{0}^{\rm refl}\cup \widetilde\Gamma_{1}^{\rm refl}\right)
\;=\; \Z^{12},\\[8pt]
\ds\Gamma_{\rm ferm}^{\rm refl}
&:=&\ds \left( \widetilde\Gamma_{2}^{\rm refl}\cup \widetilde\Gamma_{3}^{\rm refl}\right)
\times  \left( \widetilde\Gamma_{2}^{\rm refl}\cup \widetilde\Gamma_{3}^{\rm refl}\right)
\times  \left( \widetilde\Gamma_{2}^{\rm refl}\cup \widetilde\Gamma_{3}^{\rm refl}\right)
\;=\; ({\textstyle{1\over2}}+\Z)^{12}.\\
\end{array}
\end{equation}
Note that
$$
V^{s\natural}_{\rm bos}\oplus V^{s\natural}_{\rm tw, ferm}
= \bigoplus_{\gamma\in\Gamma_{\rm bos}^{\rm refl}} \HHH_\gamma,\qquad
V^{s\natural}_{\rm ferm}\oplus V^{s\natural}_{\rm tw, bos}
= \bigoplus_{\gamma\in\Gamma_{\rm ferm}^{\rm refl}} \HHH_\gamma;
$$
 our counterintuitive choice of notations will be justified in Section \ref{flipmodule}.

To introduce cocycles, we  may again invoke the results of Appendix \ref{halfcoc}, 
since  $\Gamma_0^{{\rm refl}}:=\left(\Gamma^{{\rm refl}}\right)^\ast$ is an 
even sublattice of index $4$ in the half integral lattice $\Gamma^{{\rm refl}}$. 
It  is given by
$$
\Gamma_0^{{\rm refl}}
=\left\{ {\bf Q} \in \mathbb{Z}^{12} \left| \vphantom{\int^7}\right. 
\sum_{k=1}^{12}Q_k\equiv 0 \quad {\rm mod\,\, 2}\right\}.
$$
With
$$\gamma^{{\rm refl} (0)}:= 0,\,\,
\gamma^{{\rm refl} (1)}:= (\widetilde\gamma^{(3)}, \widetilde\gamma^{(3)},\widetilde\gamma^{(3)}),\,\,
\gamma^{{\rm refl} (2)}:= (\widetilde\gamma^{(1)}, \widetilde\gamma^{(1)},\widetilde\gamma^{(1)})\,\, {\rm and}\,\,
\gamma^{{\rm refl} (3)}:=\gamma^{{\rm refl} (1)}+\gamma^{{\rm refl} (2)},
$$
the four cosets are 
$$
\Gamma^{{\rm refl}}_a=\gamma^{{\rm refl}(a)}+\Gamma_0^{{\rm refl}}\qquad \forall a \in \{0,1,2,3\}.
$$
The two lattices $(\Gamma_0^{{\rm refl}},\Gamma^{{\rm refl}})$ form a $\Z_2$ 
lattice pair in the terminology of \cite{gnors87}. In this context one may associate
the two Lie algebras $D_{12}$ and $B_{12}$ to this lattice pair.
Indeed, the set of $264$ vectors of length square $2$ in $\Gamma^{{\rm refl}}$ 
form a root system of type $D_{12}$, which together with 
the set of $24$ vectors of length 
square $1$ in $\Gamma^{{\rm refl}}$ form a root system of type $B_{12}$.
In fact, $\Gamma^{{\rm refl}}_0$ is a root lattice of type $D_{12}$, while
$\Gamma^{{\rm refl}}$ is a root lattice of type $B_{12}$.
We obtain well-defined cocycles $\eps^{\rm refl}$ on $\Gamma^{{\rm refl}}$ 
from the construction summarized in Appendix \ref{halfcoc},
and they obey  the additional symmetry and gauge requirements 
\eqref{cocycleswap}, \eqref{epsbimulti}, \eqref{specialgauge}.
\smallskip\par

By construction, $V^{s\natural}\oplus V^{s\natural}_{\rm tw}$
enjoys a natural action of $\mb{Spin}(\mathfrak{a})$ 
which respects the super vertex operator algebra 
and twisted module structures and which on 
$A(\mathfrak{a})$ factors over $\mb{SO}(\mathfrak{a})$.
Now let $\Lambda$ denote the Leech lattice and
$\mathfrak{a}=\Lambda\otimes_\Z \C\cong\C^{24}$ with the standard
bilinear form $\left(\cdot,\cdot\right)$, such that
$\mb{Co}_0 = \mb{Aut}(\Lambda)\subset \mb{SO}(\mathfrak{a})$.
Then by \cite{dmc14} (c.f.\ \cite[Prop.~7.1]{dmc15}), there
is a unique lift $\widehat{\mb{Co}}_0$, i.e.\  a subgroup
$\widehat{\mb{Co}}_0\subset \mb{Spin}(\mathfrak{a})$
such that the natural map 
$\mb{Spin}(\mathfrak{a})\rightarrow \mb{SO}(\mathfrak{a})$
induces an isomorphism $\widehat{\mb{Co}}_0\cong\mb{Co}_0$,
thus yielding a $\mb{Co}_0$-action on the super vertex operator
algebra $V^{s\natural}$ along with its canonically twisted
module $V^{s\natural}_{\rm tw}$. Without loss of generality,
one assumes $(-1)^F$ to yield the non-trivial central
element of $\widehat{\mb{Co}}_0$ by
modifying the polarization of $\mathfrak{a}$ accordingly 
if need be.
\subsection{Choosing $U(1)$ currents}\label{u1choices}
According to \cite{dmc15}, the choice of an appropriate
$U(1)$ current for the Conway Moonshine Module
allows to attach a weak Jacobi form to any symplectic 
derived equivalence of a K3 surface that fixes a suitable
stability condition on K3.
Following \cite[(9.5)]{dmc15}, for $V^{s\natural}\oplus V^{s\natural}_{\rm tw}$
we choose a $U(1)$ current $J$ with zero mode $J_0$, by
distinguishing four free Majorana fermions which are associated
to an orthonormal basis of a four-dimensional subspace 
 $\mathfrak x$ of $\mathfrak a$.
Analogously to \eqref{Dirac}, these are combined to form
Dirac fermions $\mathbf a_X^\pm$ and $\mathbf a_Z^\pm$, 
such that insertion into the bilinear form $(\cdot,\cdot)$
on $\mathfrak{a}$ yields
$( \mathbf a_X^\pm, \mathbf a_X^\mp)
=(\mathbf a_Z^\pm,\mathbf a_Z^\mp)=1$, while inserting
any other combination of $\mathbf a_X^\pm$, $\mathbf a_Z^\pm$ in
$(\cdot,\cdot)$ yields zero. \,Then\footnote{Note that our formula
differs from that given in \cite[(9.5)]{dmc15} by a factor of $-2$ due to 
the difference in the normalization of the Clifford algebra \eqref{Clifford}.}
\be\label{dmcu1choice}
\textstyle
J := \nop[\mathbf a_X^+ \mathbf a_X^-] 
+ \nop[\mathbf a_Z^+ \mathbf a_Z^-].
\ee
This introduces charges $\pm1$ for $\mathbf a_X^\pm$ and $\mathbf a_Z^\pm$
with respect to the zero mode $J_0$ of the associated field $J(z)$,
while all  fermions corresponding to states
in $\mathfrak a$ that are perpendicular to $\mathfrak x$ remain uncharged.
Below, we will see
that this choice is compatible with the identification of 
$V^{s\natural}\oplus V^{s\natural}_{\rm tw}$ with the  space of states
underlying the K3 theory of Section \ref{gtvw}, such that $J$ is 
mapped to our choice \eqref{gtvwu1choice} of 
$U(1)$ current in the left-moving $N=4$ superconformal algebra 
of \cite{gtvw14}. Since in that model, we also naturally have a 
right-moving  $N=4$ superconformal
algebra, in addition to the data given in \cite{dmc15}, we need to determine
another $U(1)$ current, to serve as the image of a choice of right-moving
$U(1)$ current under the reflection procedure to be described
in Section \ref{flip}.
The most natural  candidate
$\widehat J$  
seems to arise
by choosing another, disjoint set of four free fermions 
associated
to an orthonormal basis of a four-dimensional 
subspace $\widehat{\mathfrak x}$ 
of $\mathfrak a$,
which is perpendicular to the four-dimensional 
subspace  $\mathfrak x$.
Indeed at  first sight,  the structure of $(J,\widehat J)$ is analogous
to the one observed in \eqref{gtvwu1choice}
for the left- and right-moving $U(1)$ currents
in the K3  theory. However, that we should have to make
additional choices
is counterintuitive. Below we will see
that consistency with the identifications of \cite{dmc15}
actually requires to introduce instead, alongside the $U(1)$ current $J$, a second $U(1)$ current
$\overline J$ with zero mode $\overline J_0$, by 
setting
\be\label{dmcu1choiceright}\textstyle
\overline J := \nop[\mathbf a_X^+ \mathbf a_X^-] 
- \nop[\mathbf a_Z^+ \mathbf a_Z^-].
\ee
This introduces charges $\pm1$ for $\mathbf a_X^\pm$ and $\mathbf a_Z^\mp$
with respect to $\qu J_0$,
while all other fermions remain uncharged. 

In \cite{dmc15}, the choice
of the  four-dimensional subspace 
$\mathfrak x$ is interpreted
in terms of the choice of a complex structure along with a stability
condition on an algebraic K3 surface following \cite{hu06,hu11,hu13}.
However, in all of these references
this structure is only used to attach a new label to the 
refined geometric interpretations (c.f.\ \cite{we05}) of the 
points in the 
moduli space of SCFTs on K3, following \cite{asmo94,nawe00}. 
Indeed, even the subdivision of the four-dimensional space into two 
two-dimensional subspaces is never relevant in the work of \cite{dmc15},
other than yielding the interpretation in terms of stability conditions.
The latter introduces the very unnatural restriction to algebraic K3 surfaces,
which is unnecessary in the original interpretation of the 
moduli space and its refinements \cite{asmo94,nawe00}. 

The above-mentioned observation that the natural $U(1)$ charges for a choice of
orthonormal basis of  the four-dimensional subspace 
$\mathfrak x\subset\mathfrak a$ turn out to be
$\pm(1,1)$ and $ \pm(1,-1)$ is in accord with the observation in
\cite{nawe00} that this four-dimensional subspace corresponds
to a choice of {\em four charged Ramond ground states} of the K3  theory, with
$U(1)$ charges $\pm(1,1),\, \pm(1,-1)$.
Indeed, under the state-field correspondence, the fields associated to
$\mathbf a_X^-, \mathbf a_X^+\in \mathfrak{a}$
create the states $(\mathbf a_X^-)_{(-{1/2})}\Omega$ and 
$ (\mathbf a_X^+)_{(-{1/2})}\Omega$, both in  
$A(\mathfrak{a})^1\subset V^{s\natural}_{\rm tw}$,
which
according to \cite[\S8]{dmc15}
are {\em Ramond  states} in the Conway Moonshine Module. 
In fact, since the Ramond sector of the Conway Moonshine Module is
$V^{s\natural}_{\rm tw}=A(\mathfrak{a})_{\rm tw}^0\oplus A(\mathfrak{a})^1$,
these states actually are Ramond {\em ground} states.
\smallskip\par

While by the above, the choice of the $U(1)$ current $J$ is crucial to the main
results of \cite{dmc15}, the charges with respect to its zero mode are 
only given for the twining elliptic genera,
there. Let us determine this information, along with the charges
with respect to  $\overline J$, for all parts of the
partition function. 
For notational convenience, we introduce
\be\label{BandFdef}
\begin{array}{rcl}
 B(\tau, z,\zeta) &:=& \textstyle{1\over2}\left( {\vartheta_3(\tau,z)^2\vartheta_3(\tau,\zeta)^2\over\eta(\tau)^4}
+ {\vartheta_4(\tau,z)^2\vartheta_4(\tau,\zeta)^2\over\eta(\tau)^4}\right)\\[8pt]
&\stackrel{\eqref{theta34ptwist}}{=}& 
\textstyle
{1\over2}\left( {\vartheta_3(\tau,z+\zeta)\vartheta_3(\tau,z-\zeta)\vartheta_3(\tau)^2\over\eta(\tau)^4}
+ {\vartheta_4(\tau,z+\zeta)\vartheta_4(\tau,z-\zeta)\vartheta_4(\tau)^2\over\eta(\tau)^4}\right),\\[8pt]
F(\tau, z,\zeta) &:=& \textstyle{1\over2}\left( {\vartheta_3(\tau,z)^2\vartheta_3(\tau,\zeta)^2\over\eta(\tau)^4}
- {\vartheta_4(\tau,z)^2\vartheta_4(\tau,\zeta)^2\over\eta(\tau)^4}\right),\\[8pt]
 \widehat F (\tau, z,\zeta) 
&:=& \textstyle{1\over2}\left( {\vartheta_3(\tau,z+\zeta)\vartheta_3(\tau,z-\zeta)\vartheta_3(\tau)^2\over\eta(\tau)^4}
- {\vartheta_4(\tau,z+\zeta)\vartheta_4(\tau,z-\zeta)\vartheta_4(\tau)^2\over\eta(\tau)^4}\right),
\end{array}
\ee
with shorthand notations $B(\tau):=B(\tau, 0,0), \; F(\tau):=F(\tau, 0,0)$.
We also set $\widetilde y:=e^{2\pi i\zeta}$ for $\zeta\in\C$.
From \eqref{DMCsectordeco}, we then find that 
the bosons in $V^{s\natural}$ are counted by
\begin{eqnarray}
 Z_{{\rm NS}^0}^{\rm DM-C}(\tau,z,\zeta)
 &:=& \tr_{A(\mathfrak{a})^0} \left( y^{J_0} \widetilde y^{\overline J_0} q^{L_{(0)}-1/2} \right) \nonumber\\
 &=& \textstyle
 {1\over2} \left( {\vartheta_3(\tau,z+\zeta)\vartheta_3(\tau,z-\zeta)\over\eta(\tau)^2}
 \cdot  {\vartheta_3(\tau)^{10}\over\eta(\tau)^{10} }
 + {\vartheta_4(\tau,z+\zeta)\vartheta_4(\tau,z-\zeta)\over\eta(\tau)^2}
 \cdot {\vartheta_4(\tau)^{10}\over\eta(\tau)^{10} }\right)\label{nsbosonsdmc}\\
 &=&  B(\tau,z,\zeta) \left( B(\tau)^2 
 +  F(\tau)^2 \right) +  \widehat F(\tau,z,\zeta)\cdot 2 B(\tau) F(\tau).\nonumber
\end{eqnarray}
The fermions in $V^{s\natural}$ are counted by
\begin{eqnarray}
 Z_{{\rm NS}^1}^{\rm DM-C}(\tau,z,\zeta)
 &:=& \tr_{A(\mathfrak{a})_{\rm tw}^1} \left( y^{J_0} \widetilde y^{\overline J_0} q^{L_{(0)} -1/2} \right) \nonumber\\
 &=& \textstyle
 {1\over2} \cdot {\vartheta_2(\tau,z+\zeta)\vartheta_2(\tau,z-\zeta)\over\eta(\tau)^2}
 \cdot 
  \left( {\vartheta_2(\tau)\over\eta(\tau) }\right)^{10}\label{nsfermionsdmc}\\
 &\stackrel{\eqref{strange}, \eqref{theta2twist}}{=}
 &\textstyle 
 F(\tau,z,\zeta) \cdot 4 F(\tau)^2.\nonumber
\end{eqnarray}
Similarly, the bosons in $V^{s\natural}_{\rm tw}$ are counted by
\begin{eqnarray}
 Z_{{\rm R}^1}^{\rm DM-C}(\tau,z,\zeta)
 &:=& \tr_{A(\mathfrak{a})_{tw}^0} \left( y^{J_0} \widetilde y^{\overline J_0} q^{L_{(0)}-1/2} \right) \nonumber\\
 &=& \textstyle
 {1\over2} \cdot {\vartheta_2(\tau,z+\zeta)\vartheta_2(\tau,z-\zeta)\over\eta(\tau)^2}
 \cdot 
  \left( {\vartheta_2(\tau)\over\eta(\tau) }\right)^{10}\label{rfermionsdmc}\\
 &\stackrel{\eqref{strange}, \eqref{theta2twist}}{=}
 &\textstyle 
 F(\tau,z,\zeta) \cdot 4 F(\tau)^2.\nonumber
\end{eqnarray}
The fermions in $V^{s\natural}_{\rm tw}$ are counted by
\begin{eqnarray}
 Z_{{\rm R}^0}^{\rm DM-C}(\tau,z,\zeta)
 &:=& \tr_{A(\mathfrak{a})^1} \left( y^{J_0} \widetilde y^{\overline J_0} q^{L_{(0)}-1/2} \right) \nonumber\\
 &=& \textstyle
 {1\over2} \left( {\vartheta_3(\tau,z+\zeta)\vartheta_3(\tau,z-\zeta)\over\eta(\tau)^2}
 \cdot  {\vartheta_3(\tau)^{10}\over\eta(\tau)^{10} }
 - {\vartheta_4(\tau,z+\zeta)\vartheta_4(\tau,z-\zeta)\over\eta(\tau)^2}
 \cdot {\vartheta_4(\tau)^{10}\over\eta(\tau)^{10} }\right)\label{rbosonsdmc}\\
 &=&  B(\tau,z,\zeta) \cdot 2 B(\tau) F(\tau) + 
 \widehat F(\tau,z,\zeta) \left( B(\tau)^2 + F(\tau)^2 \right).\nonumber
\end{eqnarray}
%
\section{Reflecting  right-moving degrees of freedom}\label{flip} 

Since it enjoys an action of a left- and a right moving
Virasoro algebra at central charges $c=\qu c=6$, 
the space of states 
$$
\HHH^{\rm GTVW}:=\HHH^{\rm NS}_{\rm bos}\oplus\HHH^{\rm NS}_{\rm ferm}\oplus
\HHH^{\rm R}_{\rm bos}\oplus\HHH^{\rm R}_{\rm ferm}
$$ 
of the K3 theory  of \cite{gtvw14}, described in Section \ref{gtvw}, 
can be regarded as a representation of 
the diagonal Virasoro algebra generated 
(as a Lie algebra) by the $L_{(n)}:=L_n+\qu L_n$, $n\in\Z$.
As such, according to \cite[Prop.~11.1]{dmc15}, 
$\HHH^{\rm GTVW}$ is isomorphic to
the space of states $V^{s\natural}\oplus V^{s\natural}_{\rm tw}$
of the Conway Moonshine Module \cite{du07} of Section \ref{supernatural}.
The aim of the present work is to compare 
the known additional structures on these  two spaces of states in
greater depth.

The most apparent difference between the  structures on 
$\HHH^{\rm GTVW}$ and $V^{s\natural}\oplus V^{s\natural}_{\rm tw}$ is the lack of
right-movers in the Conway Moonshine Module, while 
$\HHH^{\rm GTVW}$ is the space of states of
a {\em left-right symmetric} theory.
In other words, left- and right-movers in our K3 theory arise on equal footing
in every respect. 
In \cite{we01}, this property was used in order to
argue that the SCFT $(\widetilde 2)^4$ is {\em mirror 
self-dual}, in fact it has this property with respect to 
several versions of mirror symmetry. 
That different incarnations  of  that quantum symmetry may be
applied to this theory is already
a harbinger of its special properties.
Prompted by the results of \cite{dmc15}, in the present
work, we argue that we are confronted with yet another  
surprising, special property of this SCFT:
without destroying mathematical consistency,
one may {\em reflect} all right-movers and view them as 
holomorphic states, instead, while leaving left-movers 
untouched. This reflection property, which we do not expect to 
be shared by many SCFTs, is responsible for the beautiful result 
\cite[Prop.~11.1]{dmc15}, which we  lift to an isomorphism  
 between modules  of $\mathfrak{u}(1)$ 
 extensions of the Virasoro algebras in Section \ref{map}.
This extension follows also from the results of \cite{cdr17},
 which have been obtained independently from ours.
\smallskip\par

The current section is devoted to a discussion of the
process of {\em reflecting} all states in $\HHH^{\rm GTVW}$ so that they  
become  holomorphic,
and of its limitations, in a more general context.
Therefore, in the following, let $\H=\H^{\rm NS}\oplus\H^{\rm R}$ 
denote the space of states of
a  SCFT at central charges $c,\, \qu c$, according to
the description in Appendix \ref{cft}. 
We wish to collect some necessary and  sufficient
conditions for
our SCFT, so that $\H^{\rm NS}$ can become a 
self-dual, $C_2$-cofinite super vertex operator algebra 
of CFT type, and $\H^{\rm R}$ can become an
admissible
twisted $\H^{\rm NS}$-module\footnote{As in Section
\ref{gtvw}, we refer the reader to the literature for the definition
of all notions concerning super vertex operator algebras
and their modules. We particularly recommend the introductory
sections of \cite{dmc14,dmc15} for a very accessible presentation.}, 
by means of an appropriate process of reflecting 
all states in $\H$.

Below, we  argue that among the necessary conditions,
we find the following restrictions on the central charges 
$c,\,\qu c$ of our SCFT:
\be\label{cconditions}
c,\,\qu c\in 6\N\qquad \mb{ and }\qquad
c-\qu c\equiv 0 \mod 24.
\ee
Examples of
SCFTs with $c=\qu c\in3\N$ are expected to arise 
from supersymmetric non-linear sigma model constructions in the context
of superstring theory. 
In such constructions,
the quantum field theory  emerges
from the study of differentiable maps from some Riemann
surface $\Sigma$, known as the {\em worldsheet},
into a compact Calabi-Yau manifold, known
as the {\em target space}. Classically, one would
restrict attention to worldsheets that are embedded
into the target space
with (locally) minimal area. 
The equations of motion governing the 
coordinates on the target space are 
wave equations, whose solutions, for boundary 
conditions corresponding to closed strings, decompose into
contributions solely  depending 
{\em holomorphically} or {\em anti-holomorphically}
on the complex coordinates of $\Sigma$, i.e.\
comprising {\em left-} and {\em right-moving} waves, 
respectively. 
In the resulting quantum field theory, the latter 
descend to the {\em left-} and {\em right-moving}
degrees of freedom  mentioned in 
 Appendix \ref{cft}. 
Therefore, a reflection which 
renders all states of a SCFT 
 holomorphic should be 
reminiscent of a complex conjugation for the 
right-moving degrees of freedom. However,
our description of SCFTs in Appendix \ref{cft} 
should  leave the reader in no doubt that the passage from string
theory, with its interpretation in terms of left- and right-moving waves, 
into mathematically well-defined 
superconformal
field theories tunnels through a number of black boxes.
In particular, as detailed in Appendix \ref{cft}, for fermionic fields,
complex conjugation entails the introduction of additional
cocycle factors, and
the many consistency conditions alluded
to in  Appendix \ref{cft}  need to be taken into account when
attempting a manipulation akin to a  {\em reflection}
on the right-moving degrees of freedom. The 
circumstances under which
one can consistently
perform such a  procedure on $\H$ are by no means trivial.
\subsection{Necessary spectral conditions}\label{spectralcond}
On the level of representations of the Virasoro algebra,
reflecting right-movers to become
holomorphic is very simple. 
It amounts to viewing $\H=\H^{\rm NS}\oplus\H^{\rm R}$, 
as assumed above, 
as a representation of a Virasoro algebra at central
charge $c+\qu c$ which is
generated, as a Lie algebra,
by the $L_{(n)}:=L_n+\qu L_n$, $n\in\Z$, and $c+\qu c$.  Unitarity of the SCFT at the outset
implies that this representation is unitary, and compatible
with the real structure on $\H$. All additional structures that
we like to impose on $\H$ in the context of superconformal field
theory or super vertex operator algebras depend
crucially on the fine structure of
these representations. 
\smallskip\par

Recall that the partition function $Z(\tau,z)$
of our SCFT, defined as in \eqref{pfdef}, is invariant under the
special M\"obius transform $(\tau,z)\mapsto(\tau+1,z)$ of 
\eqref{Moebius}. Since uniqueness of the vacuum in our theory
implies that the leading order term of $Z(\tau,z)$ is 
$q^{-c/24}\qu q^{-\qu c/24}$, we conclude
\be\label{24}
c-\qu c\equiv 0\mod 24,
\ee
as announced in \eqref{cconditions}. 
The same reasoning for each summand 
$M\cdot y^Q\qu y^{\qu Q}q^{h-c/24}\qu q^{\qu h-\qu c/24}$ 
of $Z(\tau,z)$  with $M\in\N\setminus\{0\}$ moreover implies that
all conformal spins of bosonic states $\upsilon\in\H_{\rm bos}$
are integral.
In other words, for $h,\,\qu h\in\R$,
$$
\mb{ if }\quad\upsilon\in\H_{\rm bos}\mb{ exists with }
\upsilon\neq0,\; L_0\upsilon = h\upsilon \mb{ and }
\qu L_0\upsilon=\qu h\upsilon,\quad
\mb{ then }h-\qu h\in\Z.
$$
Furthermore, semilocality, together with conformal covariance,
 forces
all conformal spins to be integral or half integral, 
where states in $\H_{\rm ferm}^{\rm NS}$ have half integral
conformal spin, i.e.\
for $h,\,\qu h\in\R$,
$$
\begin{array}{rl}
\mb{ if }\quad\upsilon\in\H_{\rm ferm}\mb{ exists with }
\hphantom{\upsilon\neq0,\upsilon\neq0,\;}\\[5pt]
\upsilon\neq0,\; L_0\upsilon = h\upsilon \mb{ and }
\qu L_0\upsilon=\qu h\upsilon,
&\mb{ then }
h-\qu h\in{1\over2}\Z;\\[5pt]
\mb{ if in addition, }\quad\upsilon\in\H^{\rm NS}_{\rm ferm},
&\mb{ then }
h-\overline h\in{1\over2}+\Z.
\end{array}
$$
For $\H^{\rm NS}$ to become a self-dual super vertex operator
algebra of CFT type with respect to the action of the 
$L_{(n)}$, and for $\H^{\rm R}$ to become
an admissible twisted $\H^{\rm NS}$-module,
by the very definition of these notions\footnote{A 
$C_2$-cofinite super vertex operator algebra of this type is
called {\em nice}, according to H\"ohn \cite{ho96}.},
all eigenvalues of $L_{(0)}$
on $\H$ must be integral or half integral, 
and those for bosonic states in $\H^{\rm NS}$
must be integral, while those for fermionic states in
$\H^{\rm NS}$ must be half integral \cite[\S 2.1, Axiom 8]{dmc14}. 
Hence, as a necessary condition on 
our SCFT we find, for $h,\,\qu h\in\R$:
\be\label{spectrum}
\begin{array}{rll}
\mb{ If }\upsilon\in\H\mb{ exists with }
\hphantom{\upsilon\neq0,\upsilon\neq0,\;}\\[5pt]
\upsilon\neq0,\; L_0\upsilon = h\upsilon \mb{ and }
\qu L_0\upsilon=\qu h\upsilon,
&\mb{ then }
h,\,\overline h\in{1\over4}\N&\mb{ and }h+\overline h\in{1\over2}\N;\\[5pt]
\upsilon\in\H^{\rm NS}_{\rm bos},
&\mb{ then }
h,\,\overline h\in{1\over2}\N&\mb{ and }h+\overline h\in\N;\\[5pt]
\mb{if in addition, }\smash{\left\{\vphantom{\vbox{\vspace*{2.5em}}}\right.}\upsilon\in\H^{\rm NS}_{\rm ferm},
&\mb{ then }
h,\,\overline h\in{1\over2}\N&\mb{ and }h+\overline h\in{1\over2}+\N;\\[5pt]
\upsilon\in\H^{\rm R}_{\rm bos},
&\mb{ then } h-\overline h\in\Z.
\end{array}
\ee
Recall that
our original SCFT was assumed to enjoy space-time
supersymmetry. 
Hence in particular, the vacuum $\Omega\in\H^{\rm NS}$, under
{\em spectral flow} \eqref{spfl}, is mapped to a non-zero state 
$$
\textstyle
\widetilde\Omega_{\rm tw}\in\H^{\rm R}\quad\mb{ with }
L_0\widetilde\Omega_{\rm tw}={c\over24}\widetilde\Omega_{\rm tw},
\quad
\qu L_0\widetilde\Omega_{\rm tw}={\qu c\over24}\widetilde\Omega_{\rm tw}.
$$
Thus, if $\H^{\rm R}$ meets the spectral requirements \eqref{spectrum}, then
$c,\,\qu c\in 6\N$ follows, which together with the above condition
\eqref{24} confirms \eqref{cconditions}.

Our assumption of space-time supersymmetry allows us to further restrict
the spectrum in \eqref{spectrum}:
with notations as in  Appendix \ref{cft}, consider
a non-zero state $\upsilon\in\H^\SSS_{h,Q;\qu h,\qu Q}$,
$\SSS\in\{{\rm NS}, {\rm R} \}$. 
Space-time supersymmetry implies
that $\upsilon\in\H_{\rm bos}$ iff $Q-\qu Q\in2\Z$ and
$\upsilon\in\H_{\rm ferm}$ iff $Q-\qu Q\in2\Z+1$, in other words, 
we have
\begin{equation}\label{stSUSYfermno}
(-1)^F=(-1)^{J_0-\qu J_0}.
\end{equation} 
The 
operator of spectral flow, in general, has $U(1)$ charges
$(Q_{\rm sf};\qu Q_{\rm sf})=({c\over6};{\qu c\over6})$, so by
the above and \eqref{cconditions}, it is bosonic. We thus
may conclude that spectral flow maps each of 
$\H_{\rm bos}$ and $\H_{\rm ferm}$ isomorphically
onto itself. According to Appendix \ref{cft}, it also induces an isomorphism
$\H^{\rm NS}_{h,Q;\qu h,\qu Q}
\cong\H^{\rm R}_{h^\prime,Q^\prime;\qu h^\prime,\qu Q^\prime}$
with $(h^\prime,Q^\prime;\qu h^\prime,\qu Q^\prime)$ as in
\eqref{spfl}. In particular, 
$\H^{\rm NS}_{h,Q;\qu h,\qu Q}\subset\H_{\rm ferm}$
is mapped isomorphically to 
$\H^{\rm R}_{h^\prime,Q^\prime;\qu h^\prime,\qu Q^\prime}\subset\H_{\rm ferm}$
with $ (h^\prime-\qu h^\prime)\in (h-\qu h)+{1\over2} +\Z$. So 
\eqref{spectrum} implies 
for $h,\,\qu h\in\R$,
\be\label{morespectrum}
\mb{ if }\;\upsilon\in\H^{\rm R} \mb{ exists with }
\upsilon\neq0,\; L_0\upsilon = h\upsilon \mb{ and }
\qu L_0\upsilon=\qu h\upsilon,
\mb{ then }
h-\qu h\in\Z.
\ee
\par

The necessary spectral  conditions on
$L_0,\,\qu L_0$ obtained so far
immediately show that the SCFTs  for which a reflection
procedure may work 
are very sparse within any of the known
moduli spaces of SCFTs, but our claim is that the K3 theory with 
space of states $\HHH^{\rm GTVW}$ is one of them. Indeed,
the free fermion description of this theory allows to break up 
$\HHH^{\rm GTVW}$  into contributions
that are constructed from three octuplets of free
Majorana fermions,  each with coupled spin structures, 
according to \eqref{NSspaceofstates} and 
\eqref{Rspaceofstates}. 
Since ground states in the sectors $\HHH_\SSS$,
$\SSS\in\{0,\,v,\,s,\,c\}$ of Section \ref{D4orbifold}
have conformal weights 
$(0;0), \; ({1\over2};0)\mbox{ or } (0;{1\over2}),\;
({1\over4};{1\over4}),\; ({1\over4};{1\over4})$, 
respectively, inspection of \eqref{NSspaceofstates}
and  \eqref{Rspaceofstates}
immediately shows that the  spectral 
conditions \eqref{spectrum}, \eqref{morespectrum}
are indeed fulfilled. We stress that our identification of 
$\HHH^{\rm GTVW}$ with the space of states of the Conway
Moonshine Module exploits the fact that the two underlying 
theories enjoy a free fermion description.
\subsection{Vertex algebra and module structure}\label{voa}
By assumption, $\H$ comes equipped with the 
$n$-point functions of a SCFT, where the
resulting maps $z\mapsto\langle\phi_1(z_1)\cdots\phi_n(z_n) \rangle$,
for $\phi_1,\, \ldots,\, \phi_n\in\H$, in general, are 
only real analytic on their domains of definition. By the delicate
consistency conditions of SCFTs, the $n$-point functions
encode all operator product expansions of the theory.
Vice versa, reflection positivity
\eqref{reflpos} determines the 
two-point functions $\langle\phi_1(z_1)\phi_2(z_2) \rangle$
entirely by means of the scalar
product $\langle\cdot,\cdot\rangle$ together with the
real structure and Virasoro representations on
$\H$, and all other $n$-point functions are 
determined by the two-point functions
together with the operator product expansion. 

The requirements for $\H$ to comprise a super vertex operator
algebra $\H^{\rm NS}$ together with a 
twisted $\H^{\rm NS}$-module $\H^{\rm R}$ after reflection are somewhat weaker: 
we only need to fix an operator product expansion 
between fields $\phi(z),\,\upsilon(w)$ in terms of
formal power series, where
$\phi\in\H^{\rm NS}$
and $\upsilon\in\H$. We  require 
expansions in $(z-w)^{\pm{1}}$ if $\phi\in\H^{\rm NS}_{\rm bos}$
or $\upsilon\in\H^{\rm NS}$ and in
$(z-w)^{\pm{1\over2}}$ if $\phi\in\H^{\rm NS}_{\rm ferm}$
and $\upsilon\in\H^{\rm R}$. All this is encoded in the rules
for assigning modes 
\be\label{modeblabla}
\textstyle
\forall \phi\in\H_{\rm bos}^{\rm NS}\colon \quad n\mapsto \phi_{(n)} \; \forall n\in\Z,\qquad
\forall \phi\in\H_{\rm ferm}^{\rm NS}\colon \quad n\mapsto \phi_{(n)} \; \forall n\in{1\over2}\Z,
\ee
as 
was detailed for the particular example of the
Conway Moonshine Module in Section \ref{supernatural}.

Nevertheless, in general one cannot expect to obtain the required
operator product expansions on the space of states 
 that arises from  $\H$ by reflection,
not even with  these weaker requirements, 
and this is due  to the real analytic
behaviour of the $n$-point functions of our SCFT.
However, if
the spectral conditions \eqref{spectrum}, \eqref{morespectrum} on the eigenvalues of
$L_0,\,\qu L_0$ hold, then
 conformal covariance of the
$n$-point functions
severely restricts the form of the power series describing
the operator product expansions in the  theory: with notations
as in  Appendix \ref{cft}, assume that $\phi_i\in\H_{h_i; \qu h_i}^{\SSS_i},\,
\phi_j\in\H_{h_j; \qu h_j}^{\SSS_j}$, where $\SSS_i,\,\SSS_j\in\{ {\rm NS,\, R}\}$.
Then all summands in the operator product expansion between
$\phi_i(z_i)$ and $\phi_j(z_j)$, by conformal covariance, have the form
\be\label{opeform}
\begin{array}{l}
\displaystyle
{\phi_k(z_i)\over (z_i-z_j)^{h_i+h_j-h_k} (\qu z_i-\qu z_j)^{\qu h_i+\qu h_j-\qu h_k} }
\qquad \mb{ with }\quad \phi_k\in \H_{h_k; \qu h_k}^{\SSS_k},\\[15pt]
\qquad\qquad \SSS_k={\rm NS} \;\mb{ if }\; \SSS_i=\SSS_j,\;  \SSS_k={\rm R}
\;\mb{ otherwise.}
\end{array}
\ee
Thus  \eqref{spectrum} implies that all
such OPEs $\phi_i(z_i)\phi_j(z_j)$ are encoded in terms of formal
power series in  $|z_i-z_j|^{\pm{1\over2}}$ and $(z_i-z_j)^{\pm{1\over2}}$
with $i\neq j$. 
This means that replacing each $|z_i-z_j|^{\pm{1\over2}}$ 
by $(z_i-z_j)^{\pm{1\over2}}$,  one 
obtains an ansatz for a ``reflected'' operator
product expansion between the fields
$\phi_i(z_i)$ and $\phi_j(z_j)$, which after
reflection should be viewed as (holomorphic)
fields in a super vertex operator
algebra.  But the spectral conditions \eqref{spectrum}, \eqref{morespectrum} do not
 ensure that the OPE between $\phi_i\in\H^{\rm NS}_{\rm bos}$ and $\phi_j\in\H^{\rm R}$, after replacing 
all $|z_i-z_j|^{\pm{1\over 2}}$  by $(z_i-z_j)^{\pm{1\over 2}}$, yields a formal power series in 
$(z_i-z_j)^{\pm1}$, as it should.
We therefore impose
one additional, very natural assumption on our original
SCFT: we require that all eigenvalues of $J_0$ and
of $\qu J_0$ are integral, i.e.\
for $Q,\,\qu Q\in\R$,
\be\label{spectrumplus}
\mb{ if }\quad\upsilon\in\H\mb{ exists with }
\upsilon\neq0,\; J_0\upsilon = Q\upsilon \mb{ and }
\qu J_0\upsilon=\qu Q\upsilon,\quad
\mb{ then }Q,\,\qu Q\in\Z.
\ee
This assumption is equivalent to the
requirement that the theory is invariant under the
{\em purely holomorphic} and {\em anti-holomorphic
two-fold spectral flows}\footnote{See for example \cite[\S3.4]{gr97} or \cite[\S3.1.1]{diss}.}.
This condition holds for every K3 theory by definition,
according to \cite[Def.~8]{we14}; there, the 
operators of two-fold spectral flow, together with the $U(1)$ 
currents, comprise the 
$\widehat{\frak{su}}(2)_{L,1}\oplus \widehat{\frak{su}}(2)_{R,1}$-subalgebra
for the left- and the right-moving $N=4$ superconformal algebras.
In particular, this condition holds for the K3 theory of \cite{gtvw14}
described in Section \ref{gtvw}.
By \cite{eoty89}, it should hold for all SCFTs that 
obey \eqref{cconditions} and arise
from a non-linear sigma model construction with a Calabi-Yau
target space.

If the additional spectral  condition \eqref{spectrumplus} holds,
then the properties \eqref{cconditions}, \eqref{spectrum} and
 \eqref{morespectrum} further
restrict the spectrum of $L_0$ and $\qu L_0$, since spectral flow
yields a multigraded isomorphism $\H^{\rm NS}\stackrel{\cong}{\longrightarrow}\H^{\rm R}$,
which obeys \eqref{spfl}: let 
$h,\,\qu h\in\R$. 
\be\label{Rspectrumplus}
\begin{array}{rl}
\mb{ If }\quad\upsilon\in\H^{\rm R}&\mb{ exists with }
\upsilon\neq0,\; L_0\upsilon = h\upsilon \mb{ and }
\qu L_0\upsilon=\qu h\upsilon,\\[5pt]
&\mb{ then }
(h;\qu h)=({c\over24};{ c\over24})+({m\over2};{\qu m\over2})
\mb{ with } m,\qu m\in \Z,\; m\equiv\qu m\mod 2.
\end{array}
\ee
With the help of \eqref{opeform} as well as the spectral conditions 
\eqref{spectrum}, \eqref{morespectrum}, \eqref{Rspectrumplus}, a 
case by case analysis for $\phi_i\in\H_{p_i}^{\rm NS}$, $\phi_j\in\H_{p_j}^{\SSS_j}$
with $\SSS_j\in\{ {\rm NS,\, R}\}$, $p_i,\,p_j\in \{ {\rm bos}, {\rm ferm} \}$  
reveals two facts: first that the OPE between $\phi_i(z_i)$ and $\phi_j(z_j)$ is encoded in 
terms of a formal power series in the 
$(z_i-z_j)^{\pm{1\over2}}, (\qu z_i-\qu z_j)^{\pm{1\over2}}$, and 
second that by replacing all 
$(\qu z_i-\qu z_j)^{\pm{1\over2}}$ by $( z_i- z_j)^{\pm{1\over2}}$,
these become formal power series in $(z_i-z_j)^{\pm1}$ if 
$\phi_i\in\H^{\rm NS}_{\rm bos}$ or $\phi_j\in\H^{\rm NS}$. 

The above yields an ansatz for an OPE after reflection
between the
fields associated with any
$\phi_i\in\H^{\rm NS}$, $\phi_j\in\H$, which are viewed as
states in a super vertex operator algebra and its admissible
modules. However, due to the occurrence of half integral
exponents in our formal power series, the construction
leaves room for ambiguities of  signs, which may
destroy the consistency of our operator product expansions. 
It is tempting to try an ansatz by which one chooses 
a basis of the vector space $\H$ such that every state
comes with a decomposition into a left and a right-moving 
contribution. In practice, in a given SCFT, this is regularly the case.
Then, operator product expansions  
can be defined by 
specifying left-moving and right-moving
contributions separately.
If reflection acts by complex conjugation on
the contributions $(\qu z_i-\qu z_j)^{\qu\nu}$ 
with $\qu\nu\in{1\over 2}\Z$
in the operator product expansions arising from right-movers,
then the reflection should be given by an anti-$\C$-linear map
on purely anti-holomorphic fields, and thus also on the right-moving
contributions to every field. However, this introduces further ambiguities
of phases for our operator product expansions, since there is no canonical way to assign
complex scalar factors to a left- or a right-moving contribution,
respectively.
Moreover, {\em associativity} of a would-be vertex algebra 
structure obtained by this procedure is by no means clear. 

Hence instead of attempting to separate left-movers from
right-movers in every state in $\H$, with notations as in 
 Appendix \ref{cft},
we choose a real basis, say,
of every  $\H_{h;\qu h}^\SSS\subset\H^\SSS$. The compatibility 
\eqref{realcomp} of our $n$-point functions with the real structure
on $\H$ together with the unitarity of the representation of the 
operator product expansion ensures that all coefficients in the formal power series 
of the OPE between two fields corresponding to 
real basis elements are real. Then, the first step of reflection on 
$\H$ can indeed be implemented by replacing all contributions 
$(\qu z_i-\qu z_j)^{\qu\nu}$ by $(z_i-z_j)^{\qu\nu}$ 
when $\qu\nu \in{1\over 2}\Z$ in the 
operator product expansion between $\phi_i(z_i)$ and $\phi_j(z_j)$ for 
real $\phi_i,\,\phi_j\in\H$. 
Since the resulting formal power series agrees with the operator
product expansion of our original SCFT if $z_i=\qu z_i$
and $z_j=\qu z_j$, i.e.\ for entries 
in $\R^n\setminus\cup_{i\neq j}\{z\in\R^n|z_i=z_j\}$
in our $n$-point functions, associativity for  this ansatz for an OPE is guaranteed.
Interestingly, additional choices of signs are required. Indeed, if 
$\nu,\qu\nu\in{1\over2}+\Z$ for contributions 
$(z_i-z_j)^{\nu}(\qu z_i-\qu z_j)^{\qu\nu}$ of an OPE in our original SCFT,
then our prescription for reflection changes the parity of this function. Thus, semilocality
of some $n$-point functions may be destroyed. Whether or not these signs can be
implemented consistently, in general, is a highly nontrivial question which so far,
has to be resolved on a case by case basis. 
These signs have the same origin as the cocycle factor $\kappa_\phi$
introduced in the formula for $\phi^\dagger$ in \eqref{Klein}.
Note however that these sign issues do not
occur as long  as the OPE involves a chiral or an antichiral field. Indeed, contributions
$(z_i-z_j)^{\nu}(\qu z_i-\qu z_j)^{\qu\nu}$ to the OPE then only yield $\nu,\,\qu\nu\in\Z$.
The above-mentioned sign ambiguity therefore does not arise when one restricts
attention to the structure of a potential bulk SCFT on $\H$, as is suggested 
in \cite{cdr17}.
Note  that by construction, if \eqref{spectrum} and \eqref{spectrumplus} hold, then
the operator product expansion between any two fields
corresponding to real states in $\H$, on restriction 
to $z_i=\qu z_i$ and $z_j=\qu z_j$, 
can be described in terms of 
the super vertex algebra formalism. 
\subsection{Reflecting: some necessary and sufficient conditions, 
and consequences}\label{reflconsequence}
As a result of the discussions in Sections \ref{spectralcond}
and \ref{voa}, we arrive at a set of some necessary and 
sufficient conditions for our reflection procedure to yield
the desired structures: assume that $\H$ is the space of states of a SCFT
as before, such that the necessary spectral 
conditions \eqref{cconditions}, \eqref{spectrum} on the
central charges $c,\, \qu c$ and
eigenvalues of $L_0,\, \qu L_0$ 
hold. To obtain a well-defined structure of a super vertex
operator algebra and admissible module on $\H$ by 
the reflection procedure described above, one needs to require
that after replacing all contributions $(\qu z_i-\qu z_j)^{\pm{1\over2}}$,
$|z_i-z_j|^{\pm{1\over2}}$ by $(z_i-z_j)^{\pm{1\over2}}$
in the OPE between $\phi_i(z_i)$ and $\phi_j(z_j)$
for any $\phi_i\in\H^{\rm NS}_{\rm bos}$,
$\phi_j\in\H^{\rm R}$, the result is a formal power series in
$(z_i-z_j)^{\pm1}$. By the discussion of Section \ref{voa},
a sufficient condition to ensure this behaviour
is invariance under the two-fold holomorphic and anti-holomorphic
spectral flows, or equivalently, \eqref{spectrumplus}. 
Furthermore, we must require that the necessary implementation
of additional signs mentioned at the end of Section \ref{voa} can be
performed consistently. Then
the reflection 
procedure as  described above is
well-defined.  All states  
become  holomorphic, and  reflection 
yields a consistent super vertex operator algebra $\H^{\rm NS}$
of CFT type, along with an admissible twisted
$\H^{\rm NS}$-module $\H^{\rm R}$. Indeed, once a consistent OPE has been
implemented on the reflected $\H$ along the lines described above,
the remaining axioms, like the {\em state-field correspondence},
the {\em vacuum} and the {\em translation axiom} are 
immediate as a heritage from the corresponding properties of the original
SCFT.  If  both the chiral and the antichiral algebra are
self-dual and $C_2$-cofinite
then this
guarantees that  reflection
yields a self-dual, $C_2$-cofinite super vertex operator algebra.
\smallskip\par

For the resulting super vertex operator algebra and its twisted module, 
one can still define the partition function as a formal power 
series in terms of its four parts by means of  
\eqref{partifuncdef}. However, in this definition,
$\qu q$ must then be replaced by  $q$, 
to take into account the
fact that $\H$ is now viewed as a Virasoro module under the
action of the $L_{(n)}
=L_n+\qu L_n,\, n\in\Z$. So indeed, 
we obtain a formal power series 
$$
Z^{\rm refl}(\tau,z,\zeta)
:=
\tr_{\H_{\rm bos}}\left(   y^{J_0}\tilde y^{\qu J_0} q^{L_{(0)}-(c+\qu c)/24} \right)
$$
in $y:=e^{2\pi iz},\,\tilde y:=e^{2\pi i\zeta}$ and $q:=e^{2\pi i\tau}$.
Since for $q=\qu q$, we have
$Z(\tau,z)=Z^{\rm refl}(\tau,z,-\qu z)$, 
the latter is a convergent power series in $q\in\R_{>0}$ and
$y\in\C$, where each summand
$$
My^Q\qu y^{\qu Q} q^{h+\qu h-(c+\qu c)/24}
$$ 
obeys
$M\in\N$. Thus
 $Z^{\rm refl}(\tau,z,\zeta)$ converges absolutely
and can be viewed as 
a function in complex variables  $\tau,\,z,\,\zeta\in\C$ with 
$\mb{Im}(\tau)>0$ and $\zeta=-\qu z$.
By the identity theorem for holomorphic
functions, it is 
uniquely determined by its values for purely 
imaginary $\tau$ and $\zeta=-\qu z$. However,
 there is no reason to expect 
 the partition function $Z^{\rm refl}(\tau,z,\zeta)$ 
 to behave like the partition functions of SCFTs under 
 ${\rm SL}(2,\Z)$,
since     
\be\label{realray}
\left\{ (\tau,z,\zeta)\in\C^3 \mid 
\mb{Re}(\tau)=0, \; \mb{Im}(\tau)>0,\; \zeta=-\qu z\right\}
\ee
is not mapped to itself under $\tau\mapsto\tau+1$.
But as the set \eqref{realray} is invariant under the map
$(\tau, z,\zeta)\mapsto(-1/\tau, z/\tau,-\zeta/\tau)$,
 $Z^{\rm refl}(\tau,z,\zeta)$ will exhibit modular
behaviour under this transformation.
In Section \ref{reflpf} we will confirm that for 
the model obtained from the K3  theory of
Section \ref{gtvw}
by reflecting right-movers to  holomorphic states on
$\HHH^{\rm GTVW}$, the resulting partition function
shares its modular behaviour under 
$S\colon(\tau, z,\zeta)\mapsto(-1/\tau, z/\tau,-\zeta/\tau)$
with that of SCFTs, while under the transformation 
$T \colon\tau\mapsto\tau+1$, the usual invariance properties
are broken. Note that invariance under $T^2\colon\tau\mapsto\tau+2$
is immediate due to our spectral assumptions \eqref{spectrum} on the 
eigenvalues of $L_0$ and $\qu L_0$ on $\H$  and
$c+\qu c\in12\N$ according to \eqref{cconditions}.
Hence $Z^{\rm refl}(\tau,z,\zeta)$ exhibits modular 
behaviour under the Hecke group 
$\textgoth{G}(2)$ - also known as the Theta Group
- that is, the level $2$, index $3$ congruence subgroup
of  $\rm{SL}(2,\Z)$ generated by $S$ and $T^2$. 
Since the  level two  principal congruence subgroup 
$\Gamma(2)\subset\textgoth{G}(2)$ 
has genus $0$, so  does the Hecke group $\textgoth{G}(2)$.
By construction, the reflected theory obeys the
spectral condition that all eigenvalues of $L_{(0)}$ lie in
${1\over2}\N$. Thus we have recovered
the modular behaviour found by H\"ohn
\cite{ho96} for ``nice'' super vertex operator algebras. 
In our setting, the modular
behaviour under $\textgoth{G}(2)$  is naturally inherited, via reflection, 
 from the 
original SCFT, and $C_2$-cofiniteness does not
enter as a necessary condition.

In addition to modular transformations, the reflected
partition function exhibits elliptic behaviour,
$$
\begin{array}{rcl}
Z^{\rm refl}(\tau,z+{\tau\over2},\zeta+{\tau\over2})
&=& q^{-(c+\qu c)/24} y^{-c/6} \widetilde y^{-\qu c/6} Z^{\rm refl}(\tau,z,\zeta),\\[5pt]
Z^{\rm refl}(\tau,z+{1\over2},\zeta+{1\over2})
&=&  Z^{\rm refl}(\tau,z,\zeta)
\end{array}
$$
as a consequence of \eqref{partitrafo}.

Recalling that for our original SCFT, $Z_{\widetilde R}(\tau,z)$ on its
own transforms like $Z(\tau,z)$ under ${\rm SL}(2,\Z)$, its reflected version 
$$
Z^{\rm refl}_{\widetilde R}(\tau,z,\zeta)
:=
\tr_{\H^{\rm R}}\left(   (-1)^F y^{J_0}\tilde y^{\qu J_0} q^{L_{(0)}-(c+\qu c)/24} \right)
$$
should also have interesting modular properties. Indeed, by the
same reasoning as for $Z^{\rm refl}(\tau,z,\zeta)$, it exhibits modular
behaviour under the Hecke group $\textgoth{G}(2)$. In addition,
\eqref{cconditions} and \eqref{Rspectrumplus} imply for 
$h,\,\qu h\in\R$, $\upsilon\in\H^{\rm R}$ with $\upsilon\neq0$,
$L_{0}\upsilon=h\upsilon$, $\qu L_0\upsilon=\qu h\upsilon$:
$$
\textstyle
\left( L_{(0)}-{c+\qu c\over24}\right) \upsilon = a\upsilon,\quad a\in\Z,
$$
where actually $a\geq0$ due to the unitarity requirements of the original
theory \cite{lvw89}. In other words, $Z^{\rm refl}_{\widetilde R}(\tau,z,\zeta)$ 
is a power series in $q,\,y^{\pm 1},\,\widetilde y^{\pm 1}$ and thus invariant under
$T$. Altogether, $Z^{\rm refl}_{\widetilde R}(\tau,z,\zeta)$ 
exhibits modular behaviour under the full modular group ${\rm SL}(2,\Z)$.
\smallskip\par

In closing this section, we  emphasize
once again that the structure of a self-dual, $C_2$-cofinite 
super vertex operator algebra of CFT type on 
$\H^{\rm NS}$ together with an admissible twisted
$\H^{\rm NS}$-module structure
on $\H^{\rm R}$, obtained by
reflecting  all states  in a SCFT with
space of states $\H=\H^{\rm NS}\oplus\H^{\rm R}$, is much
weaker than that of the original SCFT. By our prescription of the reflection, a priori,
we  solely obtain the formal power
series expansions required for the definition of the 
super vertex algebra and twisted module structures on
$\H^{\rm NS}$, $\H^{\rm R}$. 
That it should yield well-defined $n$-point functions 
as in a full-fledged SCFT, is by no means 
guaranteed and is also 
not required in the definition of a 
super vertex operator algebra and its admissible
modules.

For the Conway Moonshine Module 
$V^{s\natural} \oplus V^{s\natural}_{\rm tw}$,
this lack of SCFT-structure may well mean that its role
for Conway or Mathieu Moonshine should not be
expected to match that of the Moonshine Module 
$V^\natural$ of Frenkel, Lepowsky and Meurman
\cite{flm84,flm85,flm88} for Monstrous Moonshine.
However, 
the beautiful results of \cite{du07,dmc14} show
that the analogues of the McKay-Thompson series 
for this module are indeed normalized principal 
moduli
for genus zero subgroups of $\mb{SL}(2, \R)$. On the other
hand,
the K3 theory of Section \ref{gtvw}
built on $\HHH^{\rm GTVW}$ offers the
more powerful structures of SCFT. In particular, we have
revealed the genus zero property of 
$Z^{\rm refl}(\tau,z=0,\zeta=0)$ as a heritage
from this theory under reflection.
It would be interesting to know whether
the genus zero properties of the Conway Moonshine
Module, in general, are inherited from its underlying K3 theory.
%
\section{Reflecting the K3 theory with $\Z_2^8\colon\M_{20}$ symmetry}\label{map}
In this section, we show that the reflection procedure described
in Section \ref{flip} transforms the K3 theory with space of states
$\HHH^{\rm GTVW}$ of Section \ref{gtvw} into Duncan's Conway
Moonshine Module 
$V^{s\natural}\oplus V^{s\natural}_{\rm tw}$, whose construction
we have recalled in Section \ref{supernatural}. On the level of Virasoro 
modules for the respective natural Virasoro 
algebras at central charge $c=12$, agreement was already shown 
by Duncan and Mack-Crane in \cite[Prop.~1.11]{dmc15}. In Section
\ref{flipmodule}, we lift this result to the level of modules
of the extensions of these
Virasoro algebras by the zero modes
$J_0$ and $\qu J_0$ of two commuting $U(1)$ currents,
where it turns out that we have to reverse the role of bosons and
fermions in the Ramond sector, in comparison to \cite{dmc15}. 
In Section \ref{flipgtvw}, we show that after reflection,
$\HHH^{\rm GTVW}$ agrees with
$V^{s\natural}\oplus V^{s\natural}_{\rm tw}$ as a
super vertex operator algebra with an admissible 
module. This allows us to uncover
a considerably more elaborate structure on the space
$V^{s\natural}\oplus V^{s\natural}_{\rm tw}$, which it 
inherits from the K3 theory on $\HHH^{\rm GTVW}$. 
In Section \ref{moonshine} we discuss the conclusions
on Moonshines that we draw from our results.
\subsection{Comparison of multigraded modules}\label{flipmodule}
With notations as in \eqref{NSspaceofstates},
\eqref{Rspaceofstates}, let 
$\HHH^{\rm NS}:=\HHH^{\rm NS}_{\rm bos}\oplus \HHH^{\rm NS}_{\rm ferm}$
denote the Neveu-Schwarz sector of the K3 theory 
with $\Z_2^8\colon\M_{20}$ symmetry
of \cite{gtvw14}, 
while $\HHH^{\rm R}:=\HHH^{\rm R}_{\rm bos}\oplus \HHH^{\rm R}_{\rm ferm}$
denotes the Ramond sector.
As Virasoro modules with respect
to the Virasoro algebra at central charge $c=12$,
which on $\HHH^{\rm GTVW}$ is generated
by the diagonal $L_{(n)}$, $n\in\Z$, of
Section \ref{flip}, 
$$
\HHH^{\rm NS}\cong V^{s\natural}, \qquad
\HHH^{\rm R}\cong V^{s\natural}_{\rm tw},
$$
according to \cite[Prop.~11.1]{dmc15}. 
Duncan and Mack-Crane obtain
this beautiful result by
means of {\em triality}, which we will come back to in
Section \ref{flipgtvw}. It is a priori not clear whether 
under the triality map, the $U(1)$ charges 
introduced in Section \ref{u1choices} agree with the ones obtained from
our choices of left- and  right-moving $U(1)$-currents
 of the K3 theory.
This issue is not addressed in \cite{dmc15} and is studied in detail here. 
It amounts to a comparison between 
$\HHH^{\rm GTVW}$ and 
$V^{s\natural}\oplus V^{s\natural}_{\rm tw}$ as modules
of the extensions of the respective 
Virasoro algebras at central charge $c=12$ by two
commuting Lie algebras of type $\mathfrak{u}(1)$.
We continue to denote
the generators of the latter by $J_0,\,\qu J_0$, 
where on $\HHH^{\rm GTVW}$, the $U(1)$ currents
are chosen according to \eqref{gtvwu1choice},
while on $V^{s\natural}\oplus V^{s\natural}_{\rm tw}$, we use
\eqref{dmcu1choice}, \eqref{dmcu1choiceright}. Both
$J_0$ and $\qu J_0$ are central in the extended Lie algebras.
To prove agreement as modules
of the extended Lie algebras, it therefore
suffices to show that the multigraded traces of
$y^{J_0}\widetilde y^{\qu J_0} q^{L_{(0)}-1/2}$ over
the respective sectors of
$\HHH^{\rm GTVW}$ and 
$V^{s\natural}\oplus V^{s\natural}_{\rm tw}$ agree.
We prove this in each sector separately:
\medskip
\subsubsection{Neveu-Schwarz bosons}\label{NSboscompared}
In our K3 theory, the bosonic contributions
to $\tr_{\HHH^{\rm NS}}\left( y^{J_0}\widetilde y^{\qu J_0} q^{L_{(0)}-1/2}\right)$
from the sector $({\rm NS}, {\rm NS}, {\rm NS} )$, according to
\eqref{sectornsnsns}, are
\begin{eqnarray*}
\textstyle
{1\over4}
 \left(
   \left( {\vartheta_3(\tau)\over\eta(\tau)}\right)^{4}
   + \left( {\vartheta_4(\tau)\over\eta(\tau)}\right)^{4} \right)^2
   && \hspace*{-4em}
   \cdot \textstyle {1\over2} \left( {\vartheta_3(\tau,z)^2\vartheta_3(\tau,\zeta)^2\over\eta(\tau)^4}
   + {\vartheta_4(\tau,z)^2\vartheta_4(\tau,\zeta)^2\over\eta(\tau)^4} \right)\\
&   \stackrel{\eqref{BandFdef}}{=} & B(\tau,z,\zeta)B(\tau)^2.\\
 \end{eqnarray*}
This agrees with the contributions from 
$U_{000}$ (see \eqref{DMCsectordeco}) to \eqref{nsbosonsdmc}.

The bosonic contributions from 
the sector $({\rm NS}, {\rm R}, {\rm R} )$, according to
\eqref{sectornsrr}, are
$$
\textstyle {1\over4}
   \left( {\vartheta_2(\tau)\over\eta(\tau)}\right)^8
   \cdot {1\over2} \left( {\vartheta_3(\tau,z)^2\vartheta_3(\tau,\zeta)^2\over\eta(\tau)^4}
   + {\vartheta_4(\tau,z)^2\vartheta_4(\tau,\zeta)^2\over\eta(\tau)^4} \right)
 \stackrel{\eqref{BandFdef},\eqref{strange}}{=}
  B(\tau,z,\zeta)F(\tau)^2 .
$$
In \eqref{nsbosonsdmc}, these agree with 
the contributions from the sector $U_{0vv}$.

The bosons in
the final sector $({\rm R}, {\rm NS}, {\rm R} ) + ({\rm R}, {\rm R}, {\rm NS} )$
by \eqref{sectornsrr}  yield
\begin{eqnarray*}
 \textstyle
   \left( {\vartheta_2(\tau)\over\eta(\tau)}\right)^4
   \cdot 
   {1\over2}
   \left(
   \left( {\vartheta_3(\tau)\over\eta(\tau)}\right)^{4}
   + \left( {\vartheta_4(\tau)\over\eta(\tau)}\right)^{4} \right)
    {1\over2} \left( {\vartheta_2(\tau,z)^2\vartheta_2(\tau,\zeta)^2\over\eta(\tau)^4}
   + {\vartheta_1(\tau,z)^2\vartheta_1(\tau,\zeta)^2\over\eta(\tau)^4} \right)\qquad\\
 \stackrel{\eqref{BandFdef},\eqref{strange}, \eqref{theta34mtwist}}{=}
\widehat F(\tau,z,\zeta)\cdot 2B(\tau)F(\tau).
\end{eqnarray*}
In \eqref{nsbosonsdmc}, these  are 
the contributions from the sectors $U_{v0v}\oplus U_{vv0}$.

Altogether, we find agreement with the result of
\eqref{nsbosonsdmc}.
\medskip
\subsubsection{Neveu-Schwarz fermions}\label{NSfermcompared}
In our K3 theory,  the fermionic contributions
to $\tr_{\HHH^{\rm NS}}\left( y^{J_0}\widetilde y^{\qu J_0} q^{L_{(0)}-1/2}\right)$
from the sector $({\rm NS}, {\rm NS}, {\rm NS} )$, according to
\eqref{sectornsnsns}, are
\begin{eqnarray*}
\textstyle
{1\over4}
 \left(
   \left( {\vartheta_3(\tau)\over\eta(\tau)}\right)^{4}
   - \left( {\vartheta_4(\tau)\over\eta(\tau)}\right)^{4} \right)^2
      \textstyle
   \cdot {1\over2} \left( {\vartheta_3(\tau,z)^2\vartheta_3(\tau,\zeta)^2\over\eta(\tau)^4}
   - {\vartheta_4(\tau,z)^2\vartheta_4(\tau,\zeta)^2\over\eta(\tau)^4} \right)\\
&& \hspace*{-5em} \stackrel{\eqref{BandFdef}}{=} F(\tau,z,\zeta)F(\tau)^2.
 \end{eqnarray*}
In  \eqref{nsfermionsdmc}, these are 
the contributions from the sector $U_{ccc}$.

The fermionic contributions from 
the sector $({\rm NS}, {\rm R}, {\rm R} )$, according to
\eqref{sectornsrr}, are
$$
\textstyle {1\over4}
   \left( {\vartheta_2(\tau)\over\eta(\tau)}\right)^8
   \cdot {1\over2} \left( {\vartheta_3(\tau,z)^2\vartheta_3(\tau,\zeta)^2\over\eta(\tau)^4}
   - {\vartheta_4(\tau,z)^2\vartheta_4(\tau,\zeta)^2\over\eta(\tau)^4} \right)\\
 \stackrel{\eqref{BandFdef},\eqref{strange}}{=}
  F(\tau,z,\zeta)F(\tau)^2.
$$
In \eqref{nsfermionsdmc}, these are 
the contributions from the sector $U_{css}$.

From \eqref{sectorrnsr}, fermions in
the final sector $({\rm R}, {\rm NS}, {\rm R} ) + ({\rm R}, {\rm R}, {\rm NS} )$ 
yield
\begin{eqnarray*}
 \textstyle
   \left( {\vartheta_2(\tau)\over\eta(\tau)}\right)^4
   \cdot 
   {1\over2}
   \left(
   \left( {\vartheta_3(\tau)\over\eta(\tau)}\right)^{4}
   - \left( {\vartheta_4(\tau)\over\eta(\tau)}\right)^{4} \right)
    {1\over2} \left( {\vartheta_2(\tau,z)^2\vartheta_2(\tau,\zeta)^2\over\eta(\tau)^4}
   - {\vartheta_1(\tau,z)^2\vartheta_1(\tau,\zeta)^2\over\eta(\tau)^4} \right)\qquad\\
 \stackrel{\eqref{BandFdef},\eqref{strange}, \eqref{newstrange}}{=}
 \textstyle 
  F(\tau,z,\zeta)
 \cdot 2 F(\tau)^2.
\end{eqnarray*}
In \eqref{nsfermionsdmc}, these are 
the contributions from  $U_{scs}\oplus U_{ssc}$.

Altogether, we find agreement with the result of
\eqref{nsfermionsdmc}.
\medskip
\subsubsection{Ramond bosons}\label{Rboscompared}
In our K3 theory, the
bosonic contributions to the multigraded trace
$\tr_{\HHH^{\rm R}}\left( y^{J_0}\widetilde y^{\qu J_0} q^{L_{(0)}-1/2}\right)$
in the sector $({\rm R}, {\rm NS}, {\rm NS} )$, according to
\eqref{sectorrnsns}, amount to
\begin{eqnarray*}
\textstyle
{1\over4}
 \left(
   \left( {\vartheta_3(\tau)\over\eta(\tau)}\right)^{4}
   + \left( {\vartheta_4(\tau)\over\eta(\tau)}\right)^{4} \right)^2
   \cdot {1\over2} \left( {\vartheta_2(\tau,z)^2\vartheta_2(\tau,\zeta)^2\over\eta(\tau)^4}
   + {\vartheta_1(\tau,z)^2\vartheta_1(\tau,\zeta)^2\over\eta(\tau)^4} \right)\\
&&\hspace*{-8em} \stackrel{\eqref{BandFdef},\eqref{theta34mtwist}}{=} 
   \widehat F (\tau,z,\zeta)B(\tau)^2.
\end{eqnarray*}
In \eqref{rbosonsdmc}, these are 
the contributions from the sector $U_{v00}$.

The bosonic contributions from 
the sector $({\rm R}, {\rm R}, {\rm R} )$, by
\eqref{sectorrrr}, are
$$
\textstyle
  {1\over4} \left( {\vartheta_2(\tau)\over\eta(\tau)}\right)^8
   \cdot {1\over2} \left( {\vartheta_2(\tau,z)^2\vartheta_2(\tau,\zeta)^2\over\eta(\tau)^4}
   + {\vartheta_1(\tau,z)^2\vartheta_1(\tau,\zeta)^2\over\eta(\tau)^4} \right)\\
 \stackrel{\eqref{BandFdef},\eqref{strange},\eqref{theta34mtwist}}{=}
 \widehat F(\tau,z,\zeta)F(\tau)^2.
$$
In \eqref{rbosonsdmc}, these are 
the contributions from the sector $U_{vvv}$.

The bosons in
the final sector 
$({\rm NS}, {\rm NS}, {\rm R} ) + ({\rm NS}, {\rm R}, {\rm NS} )$,
by \eqref{sectornsnsr}, yield
\begin{eqnarray*}
 \textstyle
   \left( {\vartheta_2(\tau)\over\eta(\tau)}\right)^4
   \cdot 
   {1\over2}
   \left(
   \left( {\vartheta_3(\tau)\over\eta(\tau)}\right)^{4}
   + \left( {\vartheta_4(\tau)\over\eta(\tau)}\right)^{4} \right)
    {1\over2} \left( {\vartheta_3(\tau,z)^2\vartheta_3(\tau,\zeta)^2\over\eta(\tau)^4}
   + {\vartheta_4(\tau,z)^2\vartheta_4(\tau,\zeta)^2\over\eta(\tau)^4} \right)\qquad\\
&&\hspace*{-12em} \stackrel{\eqref{strange}}{=}
 B(\tau,z,\zeta)\cdot 2B(\tau)F(\tau).\\
 \end{eqnarray*}
In \eqref{rbosonsdmc}, these are 
the contributions from the sectors $U_{00v}\oplus U_{0v0}$.

Altogether, we find agreement with 
\eqref{rbosonsdmc}, which however counts the {\em fermions}
in the twisted module $V^{s\natural}_{\rm tw}$ of \cite{dmc15}.
Indeed, we have obtained the multigraded trace over 
$V^{s\natural}_{\rm tw, ferm}$, which according to 
\eqref{DMCsectordeco}  is given by
$U_{v00}\oplus U_{vvv}\oplus U_{00v}\oplus U_{0v0}$, where 
every state receives one or three tensor factors
from the vector representation $v$ of 
$\widehat{\mathfrak{so}}(8)_1$. This explains why
from the viewpoint of the Conway Moonshine Module, it seems
natural to dub these states fermionic. 
On the other hand, in the K3 theory, they arise from
$\HHH_{\rm bos}^{\rm R}
= \HHH_{s00} \oplus \HHH_{sss}\oplus \HHH_{00s} \oplus \HHH_{0s0}$,
thus solely receiving contributions from the vacuum representation $0$
and the spinor representation $s$ of 
$\widehat{\mathfrak{so}}(8)_1$, all of which are naturally 
interpreted as being bosonic. 
Moreover, the total $U(1)$ charge with respect to $J_0+\qu J_0$
of each of the states in $\HHH_{\rm bos}^{\rm R}$ 
is even, as is the eigenvalue of $J_0-\qu J_0$
by the spectral condition \eqref{spectrumplus}. Therefore,
according to \eqref{stSUSYfermno},
space-time supersymmetry  implies that
$(-1)^F$ acts by
multiplication with $+1$ on $\HHH_{\rm bos}^{\rm R}$.
We therefore continue to interpret these states as bosons,
that is, we choose to
interchange the roles of bosons and fermions in
$V^{s\natural}_{\rm tw}$. This solely introduces
a difference by a global factor of
$(-1)$ for the action of $(-1)^F$ on $V^{s\natural}_{\rm tw}$, 
which we will come back to in Subsection \ref{reflpf}.
\medskip
\subsubsection{Ramond fermions}\label{Rfermcompared}
In our K3 theory,  the fermionic contributions from 
the sector $({\rm R}, {\rm NS}, {\rm NS} )$, by
\eqref{sectorrnsns}, are
\begin{eqnarray*}
\textstyle
{1\over4}
 \left(
   \left( {\vartheta_3(\tau)\over\eta(\tau)}\right)^{4}
   - \left( {\vartheta_4(\tau)\over\eta(\tau)}\right)^{4} \right)^2
   \cdot {1\over2} \left( {\vartheta_2(\tau,z)^2\vartheta_2(\tau,\zeta)^2\over\eta(\tau)^4}
   - {\vartheta_1(\tau,z)^2\vartheta_1(\tau,\zeta)^2\over\eta(\tau)^4} \right)\\
&&\hspace*{-8em}   \stackrel{\eqref{BandFdef},\eqref{newstrange}}{=} 
\textstyle
 F(\tau,z,\zeta) F(\tau)^2.
\end{eqnarray*}
In \eqref{rfermionsdmc}, these are 
the contributions from the sector $U_{scc}$.

The fermions in 
the sector $({\rm R}, {\rm R}, {\rm R} )$, according to
\eqref{sectorrrr}, are counted by
\begin{eqnarray*}
\textstyle {1\over4}
   \left( {\vartheta_2(\tau)\over\eta(\tau)}\right)^8
   \cdot {1\over2} \left( {\vartheta_2(\tau,z)^2\vartheta_2(\tau,\zeta)^2\over\eta(\tau)^4}
   - {\vartheta_1(\tau,z)^2\vartheta_1(\tau,\zeta)^2\over\eta(\tau)^4} \right)\\
 &&\hspace*{-10em} \stackrel{\eqref{BandFdef},\eqref{strange},\eqref{newstrange}}{=}
 \textstyle
 F(\tau,z,\zeta) F(\tau)^2.
\end{eqnarray*}
In \eqref{rfermionsdmc}, these are 
the contributions from the sector $U_{sss}$.

The fermionic contributions in
the final sector 
$({\rm NS}, {\rm NS}, {\rm R} ) + ({\rm NS}, {\rm R}, {\rm NS} )$,
by \eqref{sectornsnsr}, yield
\begin{eqnarray*}
 \textstyle
   \left( {\vartheta_2(\tau)\over\eta(\tau)}\right)^4
   \cdot 
   {1\over2}
   \left(
   \left( {\vartheta_3(\tau)\over\eta(\tau)}\right)^{4}
   - \left( {\vartheta_4(\tau)\over\eta(\tau)}\right)^{4} \right)
    {1\over2} \left( {\vartheta_3(\tau,z)^2\vartheta_3(\tau,\zeta)^2\over\eta(\tau)^4}
   - {\vartheta_4(\tau,z)^2\vartheta_4(\tau,\zeta)^2\over\eta(\tau)^4} \right)\qquad\\
 \stackrel{\eqref{strange}}{=}
 \textstyle 
 F(\tau,z,\zeta) \cdot 2 F(\tau)^2.
\end{eqnarray*}
In \eqref{rfermionsdmc}, these are 
the contributions from the sectors $U_{ccs}\oplus U_{csc}$.

Altogether, we find agreement with the result of
\eqref{rfermionsdmc}.
\subsubsection{Reflected partition function}\label{reflpf}
From the above, we collect the  contributions
to the reflected partition function in each sector:
\begin{eqnarray}
&&\hspace*{-5em}Z_{\widetilde{\rm NS}}^{\rm refl}(\tau, z,\zeta) \nonumber\\
&:=&  \tr_{\HHH^{\rm NS}} \left( (-1)^F y^{J_0}\widetilde y^{\qu J_0} q^{L_{(0)}-1/2}\right)\nonumber\\
&=&\textstyle{1\over2}\left(
   {1\over2}\sum\limits_{k=2}^4 \left( {\vartheta_k(\tau)\over\eta(\tau)}\right)^{8}
 \cdot  {\vartheta_4(\tau,z)^2\vartheta_4(\tau,\zeta)^2\over\eta(\tau)^4} 
 + \left( {\vartheta_3(\tau)\vartheta_4(\tau)\over\eta^2(\tau)}\right)^4
 \cdot {\vartheta_3(\tau,z)^2\vartheta_3(\tau,\zeta)^2\over\eta(\tau)^4}  \right.\nonumber\\
 &&\textstyle
 \hfill\left.\;\; +
 \left( {\vartheta_2(\tau)\vartheta_3(\tau)\over\eta^2(\tau)}\right)^4
 \cdot  {\vartheta_1(\tau,z)^2\vartheta_1(\tau,\zeta)^2\over\eta(\tau)^4} 
 + 
 \left( {\vartheta_2(\tau)\vartheta_4(\tau)\over\eta^2(\tau)}\right)^4
 \cdot {\vartheta_2(\tau,z)^2\vartheta_2(\tau,\zeta)^2\over\eta(\tau)^4}\right)\!\!,\label{flipnstilde}
 \end{eqnarray}
 \vspace*{-2em}
\begin{eqnarray}
&&\hspace*{-5em}Z_{{\rm NS}}^{\rm refl}(\tau, z,\zeta) \nonumber\\
&:=&  \tr_{\HHH^{\rm NS}} \left(  y^{J_0}\widetilde y^{\qu J_0} q^{L_{(0)}-1/2}\right)\nonumber\\
&=&\textstyle{1\over2}\left(
  {1\over2}\sum\limits_{k=2}^4 \left( {\vartheta_k(\tau)\over\eta(\tau)}\right)^{8}
\cdot {\vartheta_3(\tau,z)^2\vartheta_3(\tau,\zeta)^2\over\eta(\tau)^4} 
+ \left( {\vartheta_3(\tau)\vartheta_4(\tau)\over\eta^2(\tau)}\right)^4
\cdot {\vartheta_4(\tau,z)^2\vartheta_4(\tau,\zeta)^2\over\eta(\tau)^4}  \right.\nonumber\\
&&\textstyle
\hfill\left.\;\; +   
\left( {\vartheta_2(\tau)\vartheta_3(\tau)\over\eta^2(\tau)}\right)^4
\cdot {\vartheta_2(\tau,z)^2\vartheta_2(\tau,\zeta)^2\over\eta(\tau)^4}
+
\left( {\vartheta_2(\tau)\vartheta_4(\tau)\over\eta^2(\tau)}\right)^4
\cdot {\vartheta_1(\tau,z)^2\vartheta_1(\tau,\zeta)^2\over\eta(\tau)^4} 
\right)\!\!,\label{flipns}
 \end{eqnarray}
 \vspace*{-2em}
\begin{eqnarray}
\textstyle
&&\hspace*{-5em}Z_{\widetilde{\rm R}}^{\rm refl}(\tau, z,\zeta) \nonumber\\
&:=& \tr_{\HHH^{\rm R}} \left( (-1)^F y^{J_0}\widetilde y^{\qu J_0} q^{L_{(0)}-1/2}\right)\nonumber\\
&=& \textstyle{1\over2}\left(
  {1\over2}\sum\limits_{k=2}^4 \left( {\vartheta_k(\tau)\over\eta(\tau)}\right)^{8}
\cdot {\vartheta_1(\tau,z)^2\vartheta_1(\tau,\zeta)^2\over\eta(\tau)^4} 
+ \left( {\vartheta_3(\tau)\vartheta_4(\tau)\over\eta^2(\tau)}\right)^4
\cdot {\vartheta_2(\tau,z)^2\vartheta_2(\tau,\zeta)^2\over\eta(\tau)^4}  \right.\nonumber\\
&&\textstyle
\hfill\left.\;\; +   
\left( {\vartheta_2(\tau)\vartheta_3(\tau)\over\eta^2(\tau)}\right)^4
\cdot {\vartheta_4(\tau,z)^2\vartheta_4(\tau,\zeta)^2\over\eta(\tau)^4} 
+ 
\left( {\vartheta_2(\tau)\vartheta_4(\tau)\over\eta^2(\tau)}\right)^4
\cdot {\vartheta_3(\tau,z)^2\vartheta_3(\tau,\zeta)^2\over\eta(\tau)^4} 
\right),\label{fliprtilde} 
\end{eqnarray}
 \vspace*{-2em}
\begin{eqnarray}
&&\hspace*{-5em}Z_{{\rm R}}^{\rm refl}(\tau, z,\zeta) \nonumber\\
&:=& \tr_{\HHH^{\rm R}} \left(  y^{J_0}\widetilde y^{\qu J_0} q^{L_{(0)}-1/2}\right)\nonumber\\
&=&\textstyle {1\over2}\left(
  {1\over2}\sum\limits_{k=2}^4 \left( {\vartheta_k(\tau)\over\eta(\tau)}\right)^{8}
\cdot {\vartheta_2(\tau,z)^2\vartheta_2(\tau,\zeta)^2\over\eta(\tau)^4} 
+ \left( {\vartheta_3(\tau)\vartheta_4(\tau)\over\eta^2(\tau)}\right)^4
\cdot {\vartheta_1(\tau,z)^2\vartheta_1(\tau,\zeta)^2\over\eta(\tau)^4}  \right.\nonumber\\
&&\textstyle
\hfill\left.\;\; +   
\left( {\vartheta_2(\tau)\vartheta_3(\tau)\over\eta^2(\tau)}\right)^4
\cdot {\vartheta_3(\tau,z)^2\vartheta_3(\tau,\zeta)^2\over\eta(\tau)^4} 
+ 
\left( {\vartheta_2(\tau)\vartheta_4(\tau)\over\eta^2(\tau)}\right)^4
\cdot {\vartheta_4(\tau,z)^2\vartheta_4(\tau,\zeta)^2\over\eta(\tau)^4} 
\right).\label{flipr}
\end{eqnarray}
Of these contributions to the partition function,
$Z_{\widetilde{\rm R}}^{\rm refl}(\tau, z,\zeta=0)$ 
agrees with the result of
\cite[(9.14)]{dmc15}, if there,
one inserts the identity element for $g$.
This reproduces the elliptic genus
of K3, as it should, and as was already confirmed
in \cite{eoty89}. 
It also reinforces our suggestion to interchange the roles of bosons
and fermions in $V^{s\natural}_{\rm tw}$, since otherwise,
the graded trace of 
$(-1)^F y^{J_0} \qu y^{\qu J_0} q^{L_{(0)}-1/2}$
over $V^{s\natural}_{\rm tw}$ yields the negative of the
elliptic genus of K3. Indeed, between 
\cite[(8.7)]{dmc15} and \cite[(9.10),(9.14)]{dmc15}, an additional
factor of $(-1)$ was introduced by hand.
The parameter $\zeta$ has not been introduced in \cite{dmc15}, since
there, the $U(1)$ current $\qu J$ of \eqref{dmcu1choiceright} was not
considered.
Concerning the partition functions for the other three sectors,
solely $Z_{\widetilde{\rm NS}}^{\rm refl}(\tau, z=0,\zeta=0)$ can 
be read from \cite[(8.6)]{dmc15}.

Inspecting the relations between the four parts of
the reflected partition function, 
note first of all that as usual, 
$Z_{\widetilde{\rm NS}}^{\rm refl}(\tau, z,\zeta)$ and
$Z_{{\rm NS}}^{\rm refl}(\tau, z,\zeta)$, 
on the one hand,
and $Z_{\widetilde{\rm R}}^{\rm refl}(\tau, z,\zeta)$ and
$Z_{{\rm R}}^{\rm refl}(\tau, z,\zeta)$,
on the other, 
are related by
$(z,\zeta)\mapsto (z+{1\over2},\zeta+{1\over2})$. 
This is a heritage from the space-time supersymmetry of
the K3 theory on $\HHH^{\rm GTVW}$, since there,
$(-1)^F=(-1)^{J_0-\qu J_0}$, as was 
noted in \eqref{stSUSYfermno}. 
Space-time supersymmetry of the underlying
K3 theory also ensures that
$Z_{\widetilde{\rm R}}^{\rm refl}(\tau, z,\zeta)$ and
$Z_{\widetilde{\rm NS}}^{\rm refl}(\tau, z,\zeta)$, on the one hand,
and
$Z_{{\rm R}}^{\rm refl}(\tau, z,\zeta)$ and
$Z_{{\rm NS}}^{\rm refl}(\tau, z,\zeta)$, on the other, 
are related by spectral flow 
$$
\textstyle
f(\tau,z,\zeta)\mapsto q^{1\over2}(y\tilde y)f(\tau,z+{\tau\over2},\zeta+{\tau\over2}),
$$
as they should, according to \eqref{partitrafo}.

Under modular transformations, we find the following behaviour.
Up to the  expected elliptic prefactor, 
$(\tau, z,\zeta)\mapsto(-1/\tau, z/\tau,-\zeta/\tau)$
leaves
$Z_{{\rm NS}}^{\rm refl}(\tau, z,\zeta)$ 
and $Z_{\widetilde{\rm R}}^{\rm refl}(\tau, z,\zeta)$ 
invariant, while it interchanges
$Z_{\widetilde{\rm NS}}^{\rm refl}(\tau, z,\zeta)$
with $Z_{\rm R}^{\rm refl}(\tau, z,\zeta)$. 
On the other hand, the transformation
$\tau\mapsto\tau+1$ interchanges
$Z_{\widetilde{\rm NS}}^{\rm refl}(\tau, z,\zeta)$  with
$-Z_{{\rm NS}}^{\rm refl}(\tau, z,\zeta)$, 
while
$Z_{\rm R}^{\rm refl}(\tau, z,\zeta)$ and
$Z_{\widetilde{\rm R}}^{\rm refl}(\tau, z,\zeta)$
are invariant.
This means that $Z_{\widetilde{\rm R}}^{\rm refl}(\tau, z,\zeta)$
shares its 
modular transformation properties under
$\rm{SL}(2,\Z)$ with that of the partition functions of SCFTs,
while the sum $Z^{\rm refl}(\tau, z,\zeta)$
of the four parts of the partition
function does {\em not}; the latter transforms
like the partition functions of a SCFT only under the Hecke group 
$\textgoth{G}(2)$.

Altogether, the
total partition function $Z^{\rm refl}(\tau, z,\zeta)$ 
and $Z^{\rm refl}_{\widetilde R}(\tau, z,\zeta)$ exhibit
the expected transformation properties, as we explained in
Section \ref{reflconsequence}.
%
\subsection{The Conway Moonshine Module as reflection of a particular
K3 theory}\label{flipgtvw}
By the results of Section \ref{flip}, our reflection procedure is
well-defined on the space of states $\HHH^{\rm GTVW}$
of the K3 theory with $\Z_2^8\colon\M_{20}$ symmetry
of \cite{gtvw14}, if the additional sign choices mentioned
at the end of Section \ref{voa} can be implemented 
consistently. However, our description of the K3 theory as
a lattice CFT, built on a half integral lattice (Section \ref{D4orbifold}),
guarantees that such sign choices solely amount to the choice
of cocycles. Indeed, as we shall argue below, the reflected theory
continues to allow a  lattice theory description, built on
a half integral charge lattice that meets the requirements 
used in Appendix \ref{halfcoc}. 
The existence of cocycles and thereby of consistent
sign choices after reflection thus follows from the constructions 
given in Appendix \ref{halfcoc}. 
Thus indeed, reflection is well-defined for our K3 theory. 
On the Neveu-Schwarz sector 
$\HHH^{\rm NS}$, this introduces  the structure of a
super vertex operator algebra at central charge $c=12$, while 
the Ramond sector
$\HHH^{\rm R}$ carries the structure of an admissible
$\HHH^{\rm NS}$-module. 
By inspection of the partition functions $Z_\SSS^{\rm refl},\,
Z_{\widetilde\SSS}^{\rm refl}$, $\SSS\in\{ {\rm NS,\, R}\}$,
of \eqref{flipnstilde}--\eqref{flipr},
one confirms that the $L_{(0)}$-eigenspace in the reflected
$\HHH^{\rm GTVW}$ at eigenvalue
${1\over 2}$ is trivial. Using 
the uniqueness result of \cite[Thm.~5.15]{du07}, we may 
conclude that  the reflected $\HHH^{\rm GTVW}$ agrees with 
$V^{s\natural}\oplus V^{s\natural}_{\rm tw}$ as a 
super vertex operator algebra plus admissible module, if
we show self-duality and $C_2$-cofiniteness.
Since our theory 
is described in terms of a lattice vertex operator algebra, 
self-duality is immediately checked. $C_2$-cofiniteness follows
by the techniques developed in \cite[\S12]{dlm00}. Instead
of going through the details of the proof of self-duality and 
$C_2$-cofiniteness, we will show by direct comparison that
after reflection, $\HHH^{\rm NS}$ is isomorphic to $V^{s\natural}$
as a super vertex operator algebra, while $\HHH^{\rm R}$ is isomorphic to 
$V^{s\natural}_{\rm tw}$ as an admissible module of 
$\HHH^{\rm NS}\cong V^{s\natural}$.

As already indicated in Section \ref{reflconsequence},
we may thereby induce a considerably richer structure
on $V^{s\natural}\oplus V^{s\natural}_{\rm tw}$ than
what has been investigated, so far. First, this space
is in fact a module of an $N=(4,4)$ super Virasoro algebra
at central charges $(c,\qu c)=(6,6)$,
since this is the case for $\HHH^{\rm GTVW}$. As decribed in Section \ref{supernaturalconstruction},
one obtains an elegant description of the super vertex algebra and module
structure in terms of a charge lattice $\Gamma^{\rm refl}$ ,
which after reflection of the superconformal field theory built on $\HHH^{\rm GTVW}$
governs all OPEs. Indeed, it is straightforward
to apply the first step of
the reflection procedure of Section \ref{flip} to the 
OPEs between real or imaginary
parts of all momentum-winding fields $V_\gamma$,
$\gamma\in\Gamma$, in \eqref{leftrightmovingOPE}. It amounts to replacing 
all contributions $(\qu z-\qu w)^{ \qu\Q\cdot\qu\Q^\prime}$
by $(z-w)^{ \qu\Q\cdot\qu\Q^\prime}$.
For  the charge lattice $\Gamma$
of our $\Dtorus$ theory  in Section \ref{D4orbifold}, which is equipped with the scalar product
$\bullet$ of \eqref{signatureform},  
this amounts to using 
the standard Euclidean scalar 
product on $\R^{12}$, instead. Hence  the reflection procedure changes
the signature of this lattice from $(6,6)$ to $(12,0)$.

Using the description given in Section \ref{D4orbifold} and the
notations of Section \ref{supernaturalconstruction}, after reflection,
the various sectors of the space of states are given by
\be\label{triplets}
\left( \Gamma_{\rm bos}^{\rm NS} \right)^{\rm refl}
\oplus \left( \Gamma_{\rm ferm}^{\rm NS} \right)^{\rm refl}
\oplus \left( \Gamma_{\rm bos}^{\rm R} \right)^{\rm refl}
\oplus \left( \Gamma_{\rm ferm}^{\rm R} \right)^{\rm refl}
\subset
\widetilde\Gamma_{2,2}^{\rm refl}\oplus\widetilde\Gamma_{2,2}^{\rm refl}\oplus\widetilde\Gamma_{2,2}^{\rm refl}
\ee
with
\be\label{ourlattices}
\begin{array}{rcl}
\ds
\left( \HHH_p^\SSS \right)^{\rm refl}
&=&\ds \bigoplus_{\gamma\in\left( \Gamma_p^\SSS \right)^{\rm refl}} \HHH_\gamma,\qquad
\SSS\in\{ {\rm NS,\,R}\},\; p\in\{ {\rm bos, ferm} \}, \\
\hspace*{-3em}\mbox{where }\hspace*{2em}\\
\ds\left( \Gamma_{\rm bos}^{\rm NS} \right)^{\rm refl}
&=& \ds(\widetilde\Gamma_0^{\rm refl})^3 
\cup \left(\widetilde\Gamma_0^{\rm refl}\oplus\widetilde\Gamma_2^{\rm refl}\oplus\widetilde\Gamma_2^{\rm refl}\right)
\cup \left(\widetilde\Gamma_2^{\rm refl}\oplus\widetilde\Gamma_0^{\rm refl}\oplus\widetilde\Gamma_2^{\rm refl}\right)
\cup \left(\widetilde\Gamma_2^{\rm refl}\oplus\widetilde\Gamma_2^{\rm refl}\oplus\widetilde\Gamma_0^{\rm refl}\right),
\\[8pt]
\ds\left( \Gamma_{\rm ferm}^{\rm NS} \right)^{\rm refl}
&=& \ds(\widetilde\Gamma_1^{\rm refl})^3 
\cup \left(\widetilde\Gamma_1^{\rm refl}\oplus\widetilde\Gamma_3^{\rm refl}\oplus\widetilde\Gamma_3^{\rm refl}\right)
\cup \left(\widetilde\Gamma_3^{\rm refl}\oplus\widetilde\Gamma_1^{\rm refl}\oplus\widetilde\Gamma_3^{\rm refl}\right)
\cup \left(\widetilde\Gamma_3^{\rm refl}\oplus\widetilde\Gamma_3^{\rm refl}\oplus\widetilde\Gamma_1^{\rm refl}\right),
\\[8pt]
\ds\left( \Gamma_{\rm bos}^{\rm R} \right)^{\rm refl}
&=& 
\ds
\left(\widetilde\Gamma_2^{\rm refl}\oplus\widetilde\Gamma_0^{\rm refl}\oplus\widetilde\Gamma_0^{\rm refl}\right)
\cup (\widetilde\Gamma_2^{\rm refl})^3 
\cup \left(\widetilde\Gamma_0^{\rm refl}\oplus\widetilde\Gamma_0^{\rm refl}\oplus\widetilde\Gamma_2^{\rm refl}\right)
\cup\left(\widetilde\Gamma_0^{\rm refl}\oplus\widetilde\Gamma_2^{\rm refl}\oplus\widetilde\Gamma_0^{\rm refl}\right),
\\[8pt]
\ds\left( \Gamma_{\rm ferm}^{\rm R} \right)^{\rm refl}
&=& \ds\left(\widetilde\Gamma_3^{\rm refl}\oplus\widetilde\Gamma_1^{\rm refl}\oplus\widetilde\Gamma_1^{\rm refl}\right)
\cup (\widetilde\Gamma_3^{\rm refl})^3 
\cup \left(\widetilde\Gamma_1^{\rm refl}\oplus\widetilde\Gamma_1^{\rm refl}\oplus\widetilde\Gamma_3^{\rm refl}\right)
\cup \left(\widetilde\Gamma_1^{\rm refl}\oplus\widetilde\Gamma_3^{\rm refl}\oplus\widetilde\Gamma_1^{\rm refl}\right).
\end{array}
\ee
The final step in the reflection procedure of Section \ref{flip},
namely  the introduction of appropriate signs in the resulting OPEs, now amounts
to the implementation of cocycle factors.
Their existence is guaranteed by the results of Appendix \ref{halfcoc}.
There, we also provide an explicit formula  \eqref{extrareal} for representatives 
of the cocycles  obeying the compatibility conditions \eqref{cocycleswap},  \eqref{specialgauge}
which are required by the role that the cocycles play in the OPEs, as detailed at the end
of Section \ref{bosonicfieldcontent}.  Using formula \eqref{extrareal} to
construct the cocycles both before and after reflection also 
shows that the required additional signs that
occur in the final step of the reflection procedure  are governed
by cocycles, as is expected from our prescription of the Hermitian
conjugate in \eqref{Klein}. 

In Appendix \ref{halfcoc},
we also show that there are precisely two inequivalent choices of cocycles
in the lattice theory obtained through reflection. 
However, the resulting super vertex operator algebra structure after
reflection is independent of that choice, since the two cocycles differ
solely by a relative sign which plays no role in the super vertex algebra.\label{inequivalent}
Indeed, this relative sign only affects the comparison of the OPEs for 
$\psi(z)\phi(w)$ and $\phi(w)\psi(z)$ involving fields $\psi(z)$, $\phi(w)$,
of which at least one creates a state in the Ramond sector from the
vacuum. Therefore, at least one of the OPEs $\psi(z)\phi(w)$ and $\phi(w)\psi(z)$
is not considered within the structure of the super vertex operator algebra plus
admissible module.
This uniqueness result (up to equivalence) for the induced
super vertex operator algebra plus module structure on
$\left(\HHH^{\rm GTVW}\right)^{\rm refl}$
is in accord with 
John Duncan's theorem on the uniqueness of the
Conway Moonshine Module \cite[Thm.~5.15]{du07}.
\begin{quote}{\em
In summary, the above proves our claim that the Conway Moonshine
Module is the reflection of the K3 theory with $\Z_2^8\colon\M_{20}$
symmetry.}
\end{quote} 

\noindent
This also shows that  the spaces of states 
$\HHH^{\rm GTVW}$ and $V^{s\natural}\oplus V^{s\natural}_{\rm tw}$
are isomorphic as $N=(4,4)$ super Virasoro modules, since 
for OPEs that involve a chiral or an antichiral field, 
by construction, our reflection procedure 
is only a formal manipulation. In fact, by the same
argument, we have given an independent
proof of \cite[Prop.~5.7]{cdr17}.
\smallskip\par

It is  instructive to study the transition from $\HHH^{\rm GTVW}$
to $V^{s\natural}\oplus V^{s\natural}_{\rm tw}$ 
under reflection more closely. 
Indeed,
comparison
to \eqref{DMCsectordeco}
reveals an apparent difference which on the level of
Virasoro modules was already explained
by Duncan and Mack-Crane in \cite[\S11]{dmc15}, invoking
{\em triality}. 
Here, triality amounts to a lattice automorphism of the lattice
$\widetilde\Gamma_{2,2}^{\rm refl}$, which 
maps $\widetilde\Gamma_{0}^{\rm refl}$ to itself, and
\be\label{latticetriality}
\widetilde\Gamma_{1}^{\rm refl} \;\longrightarrow\;
\widetilde\Gamma_{3}^{\rm refl} \;\longrightarrow\;
\widetilde\Gamma_{2}^{\rm refl} \;\longrightarrow\;
\widetilde\Gamma_{1}^{\rm refl} .
\ee
On the level of representations of $\widehat{\mathfrak{so}}(8)_1$,
triality thus induces isomorphisms $v\rightarrow  c\rightarrow  s\rightarrow  v$.
This indeed transforms  \eqref{NSspaceofstates}, \eqref{Rspaceofstates} into
to \eqref{DMCsectordeco}, up to interchanging the roles of bosons and
fermions in the Ramond sector, which we have already discussed in 
Section \ref{flipmodule}, above.
On the level of lattices, \eqref{latticetriality} transforms
$\left( \Gamma_{\rm bos}^{\rm NS} \right)^{\rm refl}$ and
$\left( \Gamma_{\rm bos}^{\rm R} \right)^{\rm refl}$ of
\eqref{ourlattices} into the subsets of $\Gamma_{\rm bos}^{\rm refl}$
of \eqref{theirlattices} labelling states from the Neveu-Schwarz
and Ramond sector, respectively. Analogously, the states labelled by 
$\left( \Gamma_{\rm ferm}^{\rm NS} \right)^{\rm refl}$ and
$\left( \Gamma_{\rm ferm}^{\rm R} \right)^{\rm refl}$ in \eqref{ourlattices}
 are transformed to those labelled by
$\Gamma_{\rm ferm}^{\rm refl}$ in \eqref{theirlattices}.

That triality plays an important role in the context of 
bosonization and fermionization,
which we have made repeated use of in our constructions, was 
probably first noticed by Shankar \cite{sh80}. A detailed discussion
can be found in \cite{gool84,gos85}. Moreover, in 
 \cite[Thm.~5.7]{ffr91}, it is shown
that triality yields an automorphism 
between 
the super vertex operator algebras corresponding to 
charge vectors in each of the lattices
$\widetilde{\Gamma}_0^{\rm refl}\cup\widetilde{\Gamma}_a^{\rm refl}$ with
$a\in\{1,\,2,\,3\}$. In fact, the description of these super vertex operator
algebras as lattice theories yields an alternative proof of this result, 
since for integral charge lattices, up to equivalence, cocycles are
unique
\cite{frka80}, \cite[Thm.~5.5]{ka96}  (see Appendix \ref{halfcoc}).

Let us discuss triality in some greater detail.
To do so, we may restrict our attention to a single summand
$\widetilde\Gamma_{2,2}^{\rm refl}\subset\R^4$ in \eqref{triplets}.
With respect to the standard basis of $\R^4$, the triality automorphism
\eqref{latticetriality}
can be expressed by the matrix
$$
\textstyle
\Theta = 
{1\over2} \begin{pmatrix} 1&1&1&1\\ 1&1&-1&-1\\ 
1&-1&1&-1\\ -1&1&1&-1\end{pmatrix} \quad\in\quad \mb{SO}(4),\;
$$
mapping
the cosets  $\widetilde{\Gamma}_a^{\rm refl}$  according to \eqref{latticetriality}
and inducing a lattice automorphism on $\widetilde{\Gamma}_0^{\rm refl}$.
In accordance with the results of Appendix \ref{halfcoc}, we may introduce
cocycles $\eps$ on $\widetilde\Gamma_{2,2}^{\rm refl}$ by means of 
formula \eqref{extrareal}, where here we state the 
matrix $M$ with respect to the standard basis as
\be\label{cocyclematrix}
M
:= \begin{pmatrix}
0&0&1&0\\
-1&0&0&0\\
0&1&0&0\\
 1&1&-1&0
 \end{pmatrix}.
 \ee
Using 
$$
A:=\begin{pmatrix} 1&0&1&1\\1&1&0&0\\0&1&0&0\\0&0&1&-1
\end{pmatrix}
$$
as a generating matrix of the lattice $\widetilde\Gamma_0^{\rm refl}$,
one easily checks 
$$
A^T M A \equiv (\Theta A)^T M (\Theta A) \mod 2.
$$
Since $\Theta\in\mb{SO}(4)$,  this  proves
that on restriction to the sublattice 
$\widetilde{\Gamma}_0^{\rm refl}$, triality
yields an automorphism of the OPE \eqref{leftrightmovingOPE},
thus confirming the result of \cite[Thm.~5.7]{ffr91}.
%
%

The explicit description of triality, above, also 
justifies our choices of $U(1)$ currents in
\eqref{dmcu1choice}, \eqref{dmcu1choiceright}. They are 
the images of the $U(1)$ currents in \eqref{gtvwu1choice}, 
after reflection and application of triality. To show this, it
suffices to restrict attention to the lattice
$\widetilde\Gamma_{2,2}$ which governs the pair
of two left- and two right-moving Dirac fermions 
$\chi_k,\,\chi_k^\ast,\,\qu\chi_k,\,\qu\chi_k^\ast$
with $k\in\{1,\,2\}$ as in Section \ref{SUSYD4torus}. 
Reflection leaves the purely holomorphic fields
untouched, such that a comparison with the structures
introduced in Section \ref{u1choices} allows us to
set
$$
a_X^+=\chi_1,\quad a_X^-=\chi_1^\ast,\quad
a_Z^+=\chi_2,\quad a_Z^-=\chi_2^\ast.
$$
After reflection to 
$\widetilde\Gamma_{2,2}^{\rm refl}\subset\R^4$
and with the notations of 
\eqref{newu1currents}, using bosonization like there,
the images of the $U(1)$ currents $J,\,\qu J$ in
\eqref{gtvwu1choice} read
$$
\textstyle
J^{\rm refl} =  \mathfrak{j}_1+\mathfrak{j}_2,\qquad
\qu J^{\rm refl} =  \mathfrak{j}_3+\mathfrak{j}_4.
$$
To calculate the images under triality, we simply determine
the images under $\Theta$ of the corresponding vectors
in $\R^4$,
$$
\Theta\begin{pmatrix}1\\1\\0\\0\end{pmatrix}=\begin{pmatrix}1\\1\\0\\0\end{pmatrix},
\qquad
\Theta\begin{pmatrix}0\\0\\1\\1\end{pmatrix}=\begin{pmatrix}1\\-1\\0\\0\end{pmatrix}.
$$
Thus $J^{\rm refl}$ is invariant under triality, yielding the image
$$
\textstyle
 \mathfrak{j}_1+\mathfrak{j}_2
=\nop[\mathbf a_X^+ \mathbf a_X^-] 
+ \nop[\!\mathbf a_Z^+ \mathbf a_Z^-]
$$
as claimed in \eqref{dmcu1choice}. On the other hand,
$\qu J^{\rm refl}$ is mapped to
$$
\textstyle
 \mathfrak{j}_1-\mathfrak{j}_2
= \nop[\mathbf a_X^+ \mathbf a_X^-] 
- \nop[\!\mathbf a_Z^+ \mathbf a_Z^-]
$$
as claimed in \eqref{dmcu1choiceright}.
%
\subsection{Moon Shines on K3}\label{moonshine}
As already mentioned in \cite{dmc15}, one may hope that
a detailed understanding of the Conway Moonshine
Module and its relation to K3 theories might 
help unveil the mysteries of Mathieu Moonshine. 
Our findings make this relation precise. 

Indeed, 
the Conway Moonshine Module arises by reflection from a particular
K3 theory with space of states $\HHH^{\rm GTVW}$.
As already emphasized  in Section \ref{reflconsequence}, 
the very fact that the reflection procedure yields a well-defined 
super vertex operator algebra plus admissible module
structure on $\HHH^{\rm GTVW}$ requires very special
properties of this SCFT.
It would be interesting to 
determine all K3 theories that allow such a procedure
-- we do not expect that there are many, although
further examples may arise from the 
potential bulk SCFTs of \cite{cdr17}.
However, the 
example presented in \cite[\S5.4]{cdr17}
requires the notion of a 
{\em quasi-potential bulk superconformal field theory},
introduced in \cite{cdr17}. This in particular
weakens the requirements
on   pairs of holomorphic and anti-holomorphic
conformal weights, allowing them to differ by arbitrary
rational numbers, in contrast to 
any well-defined SCFT. Indeed,
the quasi-potential $N=(2,2)$ bulk superconformal
field theory in question cannot arise as the  image under reflection
of any well-defined SCFT, and it is not a 
{\em quasi-potential  $N=(4,4)$ bulk superconformal field theory}.

Reflection always yields a 
super vertex operator algebra plus
admissible module that obeys some of the additional properties
required by H\"ohn for ``nice'' theories, namely the spectral ones. 
It would be interesting to know whether all reflected SCFTs are 
nice, i.e.\ whether  $C_2$-cofiniteness  is immediate,
and whether vice versa, all self-dual nice 
super vertex operator algebras at central charge $6N$, 
$N\in\N$, plus admissible modules
arise by reflection from some SCFTs.
According to the classification result of \cite[Thm.~3.1]{cdr17},
there are only three such super vertex operator algebras at 
central charge $12$ and only one at central charge $6$.

For the special K3 theory with space of states $\HHH^{\rm GTVW}$,
reflection  becomes straightforward
due to our description in terms of lattice vertex operator algebras.
Here, the underlying
SCFT on $\HHH^{\rm GTVW}$ induces a considerably
richer structure by including OPEs between pairs of
fields from the admissible module.
In other words, the super vertex operator language
yields a {\em forgetful description}. Further restricting attention to
OPEs in our SCFT that involve holomorphic or antiholomorphic fields, which 
 up to a formal manipulation
remain unchanged under reflection,
one obtains precisely the structure that defines the 
potential bulk SCFTs of \cite{cdr17}.

The interpretation of the Conway Moonshine Module
as image of a K3 theory under reflection elucidates the 
modular properties of its partition function. Indeed,
that $Z^{\rm refl}_{\widetilde R}(\tau,z,\zeta)$ is invariant
under the full modular group ${\rm SL}(2,\Z)$, to our 
knowledge, had not
been noticed, so far. It would also be interesting to know
whether the genus zero property for all analogues
of the McKay-Thompson series for Conway Moonshine
can be traced back to K3, as is the case for 
$Z^{\rm refl}_{\widetilde R}(\tau,z,\zeta=0)$.
\smallskip\par

In \cite{dmc15}, the identification of the Virasoro
modules $\HHH^{\rm GTVW}$ and
$V^{s\natural}\oplus V^{s\natural}_{\rm tw}$ is
used to arrive at a procedure that realizes all
possible symmetries of K3 theories, and more, 
within the Conway Moonshine Module.
Our findings allow a precise interpretation of this process. 

Indeed, the symmetries of our K3 theory are 
naturally described in terms of lattice automorphisms
of certain indefinite lattices. For our  special K3
theory with $\Z_2^8\colon\M_{20}$ symmetry and
those symmetries that respect the 
$\widehat{\mathfrak{su}}(2)_{1,L}^6\oplus\widehat{\mathfrak{su}}(2)_{1,R}^6$
structure,  this  can be done  by means of the
indefinite charge lattice
$\Gamma_{\rm bos}\cong\Gamma_{6,6}\hookrightarrow\R^{6,6}$
introduced in Section \ref{D4orbifold}. Using the ideas
of \cite{nawe00} and \cite[App.~A]{tawe13}, one may
translate this into the traditional description 
 by automorphisms of the lattice of integral cohomology 
of K3, which has signature $(4,20)$.
To do so, one relates the lattice $\Gamma_{\rm bos}$ 
back to the charge lattice of the underlying toroidal theory,
which captures the dependence on moduli. 
This amounts to dropping contributions from the first
summand $\widetilde\Gamma_{2,2}$ in \eqref{tripletbeforereflection}
which governs the ``external free fermions ''
$\chi_j,\, \chi_j^\ast,\, \qu\chi_j,\, \qu\chi_j^\ast$
with $j\in\{1,\,2\}$ introduced in Section \ref{SUSYD4torus},
and then reducing to 
$\Gamma_{4,4}\cong\left(\widetilde\Gamma_0\right)^2\cup 
\left(\widetilde\gamma^{(2)}+\widetilde\Gamma_0\right)^2$.
The result is the image of the traditional charge lattice of
\cite{na86} under the triality map described in \cite{nawe00}.
$\Z_2$-orbifolding induces a map into the even self-dual
lattice of signature $(4,20)$ on which the moduli space
of K3 theories is modelled. This map is  determined
explicitly in \cite{nawe00}.
By the transition from $\HHH^{\rm GTVW}$ to
$V^{s\natural}\oplus V^{s\natural}_{\rm tw}$,  
the latter,  a priori, also
serves as a medium to capture the symmetries
of the K3 theory.
At its heart, reflection then amounts to replacing
the charge lattices with signature $(d,d)$, $d\in\N$,
which govern the behaviour of the particular K3 theory
in question, by the positive definite lattices that are
used in the realm of super vertex operator algebras.
Though the precise mechanism certainly deserves
further investigation, we remark that within
this process, the even self-dual K3 lattice with
signature $(4,20)$ is replaced by a positive definite 
even self-dual lattice $\Lambda$, to be specified below. 
Reflection thus is an implementation of the 
beautiful strategy of Kondo's proof \cite{ko98} of
Mukai's classification result for symplectic automorphisms 
of K3 surfaces \cite{mu88}.

The lattice $\Lambda$ most naturally features in the
construction by
generating the space $\mathfrak{a}$ over $\C$, 
which underlies  the Conway
Moonshine Module. 
The authors of  \cite{dmc15} choose the Leech lattice
for $\Lambda$ and then extend
the discussion of symmetries 
to {\em all} automorphisms of the Leech lattice.
This in particular includes all possible symmetry groups
of K3 theories, by the results of \cite{ghv12}.
Only very few K3 theories lend themselves to  transition from the K3 lattice to 
$\Lambda$  through reflection, i.e.\
by means of a map from the SCFT to some
super vertex operator algebra and admissible module.
But since we have one K3 theory where this is possible,
the reflection procedure does employ  the Leech lattice  as a medium
that collects symmetries of K3 theories
from distinct points of the moduli space
of such theories. The idea thus reveals itself as an 
incarnation of {\em symmetry surfing} as
advocated in 
\cite{tawe11,tawe12}.
We therefore do not regard the Conway Moonshine
Module as
a {\em universal} object, but rather as the reflected
version of one special K3 theory which is a
particularly convenient point of reference in 
symmetry surfing. 

Note that the Conway Moonshine Module  possesses an {\em infinite}
symmetry group. Indeed, in the notations of 
Section \ref{supernatural},   the symmetry group is a
$\Z_2$-quotient of
$\mb{Spin}(\mathfrak{a})$, by \cite[Prop.~4.6]{du07}. This is a 
consequence of the forgetful description, alluded
to, above: the weaker the structure that the
symmetries are required
to preserve, the more symmetries one expects to find.
Nevertheless, $V^{s\natural}\oplus V^{s\natural}_{\rm tw}$
exhibits Conway Moonshine with respect to 
a natural action of $Co_0$,  since
by \cite[Prop.~3.1]{dmc14}, 
realizing the Leech lattice with respect to some choice of real structure
as a lattice in the real part of
$\mathfrak{a}$, 
the action 
of its automorphism group lifts to a finite subgroup
of the symmetry group. More precisely, by
 \cite[Thm.~4.11]{du07}, 
$Co_0$ is the automorphism
group that leaves invariant a choice of $N=1$ structure
on  $V^{f\natural}:=A(\mathfrak{a})^0\oplus A(\mathfrak{a})_{\rm tw}^0$
(with the notations of Sect.~\ref{supernaturalconstruction}),
factorizing through $Co_1$. In \cite{cddhkw14}, this result
is generalized to larger extended chiral algebras, yielding
Mock Modular Moonshine for various subgroups of $Co_0$.

From the viewpoint of
Mathieu Moonshine, however, we find it more
natural to realize $\mathfrak{a}$ as a complex vector space generated
by the Niemeier lattice $\Lambda$
of type $A_1^{24}$, whose symmetry
group is an extension of the largest Mathieu group $M_{24}$. 
Since $M_{24}$ has trivial Schur multiplier, and it is a simple group,
the proof of \cite[Prop.~3.1]{dmc14} can be applied to
this group just as well, showing that the lattice automorphisms
in $M_{24}$ lift to form a symmetry 
group of the 
super vertex operator algebra and admissible module structure
on $V^{s\natural}\oplus V^{s\natural}_{\rm tw}$. 
 Symmetry surfing the moduli
space of K3 theories
allows  to generate the action of the entire group $M_{24}$ 
on $V^{s\natural}\oplus V^{s\natural}_{\rm tw}$. 
In the process, one should keep in mind that
it has long become clear that the symmetry groups
of K3 theories cannot  
explain Mathieu Moonshine, since these 
groups, in general, need not even form subgroups
of $M_{24}$.
As 
in our earlier work \cite{tawe11,tawe12,tawe13}, 
we  emphasize that this problem can possibly
be cured by restricting attention to
{\em geometric} symmetry groups of K3
theories\footnote{Here and in the references \cite{tawe11,tawe12,tawe13}
by a {\em geometric symmetry} we mean a symmetry which induces a lattice
automorphism on the K3 lattice $H^\ast({\rm K3},\Z)$ which in some geometric
interpretation leaves $H^0({\rm K3},\Z)$ and $H^4({\rm K3},\Z)$
pointwise invariant.} rather than including all quantum 
symmetries.
The resulting $M_{24}$-twining elliptic genera agree with the ones obtained
by Duncan and Mack-Crane in \cite{dmc15}. Note that seven of these
twining elliptic genera
differ from the ones of Mathieu Moonshine according to
\cite[\S1.4]{dmc15}.
As was emphasized in \cite{tawe11,tawe13}, symmetry surfing
by merely employing lattice techniques cannot be expected to 
yield the $M_{24}$-modules of Mathieu Moonshine. Indeed, the
results of \cite{tawe12,gakepa16} show that the construction of the
relevant representations of $M_{24}$ by symmetry surfing must involve
a twist, which is not implemented in the Conway Moonshine Module. 
\vskip 1cm

{\textbf{Acknowledgements}}

The authors wish to thank Durham University, the London Mathematical Society 
and EPSRC for funding a Durham LMS-EPSRC Symposium on
{\em New Moonshine, Mock Modular Forms and String Theory} 
in August 2015, as well as the Institut des Hautes \'Etudes Scientifiques in 
Bures-sur-Yvette, where part of this work was carried out.
They also thank Roberto Volpato 
and in particular John Duncan for enlightening conversations. 
The authors are grateful to Thomas Creutzig, John Duncan and Wolfgang
Riedler for correspondence regarding their recent manuscript 
\cite{cdr17}.

AT thanks Jens Funke for helpful remarks. KW 
particularly thanks Andr\'e Henriques and James Tener
for enlightening discussions. She
is indebted to Yi-Zhi Huang for a helpful exchange, and she 
thanks the organisers of the 2015 Notre Dame Conference on {\em Lie Algebras,
Vertex Operator Algebras, and Related Topics}
in honour of J. Lepowsky and R. Wilson 
for the inspiring environment. 
KW is grateful to the organisers of the program on {\em Automorphic forms, mock modular forms 
and string theory} in September 2016 and to the Simons Center for Geometry 
and Physics at Stony Brook
for the support and hospitality, and for the inspiring environment during 
 some crucial steps of this work.
 She moreover thanks the Isaac Newton Institute for Mathematical 
 Sciences for its hospitality during the programme {\em Operator algebras:
 subfactors and applications} which was supported by EPSRC grant
 number EP/K032208/01, and she is grateful to the organisers.
 Finally, she thanks Durham University for its hospitality during the
 finals steps of this work.
\appendix
\section{Necessary ingredients from conformal field theory}\label{cft}
To pave the way for a meaningful comparison between
the K3 theory  studied in \cite{gtvw14}
and the Conway Moonshine Module, in this  Appendix,
we collect some of the main ingredients to (super-)conformal
field theory.
Throughout this work, by a {\em conformal field
theory} (CFT) we mean a (compact) Euclidean two-dimensional
unitary conformal field theory. 
In fact, we restrict our attention to superconformal field theories 
(SCFTs) with at least $N=2$ {\em worldsheet
supersymmetry} both for the left- and for the right-movers,
and in addition, we assume {\em spacetime supersymmetry}
to hold. 
Our presentation is by no means complete,
and we refer to the literature for further details, see e.g.\
\cite{bpz84,na87,gi88b,gr92,dms96,ka96,ga96,ga99,ga00,
boba02,frbe04,sch97,we14,we04}, and references therein.
\smallskip\par

The first main ingredient of a {\em conformal field theory} is
a complex vector {\em space of states} $\H$, 
equipped with a positive definite scalar product $\langle\cdot,\cdot\rangle$
and a compatible real structure $\H\longrightarrow\H$,
$v\mapsto v^\ast$. For a {\em superconformal 
field theory} as specified above, the space of states $\H$ is assumed to
decompose into a direct sum of simultaneous unitary
representations of two super-commuting copies of an $N=2$ 
super-Virasoro algebra at central
charges $c$ and $\overline c$, respectively,
compatible with the real structure of $\H$. 
The other even standard 
generators of these super-Lie algebras
are traditionally denoted $L_n,\,J_n;\, \qu L_n,\, \qu J_n$
with $n\in\Z$.
These two super-Lie algebras and all structures
arising from them, in the physics literature, 
are known as {\em left-moving} or {\em holomorphic},
and {\em right-moving} or {\em anti-holomorphic}, 
respectively, where the latter are denoted by 
overlined letters, in general.
By our assumptions on the worldsheet
supersymmetry of our SCFTs, 
$\H$ enjoys a $\Z_2$-grading by $(-1)^F$,
where $F$ denotes the worldsheet fermion number operator.
The eigenspaces of $(-1)^F$ with eigenvalues $\pm1$
contain the {\em worldsheet bosons} and {\em fermions},
respectively, and they  are denoted $\H_{\rm bos}$,
$\H_{\rm ferm}$, hence 
$\H=\H_{\rm bos}\oplus\H_{\rm ferm}$.
In addition, spacetime supersymmetry imposes
a second, compatible $\Z_2$-grading by
fermion boundary conditions, which decomposes
the space of states\footnote{In general, one should include
an R-NS and an NS-R sector, but these are trivial in our examples.
We therefore use the shortcut notation R for the R-R sector, and NS for the NS-NS sector
in this work.} into a {\em Neveu-Schwarz sector}
$\H^{\rm NS}$ and a {\em Ramond sector} $\H^{\rm R}$, hence
$\H=\H^{\rm NS}\oplus\H^{\rm R}$. Both $\Z_2$-gradings are
compatible with the real structure on $\H$.

The linear operators $L_0,\, J_0;\,\qu L_0,\,\qu J_0$
are assumed to restrict to pairwise commuting self-adjoint linear
operators on each of the four sectors
$$
\H^\SSS_p := \H_p\cap\H^\SSS,
\quad p\in\{\mb{\rm bos}, \mb{\rm ferm}\}, \;
\SSS \in\{{\rm NS, R}\}.
$$
They are simultaneously diagonalizable with
finite dimensional simultaneous eigen\-spa\-ces of $L_0$
and $\qu L_0$, and with one-dimensional
$$
\mb{ker}\left(L_0\right)\cap \mb{ker}\left(\qu L_0\right)
= \mb{ker}\left(L_0\right)\cap \mb{ker}\left(\qu L_0\right)\cap \mb{ker}\left(J_0\right)\cap \mb{ker}\left(\qu J_0\right)
\subset \H_{\rm bos}^{\rm NS}.
$$
The latter condition is known as the {\em uniqueness of the
vacuum}; one chooses a real
$\Omega\in \mb{ker}\left(L_0\right)\cap \mb{ker}\left(\qu L_0\right)$
with $\langle\Omega,\Omega\rangle=1$
and calls it the {\em vacuum}.
\smallskip\par

The second main ingredient of a SCFT is 
a {\em system of $n$-point functions}, that is, 
for every $n\in\N$, there are maps
$$
\begin{array}{rcc}
\mathbb H^{\otimes n}_{{\rm bos}}
&\longrightarrow& \mbox{Maps}(\C^n\setminus\cup_{i\neq j}\{z\in\C^n|z_i=z_j\},\C),\\[5pt]
{(}\mathbb H{^{\rm NS})}^{\otimes n}
&\longrightarrow& \mbox{Maps}(\C^n\setminus\cup_{i\neq j}\{z\in\C^n|z_i=z_j\},\C),\\[5pt]
\mb{{both} denoted by }\qquad
\phi_1\otimes \cdots\otimes \phi_n
&\mapsto&
(\;z\mapsto \langle \phi_1(z_1)\cdots\phi_n(z_n) \rangle\;),
\end{array}
$$
since they agree on $(\mathbb H_{\rm bos}^{\rm NS})^{\otimes n}$. 
These maps form a {\em semilocal}, 
{\em Poincar\'e covariant}, 
{\em conformally covariant} system, which 
is a {\em unitary representation} of an {\em operator
product expansion} (OPE). 
If $n\geq2$, then the $n$-point functions have extensions to
$$
(\mathbb H_{\rm bos}^{\rm NS})^{\otimes (n-2)}\otimes 
\H^{\otimes 2}
\longrightarrow \mbox{Maps}(\C^n\setminus\cup_{i\neq j}\{z\in\C^n|z_i=z_j\},\C)
$$
obeying all the above-mentioned properties. Moreover, for any choice of a contractible open
subset $U\subset \C^n\setminus\cup_{i\neq j}\{z\in\C^n|z_i=z_j\}$,
they can be extended to maps from $\mathbb H^{\otimes n}$ into 
$\mbox{Maps}(U,\C)$.
For details concerning this terminology,
along with a list of the many
consistency conditions and
properties that the above-mentioned structures are assumed to obey
in a full-fledged SCFT, we need to refer the reader to the literature,
since a full account would lead way beyond the scope of this
work. In the following,
we will, however, collect some of the consequences
of these consistency conditions which turn out to be most 
crucial to us.
\smallskip\par

For example, one assumes
that every SCFT has a well-defined
{\em partition function} $Z(\tau,z)$. That is, 
with $q:=e^{2\pi i\tau},\; y:=e^{2\pi iz}$ for all $\tau,\, z\in\C$ with $\rm{Im}(\tau)>0$,
and with $\qu q,\,\qu y\in\C$ denoting the complex conjugates of $q,\,y\in\C$,
\be\label{pfdef}
\begin{array}{rcl}
Z(\tau,z) 
&:=& 
\tr_{\H}\left( {1\over2} (1+(-1)^F)\;  y^{J_0}\qu y^{\qu J_0} q^{L_0-c/24} \qu q^{\qu L_0-\qu c/24} \right)\\[5pt]
& =& {
\tr_{\H_{\rm bos}}\left(   y^{J_0}\qu y^{\qu J_0} q^{L_0-c/24} \qu q^{\qu L_0-\qu c/24} \right)
}
\end{array}
\ee
is convergent. Moreover,
under ``integral'' M\"obius transformations
\be\label{Moebius}
\textstyle
(\tau,z)\mapsto \left(\frac{a\tau+b}{c\tau+d}, \frac{z}{c\tau+d}\right)\,,\quad 
\left(\begin{array}{cc}a&b\\c&d\end{array}\right)\in {\rm SL}(2,\mathbb Z),
\ee
$Z(\tau,z)$ transforms like the product of a weak Jacobi form of weight $0$
and index ${c\over6}$ with the complex conjugate of such a weak Jacobi
form at index ${\qu c\over6}$.
By our assumptions on supersymmetries, we may decompose $Z(\tau,z)$ according to
\be\label{partifuncdef}
\left.\hphantom{where}
\textcolor{black}{\begin{array}{rcl}
Z(\tau,z) 
&=& \textstyle
{1\over2} \left( Z_{\rm NS}(\tau,z)+Z_{\widetilde{\rm NS}}(\tau,z)+Z_{\rm R}(\tau,z)+Z_{\widetilde{\rm R}}(\tau,z)\right),\\[5pt]
\makebox[0pt]{where for  $\SSS\in\{{\rm NS,\, R}\}$,}\quad \\
Z_{\SSS}(\tau,z)
&:=& \tr_{\H^{\SSS}} \left( y^{J_0}\qu y^{\qu J_0} q^{L_0-c/24} \qu q^{\qu L_0-\qu c/24} \right),\\[5pt]
Z_{\widetilde{\SSS}}(\tau,z)
&:=&  \tr_{\H^{\SSS}} \left( (-1)^F y^{J_0}\qu y^{\qu J_0} q^{L_0-c/24} \qu q^{\qu L_0-\qu c/24} \right),
\end{array}}\right\}
\ee
and the functions $Z_{\SSS}(\tau,z)$, $Z_{\widetilde{\SSS}}(\tau,z)$ are well-defined
for all $\tau,\, z\in\C$ with $\rm{Im}(\tau)>0$, as well.
As detailed, for example, in \cite[Thm.~3.1.4]{diss}, 
since we have assumed worldsheet {\em and} 
spacetime supersymmetry,
the above four summands
of the partition function are related as follows:
\be\label{partitrafo}
\begin{array}{rclrcl}
Z_{\rm R}(\tau,z) &=& \multicolumn{4}{l}
{q^{c/24}\qu q^{\qu c/24} y^{c/6}\qu y^{\qu c/6} Z_{\rm NS}(\tau,z+{\tau\over2}),}\\[5pt]
Z_{\rm NS}(\tau,z) &=& 
\multicolumn{4}{l}{q^{c/24}\qu q^{\qu c/24} y^{c/6}\qu y^{\qu c/6} Z_{\rm R}(\tau,z+{\tau\over2}),}\\[5pt]
Z_{\widetilde{\rm NS}}(\tau,z)&=& Z_{\rm NS}(\tau,z+{1\over2}), &\qquad Z_{\widetilde{\rm R}}(\tau,z)&=&Z_{\rm R}(\tau,z+{1\over2}).
\end{array}
\ee
Moreover, $Z_{\widetilde{\rm R}}(\tau,z)$, on its own,  
transforms  like $Z(\tau, z)$ under the ``integral''
M\"obius transformations stated above.

The first two lines of 
\eqref{partitrafo}  are an immediate consequence
of our assumption of space-time supersymmetry. 
The latter implies that the theory is invariant under
{\em spectral flow} (see, for example, \cite{se86,se87},
or \cite[\S3.4]{gr97}), which induces a multigraded
isomorphism $\H^{\rm NS}\stackrel{\cong}{\longrightarrow}\H^{\rm R}$.
On the eigenspaces $\H_{h,Q;\qu h,\qu Q}^\SSS\subset\H^\SSS$,
$\SSS\in\{{\rm NS,\, R}\}$, with eigenvalues $(h,Q;\qu h,\qu Q)$
with respect to $(L_0,\,J_0;$ $\qu L_0,\qu J_0)$, spectral flow induces
$\H_{h,Q;\qu h,\qu Q}^{\rm NS}\stackrel{\cong}{\longrightarrow}
\H_{h^\prime,Q^\prime;\qu h^\prime,\qu Q^\prime}^{\rm R}$ with
\be\label{spfl}
\textstyle
(h^\prime,Q^\prime;\qu h^\prime,\qu Q^\prime)
= (h+{Q\over2}+{c\over24},Q+{c\over 6}\;;\;\qu h+{\qu Q\over2}+{\qu c\over24},\qu Q+{\qu c\over 6}).
\ee
By what was said above, each 
subspace 
$$
\H_{h;\qu h}^\SSS := \left\{ \upsilon\in\H^\SSS \mid L_0\upsilon = h\upsilon,\; 
\qu L_0\upsilon = \qu h\upsilon \right\}
= \textstyle \bigoplus\limits_{Q,\qu Q}\;  
\H_{h,Q;\qu h,\qu Q}^\SSS\subset\H^\SSS
$$ 
is finite
dimensional and obeys $(\H_{h;\qu h}^\SSS)^\ast
=\H_{h;\qu h}^\SSS$.

The assumption that the system of 
$n$-point functions in a CFT represents an OPE
{\em unitarily}
implies {\em reflection positivity}. This property
of a quantum field theory
amounts to a compatibility condition between 
the scalar product,  the real 
structure, and the $n$-point functions on $\H$,
as we shall recall now (see also \cite[\S3.5]{ga00}, for 
example, which however focusses on bosonic CFTs). 
First, for every $n$-point function
$z\mapsto\langle \phi_1(z_1)\cdots\phi_n(z_n) \rangle$,
$\phi_1,\,\ldots,\phi_n\in\H$,
that is well-defined on a domain
$U\subset\C^n\setminus\cup_{i\neq j}\{z\in\C^n|z_i=z_j\}$
with $\qu z\in U$ for all $z\in U$,
the following compatibility condition with the real structure holds:
\be\label{realcomp}
\forall z\in U\colon\qquad
\qu{\langle \phi_1(z_1)\cdots\phi_n(z_n) \rangle}
= \langle \phi_1^\ast(\qu z_1)\cdots\phi_n^\ast(\qu z_n) \rangle.
\ee

To give the full statement
of reflection positivity, we  introduce the following
notation:
for a complex vector space $V$,
let $V\!\{z\}$ denote the vector space of formal power series of the form
$$
\sum_{(r,\qu r)\in R} C_{r,\qu r} z^{r} \qu z^{\qu r}, \quad
C_{r,\qu r}\in V,\quad 
\mb{ with finite }
R\subset\left\{(h,\qu h)\in\R_{\geq0}^2\mid h-\qu h\in\textstyle{1\over2}\Z\right\}.
$$
Then, reflection positivity amounts to the existence of
an anti-$\C$-linear map $\H\longrightarrow\H\{x\}$,
$\phi\longmapsto\phi^\dagger$,
which induces a map 
$$
\phi(z)\longmapsto\left(\phi(z)\right)^\dagger = \phi^\dagger(\qu z^{-1})
\qquad \mb{ with } \left( \left( \phi(z)\right)^\dagger\right)^\dagger=\phi(z)
$$ 
on the level of the associated fields (usually called {\em Hermitian conjugation} 
in the physics literature), 
where
\be\label{daggercondition}
\qu{\langle \phi_1(z_1)\cdots\phi_n(z_n) \rangle}
= \langle \left(\phi_n( z_n)\right)^\dagger\cdots\left(\phi_1( z_1)\right)^\dagger \rangle
\ee
for $\phi_1,\,\ldots,\,\phi_n$ and $z_1,\,\ldots,\,z_n$ as in \eqref{realcomp}.
For  $\psi,\phi\in\H$,
with $(\psi(z))^\dagger=\psi^\dagger(\qu z^{-1})$ as above,
and by linear extension of $\langle\cdots\rangle$ to $\H\{x\}\otimes \H$,
\be\label{reflpos}
\langle\psi,\phi\rangle 
= \lim_{x,w\rightarrow0} \langle \psi^\dagger(\qu x^{-1})\phi(w)\rangle,
\ee
where \eqref{daggercondition} ensures the compatibility of \eqref{reflpos} with
the Hermitian product structure $\langle\psi,\phi\rangle=\qu{\langle\phi,\psi\rangle}$.
For the field $\phi(z)$, in general, $(\phi(z))^\dagger$ is
the image of $(\phi(z))^\ast$ under the conformal transformation $z\mapsto z^{-1}$.
If  $\phi\in\H$ is {\em quasi-primary}, i.e.\
$L_{-1}\phi=0$ and $\qu L_{-1}\phi=0$, and
if $L_0\phi = h\phi$ and $\qu L_0\phi = \qu h\phi$, 
then
\be\label{Klein}
\phi^\dagger=x^{2h}\qu x^{2\qu h}\kappa_\phi\phi^\ast,
\ee
where $\kappa_\phi=(-1)^{h-\qu h}$ if $\phi\in\H_{\rm bos}$.
In general, $\kappa_\phi$ 
is an operator that plays the role of an
additional cocycle factor\footnote{We remark at this point that  
already on the level of bosonic fields,  the factor $(-1)^{h-\qu h}$ is forgotten
in various standard conformal field theory texts.}, thus the notation, reminiscent of 
the {\em Kleinian transformations}\footnote{James Tener has 
explained this to us; the cocycle factor
$\kappa_\phi$ is indispensible for fermionic vertex operators 
$\phi(z)$ in order to 
consistently define adjoint intertwining operators \cite{te17}.
In the literature,
incarnations of $\kappa_\phi$ can already 
be found, for example, in 
\cite{ya13,aili15,te16}.} \cite{kl38}.
It takes into account the fact that 
the very definition of $(\phi(z))^\ast$ requires the
specification of
complex conjugation, which  on the Riemann surface $\Sigma$ 
that parametrizes
a bosonic field $\phi(z)$, $z\in\Sigma$, reverses the co-orientation. 
For fermionic fields, we effectively work on a $2\colon1$ cover of $\Sigma$,
which entails the  choice of 
a lift of the complex conjugation
 on $\Sigma$. This choice introduces  additional
phase factors\footnote{Andr\'e 
Henriques has calculated the lift of the complex conjugation 
to a $2\colon1$ cover of $\Sigma=\C^\ast$, confirming the occurrence
of $\kappa_\phi$ \cite{he17}.}
which may be consistently 
implemented by means of the
cocycle factor $\kappa_\phi$.

Note  that the two-point functions 
 $\langle\psi^\ast(\qu x^{-1})\phi(w)\rangle$ with $\psi\in\H^{\SSS}_p,\,
\phi\in\H^{\widetilde{\SSS}}_{\widetilde p},\, \SSS,\widetilde\SSS\in\{{\rm NS,\, R}\},\, \break p,\widetilde p\in\{{\rm bos, ferm}\}$,
can only be non-vanishing if $\SSS=\widetilde\SSS$ and $p=\widetilde p$. Hence, despite 
the restrictions
on the validity of \eqref{realcomp}, 
equation \eqref{reflpos} can be used to extract the scalar
product $\langle\cdot,\cdot\rangle$ on $\H$
from these two-point functions, with $\langle\psi,\phi\rangle=0$ 
for $\psi,\phi$ as above if $\SSS\neq\widetilde\SSS$ or $p\neq\widetilde p$. Unitarity then implies
that $\langle\cdot,\cdot\rangle$ is positive definite.

Reflection positivity together with conformal covariance
of the $n$-point functions ensures that the {\em state-field
correspondence} holds in our theory, i.e.\ that there is a linear
map associating to every state $\upsilon\in\H$ a {\em field}
$\upsilon(z)$ which {\em creates $\upsilon$ from the vacuum}.
\smallskip\par

As a main result of this work, we show that for  particular SCFTs, one can
consistently  {\em reflect} all 
 fields, transforming them into  holomorphic  fields
to obtain a superconformal vertex operator algebra 
along with a twisted module from
the original superconformal field theory.
To do so, we need to pay special attention to the above-mentioned
 consequences of
unitarity. Let us  illustrate
this for the left- and  right-moving components
$(\psi,\, \qu\psi)$ of a
free Majorana fermion, like 
in the Ising model, where
we follow the normalisations used in 
\cite{bpz84,dms96}:
$$
\psi(z)\psi(w) \sim {1\over z-w},\qquad 
\qu\psi(\qu z)\qu\psi(\qu w) \sim {1\over\qu z-\qu w}.
$$ 
By definition, the bosonic field
$\eps(z,\qu z):=i\nop[\psi(z)\qu\psi(\qu z)]$ obeys
$$
\langle \eps(z,\qu z)\eps(w,\qu w)\rangle = {1\over|z-w|^2},
$$
and $\eps^\dagger=x\qu x\eps$. As is customary,
we choose our real structure on the 
space of states such  that $\psi^\ast=\psi$,
but then $\qu\psi^\ast=-\qu\psi$ follows. 
This can also be seen as a consequence of 
the `reality' condition \cite{gi88b,gr92}
for a Majorana spinor $(\xi_1,\xi_2)$, 
$\xi_1^+ = i\xi_2$, $\xi_2^+ = i\xi_1$,
and thus $\xi_2^\ast=\xi_2$ implies 
$\xi_1^\ast=-\xi_1$. To have a purely real
left-moving field $\psi(z)$, we have chosen the
convention $(\psi,\qu\psi)=i(\xi_1,\xi_2)$.

In contrast, for two  
left-moving free Majorana fermions $\psi_k$, $k\in\{1,2\}$,
with coupled spin structures, the assumption that
$\psi_k^\ast=\psi_k$ for both $k\in\{1,2\}$
is consistent with the required 
$\eps_{12}^\dagger=x^2\eps_{12}$
for $\eps_{12}(z):=i\nop[\psi_1(z)\psi_2(z)]$. In other words, 
the {\em real structure} $v\mapsto v^\ast$ of the underlying spaces 
of states is not compatible with a map $\psi\mapsto\psi_1$
and $\qu\psi\mapsto\psi_2$. One readily checks that for the
left-moving component $\psi$ of the free Majorana fermion, with
our conventions, $\psi^\dagger = i x\psi$.
More generally, as is explained at the end of Section \ref{voa},
reflection indeed entails the occurrence of additional cocycle factors,
as may well be expected by the above discussion.

\section{Cocycle construction for certain half integral lattices}\label{halfcoc}
In this Appendix, 
we  provide a construction of consistent $2$-cocycles
obeying a number of additional conditions,
for a certain type of lattice which is central to our work.
Our presentation extends 
to certain
half integral lattices
the classical results of 
\cite{frka80,se81,gool84}, which apply to lattice vertex
operator algebras built on integral lattices. 
Our analysis is inspired by 
\cite{gnos86,gnors87}, though we found it useful
to include a proof of consistency for $2$-cocycles on the
lattices that are relevant to our work. Indeed, to extend
the solutions offered, for example, in \cite{kllsw87}
for the (integral!) lattices governing certain current
algebras to the half integral lattices that govern the fermionic
fields, additional consistency requirements are necessary.
In \cite{bpz16}, a solution similar to ours
is presented for 
the lattice we denote $\widetilde\Gamma_{2,2}$ in  \eqref{gammat22}.
\smallskip\par

Let $\Gamma\subset\R^D$ denote a lattice of rank $D$.
We begin by recalling the general definition of $2$-cocycles on $\Gamma$, following
the exposition in \cite[\S5.5]{ka96}.
Let $Z\subset\C^\ast$ denote a multiplicative finite subgroup
with $-1\in Z$.
We call a map
$$
\eps\colon \Gamma\times\Gamma \longrightarrow Z
$$
a {\em $2$-cocycle on $\Gamma$ with values in $Z$}, if the following coboundary
condition holds:
\be\label{cocycledef}
\forall \alpha,\, \beta,\,\gamma\in \Gamma\colon\qquad
\eps(\alpha,\beta)\eps(\alpha+\beta,\gamma) = 
\eps(\alpha,\beta+\gamma)\eps(\beta,\gamma).
\ee
Two such $2$-cocycles $\eps,\,\widetilde\eps$ are said to be {\em equivalent} if 
there is a map $\eta\colon\Gamma\rightarrow Z$, $\alpha\mapsto\eta_\alpha$,
such that
\be\label{gaugetransfo}
\forall \alpha,\, \beta\in \Gamma\colon\qquad
\widetilde\eps(\alpha,\beta) = \eta_\alpha\eta_\beta\eta_{\alpha+\beta}^{-1}\eps(\alpha,\beta).
\ee
In other words, the equivalence classes of $2$-cocycles with values in $Z$ 
are the elements of $H^2(\Gamma,Z)$,
the second group cohomology of $\Gamma$ with values
in the trivial $\Gamma$-module $Z$.
Note also that every $2$-cocycle $\eps$ with values in $Z$ defines a central 
extension $\widetilde\Gamma$ of $\Gamma$ by $Z$, where as a set,
$\widetilde\Gamma=\Gamma\times Z$, and one has the group law
$(\alpha,\lambda)\cdot(\beta,\mu):=(\alpha+\beta,\eps(\alpha,\beta)\lambda\mu)$
for $(\alpha,\lambda),\,(\beta,\mu)\in\Gamma\times Z$. This induces an
isomorphism between $H^2(\Gamma,Z)$ and the equivalence classes of
central extensions  of $\Gamma$ by $Z$.

Given a $2$-cocycle $\eps$ with values in $Z$,
it is convenient to introduce the {\em symmetry factor}
$$
S\colon \Gamma\times\Gamma \longrightarrow Z,
\qquad
(\alpha,\beta)\mapsto S(\alpha,\beta) := \eps(\alpha,\beta)\eps(\beta,\alpha)^{-1}.
$$
One immediately checks that $S$ satisfies the following conditions:
\be\label{conditionsonS}
\left.
\begin{array}{rrcl}
\forall \alpha,\, \beta,\,\gamma\in \Gamma\colon\qquad&
S(\alpha,\alpha) &=& 1,\\[5pt]
&S(\alpha,\beta)S(\beta,\alpha)&=& 1,\\[5pt]
&
S(\alpha+\beta,\gamma) &=& S(\alpha,\gamma)S(\beta,\gamma).
\end{array}\right\}
\ee
According to \cite[Lemma 5.5]{ka96}, the above yields
a $1\colon1$ correspondence
between symmetry factors $S$ obeying \eqref{conditionsonS} and  equivalence
classes of $2$-cocycles with values in $Z$. Note that \eqref{conditionsonS}
allows us to express $S$ in terms of its values on a choice
$\alpha_1,\,\ldots,\,\alpha_D$ of generators of $\Gamma$,
since
\be\label{Sfromgenerators}
\mbox{for } \alpha=\sum_{j=1}^D a_j\alpha_j, \; \beta=\sum_{k=1}^D b_k\alpha_k\in\Gamma\colon\qquad
S(\alpha,\beta) = \prod_{j,k=1}^D S(\alpha_j,\alpha_k)^{a_j b_k}.
\ee
The discussion, so far, 
is independent of any quadratic form that we may choose on $\Gamma$.
In our applications, however, we are interested in special choices of
$2$-cocycles, where in addition to the above, we assume that
the lattice $\Gamma$ is equipped
with a non-degenerate symmetric bilinear form $\langle\cdot,\cdot\rangle$ 
with $\langle\alpha,\alpha\rangle\in\Z$ for all
$\alpha\in\Gamma$. Furthermore, for
our $2$-cocycles, we  require 
the following:
\be\label{cocycleswap}
\forall \alpha,\, \beta\in \Gamma\colon\qquad
\mbox{if } \langle\alpha,\beta\rangle\in\Z, \mbox{ then }
\eps(\alpha,\beta) =
(-1)^{\langle\alpha,\beta\rangle+\langle\alpha,\alpha\rangle\cdot\langle\beta,\beta\rangle}
\eps(\beta,\alpha).
\ee
In other words, the associated symmetry factor $S$ must
obey
\be\label{Sasdef}
\forall \alpha,\, \beta\in \Gamma\colon\qquad
S(\alpha,\beta) =
(-1)^{\langle\alpha,\beta\rangle+\langle\alpha,\alpha\rangle\cdot\langle\beta,\beta\rangle}
\qquad
\mbox{ if } \langle\alpha,\beta\rangle\in\Z.
\ee
Since for all $\alpha\in\Gamma$,
we have assumed that $\langle\alpha,\alpha\rangle\in\Z$,
  equation \eqref{Sasdef} ensures 
$S(\alpha,\alpha)=1$, as required by \eqref{conditionsonS}.

If the lattice $\Gamma$ is {\em integral}, i.e.\ if the
bilinear form $\langle\cdot,\cdot\rangle$ takes values in $\Z$ only, then
by the above, there is a unique equivalence class of $2$-cocycles with
values in $Z$ that obeys the additional condition
\eqref{cocycleswap}. We now show how this statement
must be modified for certain half integral lattices.

\noindent
From now on,
we restrict our attention to lattices where
 the following holds:
the lattice 
$$
\Gamma_0:=\Gamma^\ast
=\left\{ \alpha\in\R^D \mid \langle\alpha,\beta\rangle\in\Z \;\forall \beta\in\Gamma \right\}
$$ 
is an even sublattice $\Gamma_0\subset\Gamma$ 
of index $4$, such that for each of the cosets 
$\Gamma_{a}$ in $\Gamma/\Gamma_0$, 
$a\in\{0,\,\ldots,\,3\}$,
the lattice
$\Gamma_0\cup\Gamma_{a}$ is  integral  (and thus
self-dual if $a\neq0$). We write 
$\Gamma_{a}=\gamma^{(a)}+\Gamma_0$, $a\in\{0,\,\ldots,\,3\}$,
with $\gamma^{(0)}:=0$.
In other words, $(\Gamma_0,\Gamma)$ is a $\Z_2$
lattice pair with $\Gamma/\Gamma_0\cong\Z_2^2$ in
the terminology of \cite{gnos86,gnors87}.
\medskip\par

We claim that $\Gamma$ possesses precisely two distinct equivalence classes
of $2$-cocycles with values in $Z$ that obey \eqref{cocycleswap}. Indeed,
by our assumptions on $\Gamma$, 
we have 
$\langle\alpha,\beta\rangle\in{1\over2}\Z$ for all $\alpha,\,\beta\in\Gamma$,
hence
given such a $2$-cocycle $\eps$ with 
symmetry factor $S$, 
 by \eqref{conditionsonS} 
and \eqref{Sasdef},
\be\label{Saschoice}
S(\alpha,\beta) = \qu{S(\beta,\alpha)} \in\{\pm i\} \qquad
\mbox{ if } \langle\alpha,\beta\rangle\in{1\over2}+\Z.
\ee
Moreover, by the assumptions on our lattice $\Gamma$, we can
choose generators $\alpha_1,\,\ldots,\,\alpha_D$ such that
$$
\alpha_1=\gamma^{(1)},\;
\alpha_2=\gamma^{(2)},\quad
\alpha_j \in \Gamma_0\quad\forall j\in\{3,\,\dots,\, D\}.
$$
Then $\langle\alpha_a,\alpha_j\rangle\in\Z$
for $a\in\{1,\,2\}$ and  all $j\geq3$, since $\Gamma_0\cup\Gamma_a$ is an 
integral lattice by assumption. Moreover, $\langle\alpha_1,\alpha_2\rangle\in{1\over2}+\Z$,
since our assumptions imply that $\Gamma$ is {\em not} an integral lattice.
In fact, by replacing $\alpha_2$ by $-\alpha_2$ if need be, we may assume that
\be\label{alpha12parity}
\langle\alpha_1,\alpha_2\rangle\in{1\over2} +2\Z.
\ee
Now $S(\alpha_j,\alpha_k)$ is uniquely determined by
\eqref{Sasdef} for all 
$(j,k)\notin\left\{(1,2), (2,1)\right\}$, and $S(\alpha_1,\alpha_2)=\qu{S(\alpha_2,\alpha_1)}
=\pm i$ by \eqref{Saschoice}. It hence follows from \eqref{Sfromgenerators} 
that there are at most
two distinct solutions for the symmetry factor $S$, and thus, there are
at most two inequivalent $2$-cocycles on $\Gamma$ that obey the additional
condition \eqref{cocycleswap}. 

Note however that the existence of any such $2$-cocycles 
is not immediate. Indeed, if $\alpha,\,\alpha^\prime,\beta\in\Gamma$
are such that $S(\alpha,\beta)$ and $S(\alpha^\prime,\beta)$ obey
\eqref{Sasdef} and \eqref{Saschoice}, then \eqref{conditionsonS}
implies $S(\alpha+\alpha^\prime,\beta) = S(\alpha,\beta)(\alpha^\prime,\beta)$,
and one needs to check that $S(\alpha+\alpha^\prime,\beta)$
obeys \eqref{Sasdef} and \eqref{Saschoice} as well. This is immediate
if $\langle\alpha+\alpha^\prime,\beta\rangle\in {1\over2}+\Z$.
If on the other hand, $\langle\alpha,\beta\rangle,\,\langle\alpha^\prime,\beta\rangle \in \Z$,
then one proves that $\langle\alpha,\alpha^\prime\rangle \langle\beta,\beta\rangle\in\Z$
by showing that either $\langle\beta,\beta\rangle\in2\Z$ or 
$\alpha,\,\alpha^\prime,\,\beta\in \Gamma_0\cup\Gamma_a$ for some
$a\in\{1,\,2,\,3\}$. From this, the claim  
$$
S(\alpha+\alpha^\prime,\beta)
=(-1)^{\langle\alpha+\alpha^\prime,\beta\rangle 
+ \langle\alpha+\alpha^\prime,\alpha+\alpha^\prime\rangle
\langle\beta,\beta\rangle},
$$ 
as required by \eqref{Sasdef}, follows.
Finally, if $\langle\alpha,\beta\rangle,\,\langle\alpha^\prime,\beta\rangle \in {1\over2}+\Z$,
by what was already shown, one may assume without loss of generality that
$\alpha,\,\alpha^\prime,\,\beta\in\mbox{span}_\Z\left\{\alpha_1,\,\alpha_2\right\}$.
One then checks that \eqref{Sasdef} holds for $S(\alpha+\alpha^\prime,\beta)$
by a direct calculation for both choices $S(\alpha_1,\alpha_2)=i$
and $S(\alpha_1,\alpha_2)=-i$. Since as explained above, symmetry factors $S$
are in $1\colon1$ correspondence to equivalence classes of $2$-cocycles on 
$\Gamma$, the claim follows.
\medskip\par

Given a choice of symmetry factor $S$ that obeys \eqref{Sasdef},
for any lattice $\Gamma$ of rank $D$
with symmetric bilinear form $\langle\cdot,\cdot\rangle$
and generators $\alpha_1,\,\ldots,\,\alpha_D$,
following \cite{gnos86,gnors87,ka96}, we obtain $2$-cocycles
\be\label{epsfromgenerators}
\mbox{for } \alpha=\sum_{j=1}^D a_j\alpha_j, \; \beta=\sum_{k=1}^D b_k\alpha_k\in\Gamma\colon\qquad
\eps(\alpha,\beta) := \prod_{j,k=1, j>k}^D S(\alpha_j,\alpha_k)^{a_j b_k}
\ee
that obey the additional condition \eqref{cocycleswap}. 
We  introduce a $D\times D$ matrix $M=(M_{jk})$ 
such that 
$$
\forall j,\,k\in\left\{1,\,\ldots,\,D\right\}\colon\qquad
\exp\left( \pi i M_{jk}\right)
=\left\{
\begin{array}{ll}
S(\alpha_j,\alpha_k)
 & \mbox{ if } j>k,\\
1 & \mbox{ if } j\leq k,
\end{array}
\right. 
$$
then with notations as in \eqref{epsfromgenerators}, we obtain 
\be\label{epsfromM}
{\boldsymbol a}:=(a_1,\,\ldots,\,a_D)^T, \; 
{\boldsymbol b}:=(b_1,\,\ldots,\,b_D)^T,\qquad
\eps(\alpha,\beta) = \exp\left(  \pi i {\boldsymbol a}^T M {\boldsymbol b}\right) .
\ee
The $2$-cocycles thus inherit the {\em bimultiplicative}
behaviour from $S$,
\be\label{epsbimulti}
\forall \alpha,\,\alpha^\prime,\, \beta,\, \beta^\prime \in\Gamma\colon\qquad
\eps(\alpha+\alpha^\prime,\beta) = \eps(\alpha,\beta) \eps(\alpha^\prime,\beta) ,\quad
\eps(\alpha,\beta+\beta^\prime) = \eps(\alpha,\beta) \eps(\alpha,\beta^\prime) .
\ee
In our applications, however, we require a special gauge for our choice 
of representative $\widetilde\eps$ in an equivalence class of $2$-cocycles as above,
\be\label{specialgauge}
\forall\alpha\in\Gamma,\,\beta\colon\quad
\widetilde\eps(\alpha,0)=\widetilde\eps(0,\alpha)=\widetilde\eps(\alpha,-\alpha)=1,
\qquad
\widetilde\eps(\alpha,\beta) =\qu{\widetilde\eps(-\beta,-\alpha) }.
\ee
As is explained in \cite{gnos86,gnors87,ka96}, this condition can
always be met. Indeed, given the
representative $\eps$ of our $2$-cocycle constructed in \eqref{epsfromgenerators}, 
any choice of map
$$
\eta\colon\Gamma\longrightarrow Z,\quad \alpha\mapsto\eta_\alpha
\quad\mbox{ with } \eta_0=1, \quad 
\forall \alpha\in\Gamma\colon\;\eta_\alpha\eta_{-\alpha}=\eps(\alpha,\alpha)
$$
and some $Z\subset\C^\ast$ as above
yields a representative $\widetilde\eps$ under the gauge transformation
\eqref{gaugetransfo} which obeys \eqref{specialgauge}. Using \eqref{epsfromM},
and for all $\alpha\in\Gamma$ and $\boldsymbol a\in\R^D$ as in
\eqref{epsfromgenerators} and \eqref{epsfromM},
the special choice
$$
\eta_\alpha := \exp\left( {\textstyle{\pi i\over2}}{\boldsymbol a}^T {M} {\boldsymbol a}\right)
$$
yields representatives $\widetilde\eps$ that obey the special gauge
\eqref{specialgauge} and inherit the bimultiplicativity property
\eqref{epsbimulti}. Indeed,
with the notations of \eqref{epsfromgenerators} and \eqref{epsfromM},
\be\label{extrareal}
\widetilde\eps(\alpha,\beta) 
= \exp\left( {\textstyle{\pi i\over2}}{\boldsymbol a}^T (M-M^T) {\boldsymbol b}\right).
\ee
At this point we would like to emphasize that our entire analysis is
independent of the signature of the lattice $\Gamma$.
The explicit formulas, of course, are dependent on the details of each
lattice $\Gamma$.

It is also worth mentioning that although we proved the existence and gave a construction of two inequivalent sets of cocyles for our special K3 theory earlier in this appendix, the choice one makes in the context of the present work is not crucial. Actually,
there are two inequivalent choices of cocycles after reflecting the K3 theory too, leading to two 
super vertex operator algebras with admissible
module and the cocycles can be fixed consistently before and after reflection. 
On the level of the structure of a super vertex algebra with admissible module
alone, one cannot distinguish between the two choices,
as we mention in Section \ref{flipgtvw}.
%
%
\section{Theta function identities}\label{theta}
In this Appendix, we fix our conventions for the various modular functions that we shall use,
and we introduce some helpful identities. 

We shall always use the parametrisation
$q:=e^{2\pi i\tau}$ and $y:=e^{2\pi iz}$, sometimes along with
$\widetilde y:=e^{2\pi i\zeta}$. The Dedekind eta function is defined as 
$$
\eta(\tau):=q^{1/24}\prod_{n=1}^\infty\left(1-q^n\right) \ ,
$$
while the Jacobi theta functions have product formula presentations of the form
\begin{align}\label{anh_th_deffkt}
\vartheta_1(\tau,z) & := 
i\sum_{n=-\infty}^\infty (-1)^n q^{{1\over2}(n-{1\over2})^2} y^{n-{1\over2}}
&=& iq^{{1\over8}} y^{-{1\over2}} \prod_{n=1}^\infty (1-q^n)(1-q^{n-1}y)(1-q^{n}y^{-1}),\nonumber\\
\vartheta_2(\tau,z) & := 
\sum_{n=-\infty}^\infty q^{{1\over2}(n-{1\over2})^2} y^{n-{1\over2}}
&=&q^{{1\over8}} y^{-{1\over2}} \prod_{n=1}^\infty (1-q^n)(1+q^{n-1}y)(1+q^{n}y^{-1}),\nonumber\\
\vartheta_3(\tau,z) & := 
\sum_{n=-\infty}^\infty q^{{n^2\over2}} y^{n}
&=&\prod_{n=1}^\infty (1-q^n)(1+q^{n-{1\over2}}y)(1+q^{n-{1\over2}}y^{-1}),
\\
\vartheta_4(\tau,z) & := 
\sum_{n=-\infty}^\infty (-1)^n q^{{n^2\over2}} y^{n}
&=&\prod_{n=1}^\infty (1-q^n)(1-q^{n-{1\over2}}y)(1-q^{n-{1\over2}}y^{-1}).    \nonumber
\end{align}
We always use the shorthand $\vartheta_k(\tau):=\vartheta_k(\tau,0)$, $k=1,\ldots,4$.

The following identities can be proved using the Jacobi 
triple  identity, and they are standard:
\begin{eqnarray}
\vartheta_2(\tau) \vartheta_3(\tau) \vartheta_4(\tau) &=& 2\eta(\tau)^3,\label{triple}\\
\vartheta_2(\tau)^4- \vartheta_3(\tau)^4+ \vartheta_4(\tau)^4 &=& 0,\label{strange}\\
\vartheta_2(\tau)^2 &=& 2 \vartheta_2(2\tau) \vartheta_3(2\tau) ,\label{standardtdouble}\\
\vartheta_3(\tau)^2 &=& \vartheta_3(2\tau)^2+ \vartheta_2(2\tau)^2 ,\label{standardpdouble}\\
\vartheta_4(\tau)^2 &=& \vartheta_3(2\tau)^2- \vartheta_2(2\tau)^2 .\label{standardmdouble}
\end{eqnarray}
Using standard theta function techniques, one finds the following
generalizations of \eqref{strange}, see e.g.\ \cite{to66} or \cite[(A3.1)]{diss}:
\begin{equation}\label{oldstrange}
\begin{array}{rcl}
\vartheta_2(\tau)^2 \vartheta_2(\tau,z)^2 -  \vartheta_3(\tau)^2 \vartheta_3(\tau,z)^2 +\vartheta_4(\tau)^2 \vartheta_4(\tau,z)^2 
&=& 0,\\[3pt]
\vartheta_2(\tau)^2 \vartheta_1(\tau,z)^2 +  \vartheta_4(\tau)^2 \vartheta_3(\tau,z)^2 -\vartheta_3(\tau)^2 \vartheta_4(\tau,z)^2 
&=& 0.
\end{array}
\end{equation}
Moreover, by \cite{to66} or \cite[(A3.3),(A3.5),(A3.6)]{diss}, we have
\begin{equation}\label{olddouble}
\begin{array}{rcl}
2\vartheta_2(2\tau,z)\vartheta_3(2\tau,z)
&=& \vartheta_2(\tau) \vartheta_2(\tau,z) , \\[5pt]
\vartheta_3(\tau)^2\vartheta_2(2\tau,2z)
&=& \vartheta_2(2\tau) \vartheta_3(\tau,z)^2 
- \vartheta_3(2\tau) \vartheta_1(\tau,z)^2, \\[5pt]
\vartheta_3(\tau)^2\vartheta_3(2\tau,2z)
&=& \vartheta_3(2\tau) \vartheta_3(\tau,z)^2 
+ \vartheta_2(2\tau) \vartheta_1(\tau,z)^2,
\end{array}
\end{equation}
and
\begin{equation}\label{olddoubleproduct}
\begin{array}{rcl}
\vartheta_3(2\tau,2z)\vartheta_3(2\tau,2\zeta)
+ \vartheta_2(2\tau,2z)\vartheta_2(2\tau,2\zeta) &=& \vartheta_3(\tau,z+\zeta)\vartheta_3(\tau,z-\zeta),\\[3pt]
\vartheta_3(2\tau,2z)\vartheta_3(2\tau,2\zeta)
- \vartheta_2(2\tau,2z)\vartheta_2(2\tau,2\zeta) &=& \vartheta_4(\tau,z+\zeta)\vartheta_4(\tau,z-\zeta).
\end{array}
\end{equation}
We deduce
\begin{eqnarray*}
\textstyle
&& \left( \vartheta_2(\tau,z)^2 \vartheta_2(\tau,\zeta)^2 -  \vartheta_1(\tau, z)^2 \vartheta_1(\tau,\zeta)^2\right)
\vartheta_2(\tau)^4\\
&&\quad
\begin{array}{cl}
\stackrel{\eqref{oldstrange}}{=} 
& \left( \vartheta_3(\tau)^2\vartheta_3(\tau,z)^2  -  \vartheta_4(\tau)^2 \vartheta_4(\tau,z)^2\right)
\left( \vartheta_3(\tau)^2\vartheta_3(\tau,\zeta)^2  -  \vartheta_4(\tau)^2 \vartheta_4(\tau,\zeta)^2\right)\\[3pt]
&\quad
-\left( \vartheta_3(\tau)^2\vartheta_4(\tau,z)^2  -  \vartheta_4(\tau)^2 \vartheta_3(\tau,z)^2\right)
\left( \vartheta_3(\tau)^2\vartheta_4(\tau,\zeta)^2  -  \vartheta_4(\tau)^2 \vartheta_3(\tau,\zeta)^2\right)\\[5pt]
=& \vartheta_3(\tau)^4\vartheta_3(\tau,z)^2 \vartheta_3(\tau,\zeta)^2  
+ \vartheta_4(\tau)^4\vartheta_4(\tau,z)^2 \vartheta_4(\tau,\zeta)^2  \\[3pt]
&\quad - \vartheta_3(\tau)^4\vartheta_4(\tau,z)^2 \vartheta_4(\tau,\zeta)^2  
- \vartheta_4(\tau)^4\vartheta_3(\tau,z)^2 \vartheta_3(\tau,\zeta)^2  
\\[5pt]
=&
\left( \vartheta_3(\tau,z)^2 \vartheta_3(\tau,\zeta)^2  - \vartheta_4(\tau,z)^2 \vartheta_4(\tau,\zeta)^2  \right)
 \left(\vartheta_3(\tau)^4 - \vartheta_4(\tau)^4\right)
\\[3pt]
\stackrel{\eqref{strange}}{=} 
&\left( \vartheta_3(\tau,z)^2 \vartheta_3(\tau,\zeta)^2 -  \vartheta_4(\tau, z)^2 \vartheta_4(\tau,\zeta)^2\right)
\vartheta_2(\tau)^4,
\end{array}
\end{eqnarray*}
implying the following very useful generalization of
\eqref{strange},
\begin{eqnarray}\label{newstrange}
\vartheta_2(\tau,z)^2 \vartheta_2(\tau,\zeta)^2 -  \vartheta_1(\tau, z)^2 \vartheta_1(\tau,\zeta)^2\nonumber\\
&&\hspace*{-8em}= \vartheta_3(\tau,z)^2 \vartheta_3(\tau,\zeta)^2 -  \vartheta_4(\tau, z)^2 \vartheta_4(\tau,\zeta)^2.
\end{eqnarray}
Moreover,
\begin{eqnarray*}
\textstyle
\vartheta_3(\tau,z+\zeta) \vartheta_3(\tau,z-\zeta)\vartheta_3(\tau)^4\\
&&\hspace*{-10em}
\stackrel{\eqref{olddouble},\eqref{olddoubleproduct}}{=} 
 \left( \vartheta_2(2\tau)^2+ \vartheta_3(2\tau)^2 \right)
\left( \vartheta_3(\tau,z)^2\vartheta_3(\tau,\zeta)^2  +  \vartheta_1(\tau,z)^2 \vartheta_1(\tau,\zeta)^2\right),
\end{eqnarray*}
from which by \eqref{standardpdouble} we find
\begin{equation}\label{theta3twist}
\vartheta_3(\tau,z+\zeta) \vartheta_3(\tau,z-\zeta)\vartheta_3(\tau)^2
=   \vartheta_3(\tau,z)^2 \vartheta_3(\tau,\zeta)^2 +  \vartheta_1(\tau, z)^2 \vartheta_1(\tau,\zeta)^2.
\end{equation}
Similarly,
\begin{eqnarray*}
\textstyle
&& \vartheta_4(\tau,z+\zeta) \vartheta_4(\tau,z-\zeta)\vartheta_3(\tau)^4\vartheta_4(\tau)^2\\
&&\quad
\begin{array}{cl}
\stackrel{\eqref{olddouble},\eqref{olddoubleproduct}}{=} 
& \left( \vartheta_3(2\tau)^2- \vartheta_2(2\tau)^2 \right)
\left( \vartheta_3(\tau,z)^2\vartheta_3(\tau,\zeta)^2  -  \vartheta_1(\tau,z)^2 \vartheta_1(\tau,\zeta)^2\right)
\vartheta_4(\tau)^2
\\[3pt]
&\quad
+ 2 \vartheta_2(2\tau)\vartheta_3(2\tau)
\left( \vartheta_3(\tau,z)^2\vartheta_1(\tau,\zeta)^2  +  \vartheta_1(\tau,z)^2 \vartheta_3(\tau,\zeta)^2\right)
\vartheta_4(\tau)^2
\\[5pt]
\stackrel{\stackrel{\scriptstyle\eqref{standardtdouble},\eqref{standardmdouble},}{\eqref{oldstrange}}}{=} 
& 
\left( \vartheta_3(\tau,z)^2\vartheta_3(\tau,\zeta)^2  -  \vartheta_1(\tau,z)^2 \vartheta_1(\tau,\zeta)^2\right)
\vartheta_4(\tau)^4 \\[3pt]
&\quad 
+ \vartheta_3(\tau,z)^2 \left( \vartheta_3(\tau)^2\vartheta_4(\tau,\zeta)^2  -  \vartheta_4(\tau)^2 \vartheta_3(\tau,\zeta)^2\right)
\vartheta_4(\tau)^2\\[3pt]
&\quad 
+ \vartheta_1(\tau,z)^2\left( \vartheta_3(\tau)^2\vartheta_4(\tau,\zeta)^2  -  \vartheta_2(\tau)^2 \vartheta_1(\tau,\zeta)^2\right)
\vartheta_2(\tau)^2
\\[5pt]
= &
- \vartheta_1(\tau,z)^2 \vartheta_1(\tau,\zeta)^2
\left( \vartheta_4(\tau)^4+\vartheta_2(\tau)^4  \right)
 \\[3pt]
&\quad 
+ \vartheta_4(\tau,\zeta)^2 \left( \vartheta_4(\tau)^2\vartheta_3(\tau,z)^2  +  \vartheta_2(\tau)^2 \vartheta_1(\tau,z)^2\right)
\vartheta_3(\tau)^2
\\[5pt]
\stackrel{\eqref{strange},\eqref{oldstrange}}{=} 
&\left( \vartheta_4(\tau,z)^2 \vartheta_4(\tau,\zeta)^2  - \vartheta_1(\tau,z)^2 \vartheta_1(\tau,\zeta)^2  \right)
 \vartheta_3(\tau)^4 ,
\end{array}
\end{eqnarray*}
from which by \eqref{newstrange} we find
\begin{eqnarray}\label{theta4twist}
\vartheta_4(\tau,z+\zeta) \vartheta_4(\tau,z-\zeta)\vartheta_4(\tau)^2
&=&   \vartheta_4(\tau,z)^2 \vartheta_4(\tau,\zeta)^2 -  \vartheta_1(\tau, z)^2 \vartheta_1(\tau,\zeta)^2\nonumber\\
&=&   \vartheta_3(\tau,z)^2 \vartheta_3(\tau,\zeta)^2 -  \vartheta_2(\tau, z)^2 \vartheta_2(\tau,\zeta)^2.
\end{eqnarray}
From \eqref{theta3twist} and \eqref{theta4twist} we obtain
\begin{eqnarray}\label{theta34ptwist}
\vartheta_3(\tau,z+\zeta) \vartheta_3(\tau,z-\zeta)\vartheta_3(\tau)^2
&+& \vartheta_4(\tau,z+\zeta) \vartheta_4(\tau,z-\zeta)\vartheta_4(\tau)^2\nonumber\\
&=& \vartheta_3(\tau,z)^2 \vartheta_3(\tau,\zeta)^2 +  \vartheta_4(\tau, z)^2 \vartheta_4(\tau,\zeta)^2,
\end{eqnarray}
\begin{eqnarray}\label{theta34mtwist}
\vartheta_3(\tau,z+\zeta) \vartheta_3(\tau,z-\zeta)\vartheta_3(\tau)^2
&-& \vartheta_4(\tau,z+\zeta) \vartheta_4(\tau,z-\zeta)\vartheta_4(\tau)^2\nonumber\\
&=& \vartheta_2(\tau,z)^2 \vartheta_2(\tau,\zeta)^2 +  \vartheta_1(\tau, z)^2 \vartheta_1(\tau,\zeta)^2.
\end{eqnarray}
Again similary,
\begin{eqnarray}\label{theta2twist}
\textstyle
&& \vartheta_2(\tau,z+\zeta) \vartheta_2(\tau,z-\zeta)\vartheta_2(\tau)^2\nonumber\\
&&\quad
\begin{array}{cl}
\stackrel{\eqref{olddouble}}{=} 
& 
4 \cdot \vartheta_2(2\tau,z+\zeta) \vartheta_3(2\tau,z+\zeta)\vartheta_2(2\tau,z-\zeta)\vartheta_3(2\tau,z-\zeta)\\[5pt]
=& \left( \vartheta_3(2\tau,z+\zeta)\vartheta_3(2\tau,z-\zeta) + \vartheta_2(2\tau,z+\zeta)\vartheta_2(2\tau,z-\zeta) \right)^2
\\[3pt]
&\quad - \left( \vartheta_3(2\tau,z+\zeta)\vartheta_3(2\tau,z-\zeta) - \vartheta_2(2\tau,z+\zeta)\vartheta_2(2\tau,z-\zeta) \right)^2
\\[5pt]
\stackrel{\eqref{olddoubleproduct}}{=} &
\vartheta_3(\tau,z)^2 \vartheta_3(\tau,\zeta)^2 -  \vartheta_4(\tau, z)^2 \vartheta_4(\tau,\zeta)^2.
\end{array}
\end{eqnarray}
%
%
%
\newcommand{\etalchar}[1]{$^{#1}$}
\def\polhk#1{\setbox0=\hbox{#1}{\ooalign{\hidewidth
  \lower1.5ex\hbox{`}\hidewidth\crcr\unhbox0}}} \def\cprime{$^\prime$}
  \newcommand{\noopsort}[1]{}
\providecommand{\bysame}{\leavevmode\hbox to3em{\hrulefill}\thinspace}

\end{document}